\documentclass[twocolumn,amsmath,amssymb,nofootinbib,a4paper]{revtex4} 

\usepackage[utf8]{inputenc}
\usepackage{graphicx} 
\usepackage{amsmath}
\usepackage{amsfonts,amsbsy}
\usepackage{amssymb}
\usepackage{subdepth}

\usepackage{bm,pbox}
\usepackage{bbm}

\newcommand{\diagram}[2][{}]{\pbox{\textwidth}{\includegraphics[#1]{{#2}}}}
\newcommand{\tr}{{\text{tr}}}    

\usepackage{adjustbox}
\usepackage{tkz-linknodes}
\usepackage{tikz}
\usetikzlibrary{matrix,fit,arrows,decorations}
\usetikzlibrary{arrows}
\usetikzlibrary{shapes.symbols,positioning} 
\usetikzlibrary{calc}
\usetikzlibrary{shadows}
\usepackage{todonotes}

\usepackage{xcolor}
\definecolor{lcolor}{rgb}{0.5,0,0}
\definecolor{citcolor}{rgb}{0,0.3,0.0}
\usepackage[breaklinks,colorlinks,urlcolor=blue,citecolor=citcolor,linkcolor=lcolor]{hyperref}

\newcommand{\rt}{{\ib{r}}}
\newcommand{\rtp}{{\ib{r}'}}
\newcommand{\rtpp}{{\ib{r}''}}
\newcommand{\ut}{{\ib{u}}}
\newcommand{\vt}{{\ib{v}}}

\newcommand{\xt}{{\ib{x}}}
\newcommand{\zt}{{\ib{z}}}
\newcommand{\yt}{{\ib{y}}}

\newcommand{\as}{\alpha_{\mathrm{s}}}
\newcommand{\qs}{Q_\mathrm{s}}
\newcommand{\rs}{R_\mathrm{s}}
\newcommand{\ud}{\, \mathrm{d}}
\newcommand{\nc}{{N_\mathrm{c}}}
\newcommand{\cf}{{C_\mathrm{F}}}
\newcommand{\cd}{{C_\mathrm{d}}}

\DeclareMathAlphabet{\ib}{OML}{cmm}{b}{it}
\newcommand{\dd}[2]{i \nabla^{{#1}}_\ib{{#2}}}

\newcommand{\uu}[1]{U_\ib{{#1}}}
\newcommand{\uud}[1]{U^\dagger_\ib{{#1}}}

\newcommand{\nr}[1]{(\ref{#1})}

\newcommand{\eq}{Eq.~}

\newcommand{\eqs}{Eqs.~}

\begin{document}

\title{JIMWLK evolution of the odderon}

\author{T. Lappi}
\email{tuomas.v.v.lappi@jyu.fi}
\affiliation{
Department of Physics, %
 P.O. Box 35, 40014 University of Jyv\"askyl\"a, Finland
}
\affiliation{
Helsinki Institute of Physics, P.O. Box 64, 00014 University of Helsinki,
Finland
}

\author{A. Ramnath}
\email{anramnat@student.jyu.fi}
\affiliation{
Department of Physics, %
 P.O. Box 35, 40014 University of Jyv\"askyl\"a, Finland
}

\author{K. Rummukainen}
\email{kari.rummukainen@helsinki.fi}

\affiliation{Department of Physics and Helsinki Institute of Physics,
  University of Helsinki, 00014 University of Helsinki, Finland}

\author{H. Weigert} \email{heribert.weigert@uct.ac.za} \affiliation{
  University of Cape Town; Dept. of Physics, Private Bag X3,
  Rondebosch 7701, South Africa }

\begin{abstract}
  We study the effects of a parity-odd ``odderon'' correlation in
  JIMWLK renormalization group evolution at high energy.  Firstly we
  show that in the eikonal picture where the scattering is described
  by Wilson lines, one obtains a strict mathematical upper limit for
  the magnitude of the odderon amplitude compared to the parity even
  pomeron one. This limit increases with $\nc$, approaching infinity
  in the infinite~$\nc$ limit. We use a systematic extension of the
  Gaussian approximation including both 2- and 3-point
  correlations which enables us to close the system of equations even
  at finite-$\nc$. In the large-$\nc$ limit we recover an evolution
  equation derived earlier. By solving this equation numerically we
  confirm that the odderon amplitude decreases faster in the nonlinear
  case than in the linear BFKL limit. We also point out that, in the
  3-point truncation at finite~$\nc$, the presence of an odderon
  component introduces azimuthal angular correlations $\sim \cos (
  n\varphi)$ at all $n$.
\end{abstract}

\maketitle

\section{Introduction}

High energy hadronic collisions at modern collider energies involve a
dense system of gluons. At high enough energy the typical phase space
density becomes nonperturbatively large, i.e. of the order of the
inverse QCD coupling constant $1/\as$. In this limit it is better to
describe these gluonic degrees of freedom as a classical color field
than as a collection of individual particles, in what is known as the
Color Glass Condensate (CGC)
picture~\cite{Weigert:2005us,Gelis:2010nm}. In practice, the important
degree of freedom here is the Wilson line, a path ordered exponential
in the color field. It gives the scattering amplitude of a colored
high energy particle passing through the CGC target. Increasing the
collision energy opens up phase space for the emission of even more
gluons, which in this case are treated as quantum fluctuations on top
of the classical field. These fluctuations can be systematically
integrated out and included in the classical field. This procedure
leads to renormalization group equations that describe the evolution
of the Wilson lines as a function of collision energy.

The complete system of evolution equations is known as the JIMWLK
equation~\cite{Jalilian-Marian:1997xn, Jalilian-Marian:1997jx,
  Jalilian-Marian:1997gr, Jalilian-Marian:1997dw,
  JalilianMarian:1998cb, Weigert:2000gi, Iancu:2000hn, Iancu:2001md,
  Ferreiro:2001qy, Iancu:2001ad, Mueller:2001uk}, or equivalently as
the Balitsky hierarchy~\cite{Balitsky:1995ub, Balitsky:1998kc,
  Balitsky:1998ya, Balitsky:2001re}. It describes the evolution of the
whole probability distribution of Wilson lines. While this equation
can be solved, at least at leading order,
numerically~\cite{Blaizot:2002xy, Rummukainen:2003ns, Dumitru:2011vk,
  Lappi:2012vw}, most phenomenological applications rely on simpler
approximations. This is typically done by an evolution equation for an
expectation value of Wilson lines that can be derived, in some
approximation, from the equation for the full probability
distribution. The usual approximation here is to use the large-$\nc$
limit, which allows one to truncate the Balitsky hierarchy and obtain
an evolution equation for the two-point function of Wilson lines known
as the Balitsky-Kovchegov~\cite{Balitsky:1995ub, Kovchegov:1999yj}
(BK) equation. A related approximation, which has an identical
dynamical content but can be used to construct the Wilson line
expectation values at finite $\nc$, is provided by the Gaussian
approximation.

The Gaussian approximation relates all Wilson line correlators to a
single two-point correlator. The purpose of this paper is to take the
first step beyond the Gaussian approximation and introduce an
intrinsic three-point correlation function of color charges using a
method that can be extended to include all $n$-point functions up to
any fixed finite number of points $m$ in what we refer to as an
(exponential) $m$-point truncation. When this is used to evaluate the
evolution equation for the two-point function of Wilson lines (the
dipole operator), it turns out that the new three-point function only
appears in a specific coordinate limit. It can, in fact, be rewritten
as an imaginary part of the earlier two-point function. Physically
this new degree of freedom corresponds to the odderon: an interaction
by the exchange of a parity-odd particle. Similar modifications to the
Gaussian average have been considered before in the context of the
McLerran-Venugopalan (MV~\cite{McLerran:1994ni, McLerran:1994ka,
  McLerran:1994vd}) model (see e.g.~\cite{Jeon:2005cf, Dumitru:2011zz,
  Dumitru:2011ax}). Here, we will go beyond the work in these papers
and derive evolution equations in rapidity for the odderon amplitude
in the exponential 3-point approximation, extending earlier
large-$\nc$ results \cite{Kovchegov:2003dm,Hatta:2005as} to finite
$\nc$. We will then numerically solve these evolution equations in a
truncation in the harmonic number in the azimuthal direction. 
To determine the consistency of truncated JIMWLK evolution we
complement our discussion with a numerical simulation of parity-even
correlations using full untruncated JIMWLK evolution in the Langevin
framework, reproducing the same qualitative behavior. 

This paper is
structured in the following way. First, in Sec.~\ref{sec:obs}, we
motivate this study by an example of a phenomenological context in
which the odderon amplitude appears directly. Then, in
Sec.~\ref{sec:group-theory-constr-corr}, we point out that the origin
of the dipole amplitude as a correlation function of Wilson lines that
live on the SU(3) group manifold places stringent mathematical bounds
on the size of the odderon.  We then quantify these bounds for
specific parametric forms of the initial conditions in
Sec.~\ref{sec:constr-init-cond}. On the same basis we argue that, in the
JIMWLK context, the odderon can not affect observables that do not
break rotational symmetry in the transverse plane.  Sec.~\ref{sec:gt}
presents the derivation of the evolution equations for the odderon
component from an exponential $n$-point truncation. We solve these
truncated equations in Sec.~\ref{sec:3pt} with a further approximation
to the lowest nontrivial $\cos n \theta$ azimuthal harmonic. Then in
Sec.~\ref{sec:jimwlk} we construct initial conditions for the JIMWLK
equation that include an odderon component and study its evolution in
a full (fixed coupling) JIMWLK simulation.

\section{Observables and cross sections}\label{sec:obs}

The simplest application of JIMWLK evolution is calculating the total
cross section in experiments like Deep Inelastic Scattering (DIS),
where a  space-like virtual photon is scattered on a nuclear
target. At small $x$ and leading order in perturbation theory, the cross
section is dominated by the $q\Bar q$ component of the photon wave
function, which interacts eikonally with the target. That is to say that
the interaction is driven by an average of the dipole operator
\begin{equation}\label{eq:defdipole}
\hat{D}_{\xt,\yt} = 
 \frac{1}{\nc}  \tr{ U_\xt U^\dag_\yt}.
\end{equation}
Diagrammatically the total cross section can be cast as
\begin{align}
  \label{eq:DIStot}
          \Biggl[ &
  \diagram[height=.9cm]{figs_final/qqbint-L}-\diagram[height=.9cm]{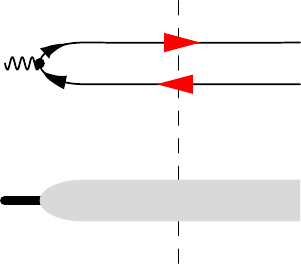}
  \Biggr] 
%  \diagram[height=.9cm]{vmwf-R}
%\,
%  \diagram[height=.9cm]{vmwf-L}
  \Biggl[
  \diagram[height=.9cm]{figs_final/qqbint-R}-\diagram[height=.9cm]{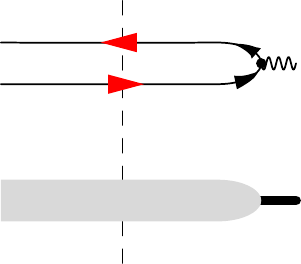}
  \Biggr]
 \\ & \notag
= 2\diagram[height=.9cm]{figs_final/qqbnoint-L}\diagram[height=.9cm]{figs_final/qqbnoint-R}
-
 \diagram[height=.9cm]{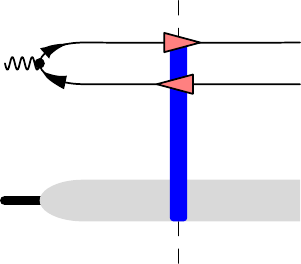}\diagram[height=.9cm]{figs_final/qqbnoint-R}
-\diagram[height=.9cm]{figs_final/qqbnoint-L}\diagram[height=.9cm]{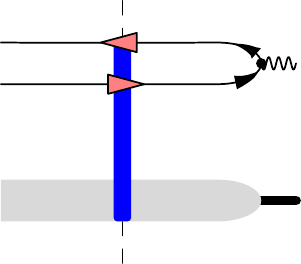}
\end{align}
and allows access only to the real part of the dipole correlator
contained in the last two terms of Eq.~\eqref{eq:DIStot}, since:
\begin{align}
  \label{eq:DIScorr}
 \langle \tr( U_{\bm x} U^\dagger_{\bm y}) \rangle(Y) 
 + 
 \langle \tr( U_{\bm x}^\dagger U_{\bm y})\rangle(Y) 
 = 
 2 \langle \text{\sffamily Re}\,  \tr( U_{\bm x} U^\dagger_{\bm y}) \rangle(Y)
 \ .
\end{align}
 As indicated, the average will depend on
$Y=(\ln1/x)$ with the $Y$ dependence governed by JIMWLK evolution.

The dipole operator \emph{does} give rise to imaginary parts in a
\emph{generic} average $\langle\hat{D}_{\xt,\yt}\rangle(Y)$, i.e. over
an ensemble not explicitly tailored to have a vanishing imaginary
part, but one needs more detailed experiments to access this
information. (We will argue in Sec.~\ref{sec:group-theory-constr-corr}
that this is in fact an absolute statement, at least within the JIMWLK
context.) The Single Transverse Spin Asymmetry (STSA) is such an
observable. Kovchegov and Sievert~\cite{Kovchegov:2013cva} have in
fact suggested a new mechanism to generate a contribution to STSA at
small $x$ that is triggered by this imaginary part. The contribution
suggested by Kovchegov and Sievert takes the diagrammatic form
\begin{align}
  \label{eq:stsa}
          \Biggl[ 
  \diagram[height=.9cm]{figs_final/q-qg-int-before-L} &
  +\diagram[height=.9cm]{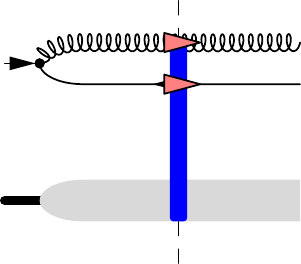}
  \Biggr] 
  \Biggl[
  \diagram[height=.9cm]{figs_final/q-qg-int-before-R}
  +\diagram[height=.9cm]{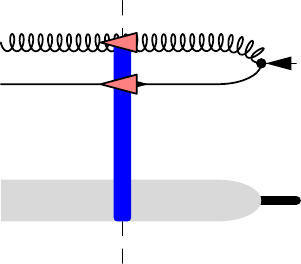}
  \Biggr]
  \\ \notag 
  & = 
  \diagram[height=.9cm]{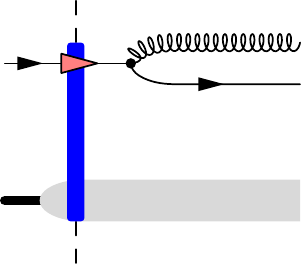}\diagram[height=.9cm]{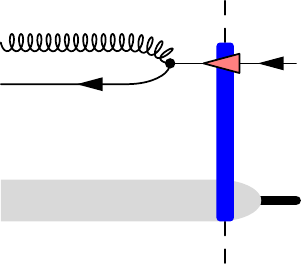}
+
\diagram[height=.9cm]{figs_final/q-qg-int-before-L}\diagram[height=.9cm]{figs_final/q-to-qg-int-R}
\\ \notag 
& +
\diagram[height=.9cm]{figs_final/q-to-qg-int-L}\diagram[height=.9cm]{figs_final/q-qg-int-before-R}
+
\diagram[height=.9cm]{figs_final/q-to-qg-int-L}\diagram[height=.9cm]{figs_final/q-to-qg-int-R}
\end{align}
where the incoming quark is taken (probabilistically) from the
incoming projectile.

In Eq.~\eqref{eq:stsa} the momentum and spin of the quark in the final
state are tagged and the color in the initial state is summed
over. Tagging the quark momentum leads to different coordinates on the
corresponding Wilson lines in amplitude and complex conjugate
amplitude. The gluon momentum is integrated over so that the gluon
Wilson line in the last term cancels between the two sides of the
cut. The color correlators from the r.h.s. of Eq.~(\ref{eq:stsa})
result in the following terms:
\begin{align}
\label{eq:stsaUcorr}
  \mathcal{I}^{(q)} = \biggl\langle &
    \frac{\tr(U_{\bm z}U^\dagger_{\bm y})}{\nc} 
-\frac1{d_A} \Tilde U^{a b}_{\bm x} 2\tr(t^a U_{\bm z} t^b U^\dagger_{\bm w}) 
\notag \\ & -\frac1{d_A} \Tilde U^{a b}_{\bm x} 2\tr(t^a U_{\bm u} t^b U^\dagger_{\bm y}) 
+ \frac{\tr(U_{\bm u} U^\dagger_{\bm w})}{\nc} 
\biggr\rangle . 
\end{align}
The STSA is driven by the contributions that are anti-symmetric under
exchange of the quark and anti-quark coordinates $\bm z
\leftrightarrow \bm y$ and thus the imaginary part of, for example,
the first term.

\section{Group theory constraints on the real and imaginary parts of
  $q\Bar q$ correlators}
\label{sec:group-theory-constr-corr}

Most readers familiar with JIMWLK and BK simulations will be prepared
to accept that the $q\Bar q$ and $q\Bar q g$ correlators in
Eq.~\eqref{eq:stsaUcorr}, with normalization factors included, are
\emph{real} and interpolate between $1$ at distances much smaller than
the inverse saturation scale and $0$ at pairwise separations much
larger than the inverse saturation scale. This behavior is indeed
respected by JIMWLK evolution in all its forms, provided it is
satisfied by the initial condition.

This situation changes if one allows imaginary parts to arise. We will
illustrate the situation with a discussion of the $q\Bar q$ dipole
correlator and its underlying configurations that appear explicitly in
a Langevin simulation of JIMWLK evolution. To this end, note that
these configurations appear as explicit $\mathsf{SU}(\nc)$ matrices
$U_{\bm x}$.\footnote{All arguments here assume that the coupling to
  the target is described fully through Wilson lines, i.e. the absence
  of sub-eikonal corrections, as is the case for all JIMWLK
  evolution.} This remains true for the products entering the $q\Bar
q$ correlators: $U_{\bm x} U_{\bm y}^\dagger\in \mathsf{SU}(\nc)$ is
unitary and therefore has $\nc$ eigenvalues of the form $e^{i
  \phi_i(\bm x,\bm y)}$, $i=1,\ldots, N_c \in\mathbbm{N}$.  All of them
are functions of both coordinates and live on the unit circle.  The
determinant condition $\det(U_{\bm x} U_{\bm y}^\dagger) = 1$ then
enforces that the phases of the eigenvalues sum to an integer multiple
of $2\pi$. Suppressing the coordinate dependence on the $\phi_i(\bm
x,\bm y)$ we have, for each pair of points:
\begin{align}
  \label{eq:sun-det-cond}
  1 = \det(U_{\bm
  x} U_{\bm y}^\dagger) = e^{i\sum\limits_{i=1}^{\nc} \phi_i}
  \Leftrightarrow \sum\limits_{i=1}^{\nc} \phi_i = 2\pi n; n\in \mathbb Z \ .
\end{align}
The trace of the dipole operator is therefore fully determined by
$\nc-1$ phases $\phi_i\in[0,2\pi[$~.

Using the constraint~\eqref{eq:sun-det-cond} to remove $\phi_{\nc}$,
one finds an expression for the trace of our group element $U_{\bm x}
U_{\bm y}^\dagger$ that reads
\begin{align}
  \label{eq:sun-trace-formula}
  \frac1{\nc}\tr(U_{\bm x} U_{\bm y}^\dagger)
  =
  \frac1{\nc}\Biggl(
  \sum\limits_{i=1}^{\nc-1} e^{i \phi_i} +  e^{- i\sum\limits_{i=1}^{\nc-1}\phi_i}
  \Biggr)
\end{align}
This trace falls into a simply connected closed subset of
the complex plane, bounded by the curve
\begin{align}
  \label{eq:trace-boundary}
  h_{\nc}(\theta) =
        \frac1{\nc}\Bigl((\nc-1) e^{i \theta} +  e^{- i(\nc-1)\theta}
  \Bigr)
\end{align}
where $\theta\in[0,2\pi[$~. (See~\cite{math-ph-0609082} for a recent
discussion of these textbook results.) Equation~\nr{eq:trace-boundary}
has a very simple geometric interpretation: The center of a small
circle (represented by the second term) is traveling along the
perimeter of a large circle (represented by the first
term).\footnote{This description is adapted from the formula --
  alternatively, one might describe the boundary as the curve traced
  by a point on a circle of radius $1/\nc$ that rolls on the inside of
  the unit circle, starting with both circles touching at $1$.} While
the large circle is traversed once in a counterclockwise direction,
the small traveling circle is traversed clockwise $\nc$ times. The
line traced out by $h_{\nc}(\theta)$ in this manner is called a
hypocycloid.  The curve is fully contained in the unit circle and, for
fixed $\nc$, has cusps at the $\nc$-th roots of unity -- these are the
only points where the curve touches the unit circle. These points
correspond to specific group elements that form the center of the
group $\{e^{i 2\pi n/\nc}\mathbbm 1|n\in\mathbbm Z/\nc\}$. In
Fig.~\ref{fig:hypocyloid-boundary-const} both the geometric origin and
the cusp structure are illustrated for a few values of $\nc$.  For
$\nc\to\infty$, the hypocycloid will approximate the unit circle. The
value $\nc =2$ does not allow for an imaginary part at all -- the
underlying reason is that the group is pseudo-real, i.e. $U^\dagger =
\epsilon U \epsilon$ ($\epsilon = [\epsilon^{ i j}]$) is isomorphic to
$U$. In this vein, $\nc = 3$ is the first case that allows an
imaginary part and has, at the same time, the strongest limitations on
its size from the group structure alone.  \newlength{\mylength}
\setlength{\mylength}{.25\textwidth-2.6em}
\begin{figure}[tb]
  \centering
  \begin{tikzpicture}
  \matrix(m)[matrix of math nodes,column sep=.6em]
   { \nc=3 & \nc = 4 \\ 
    \diagram[width=\mylength]{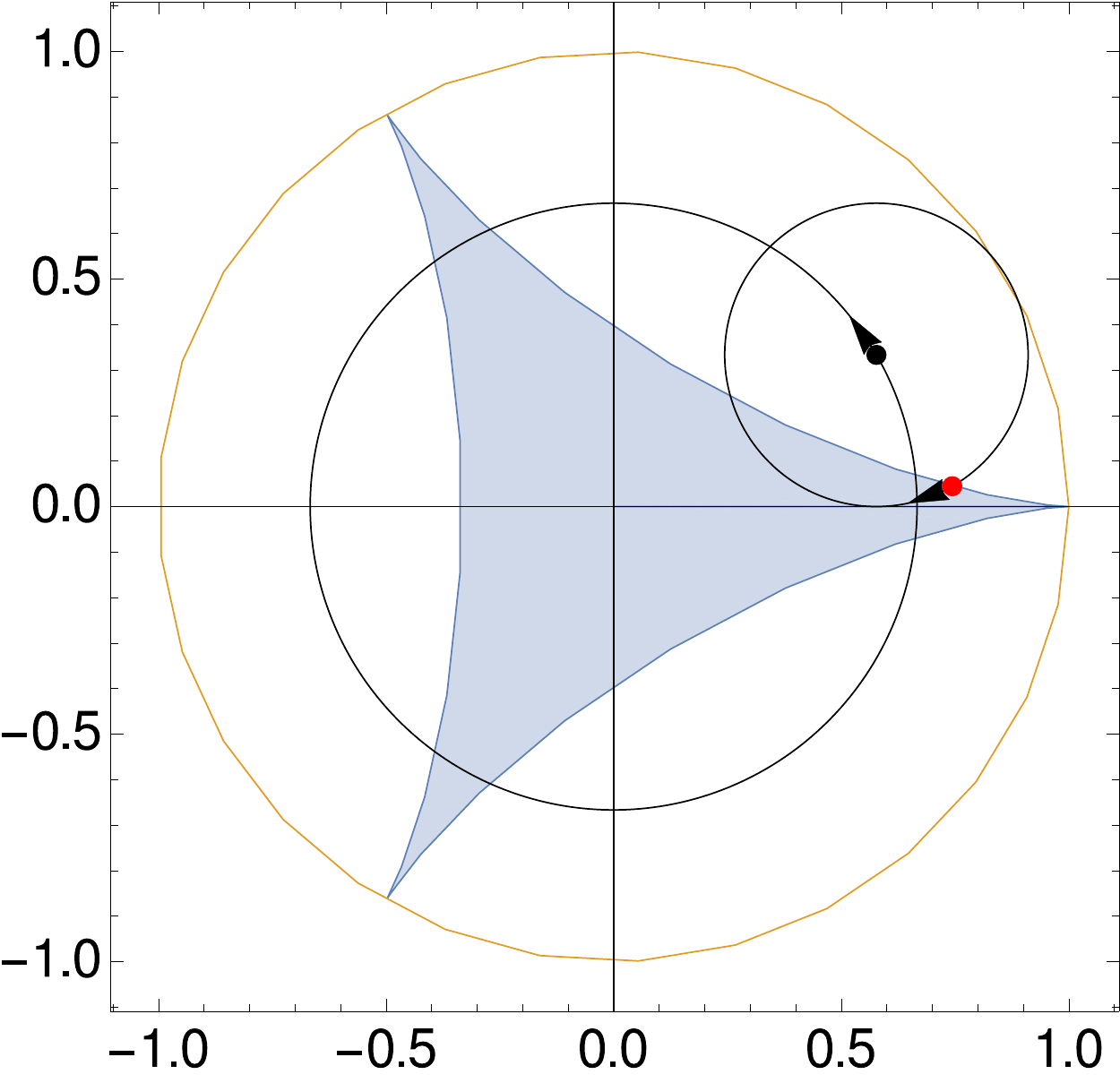} &
    \diagram[width=\mylength]{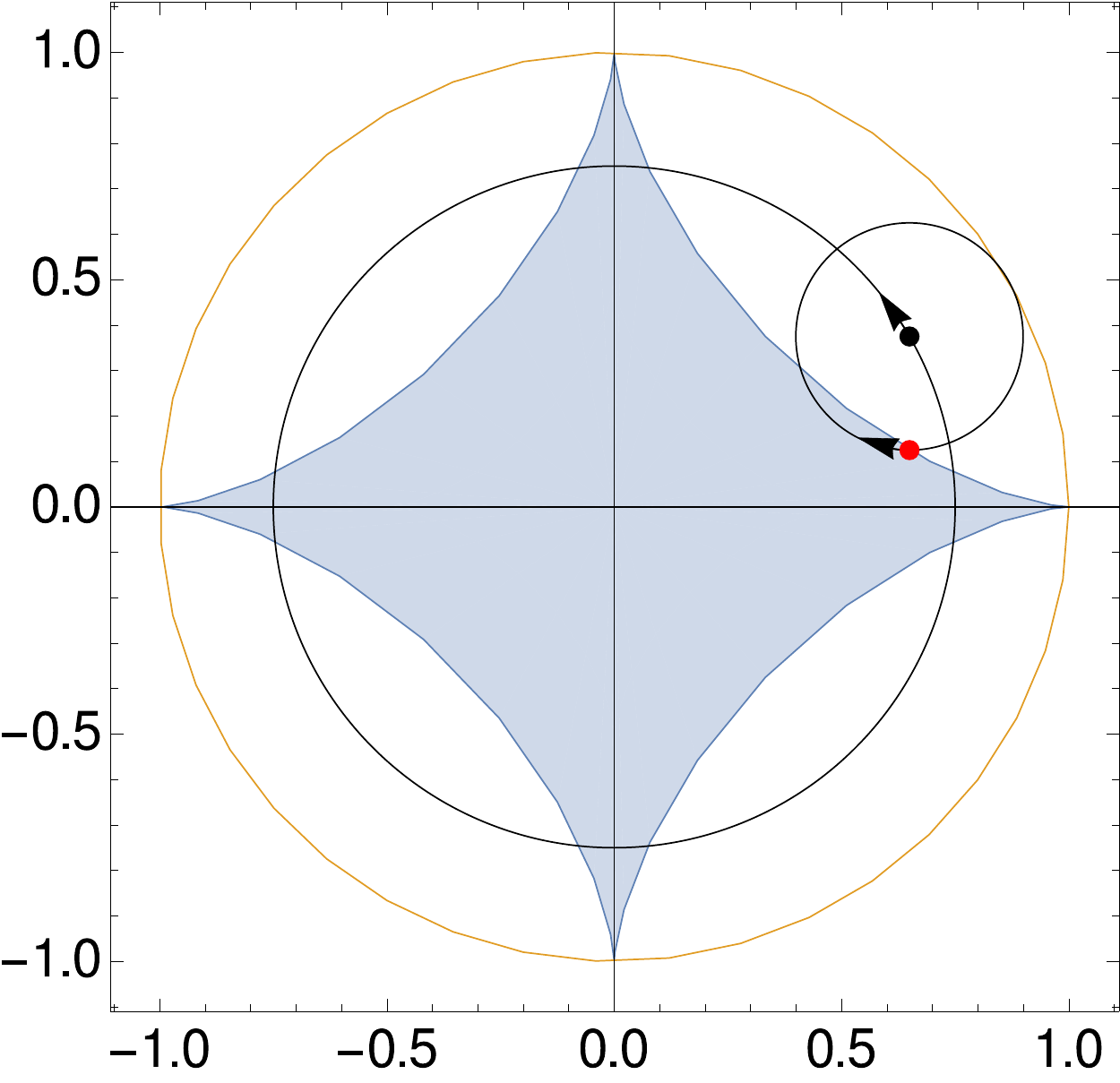} \\
    \nc = 5  & \nc = 10  \\
    \diagram[width=\mylength]{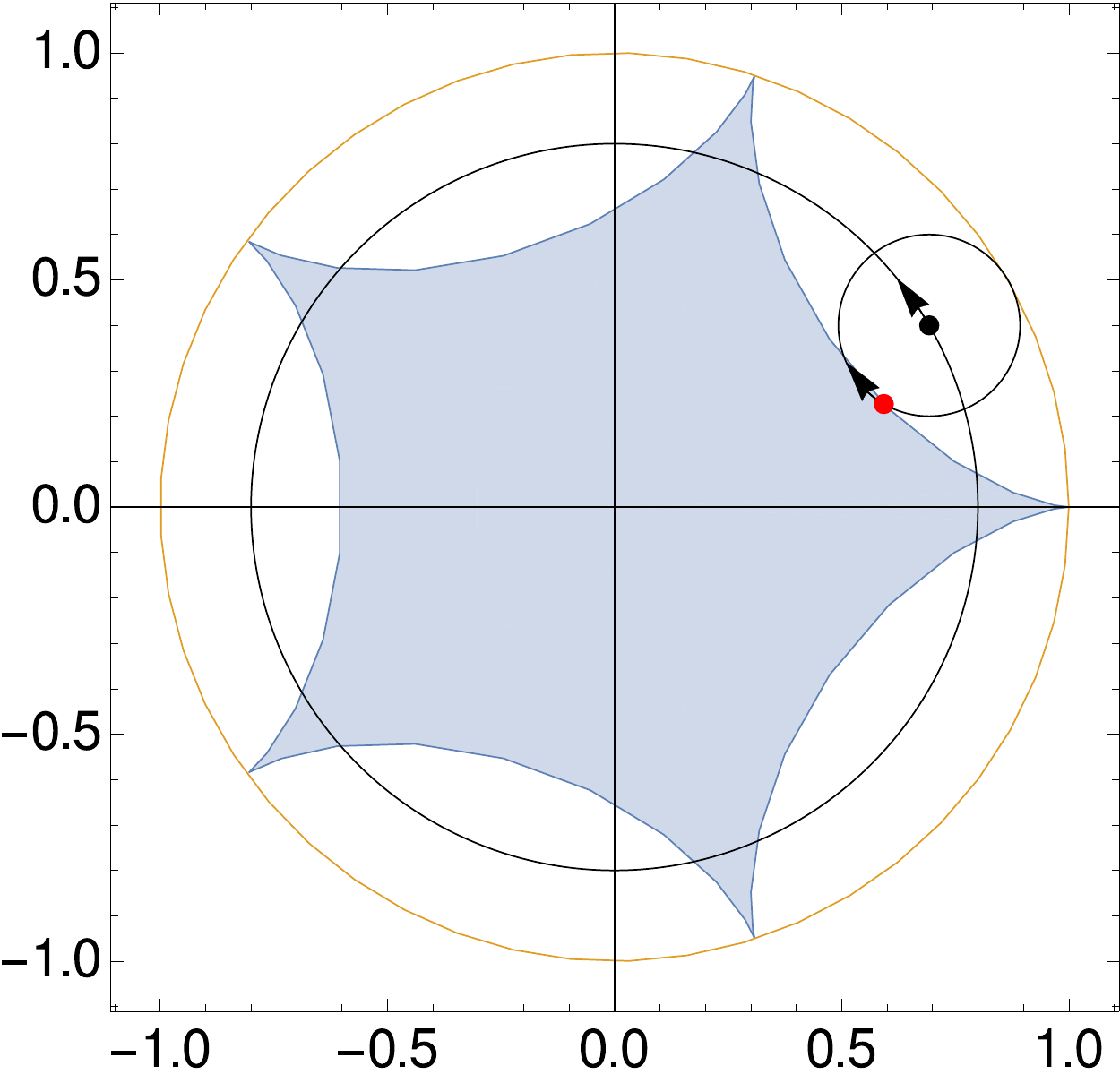} &
%    \ldots &
    \diagram[width=\mylength]{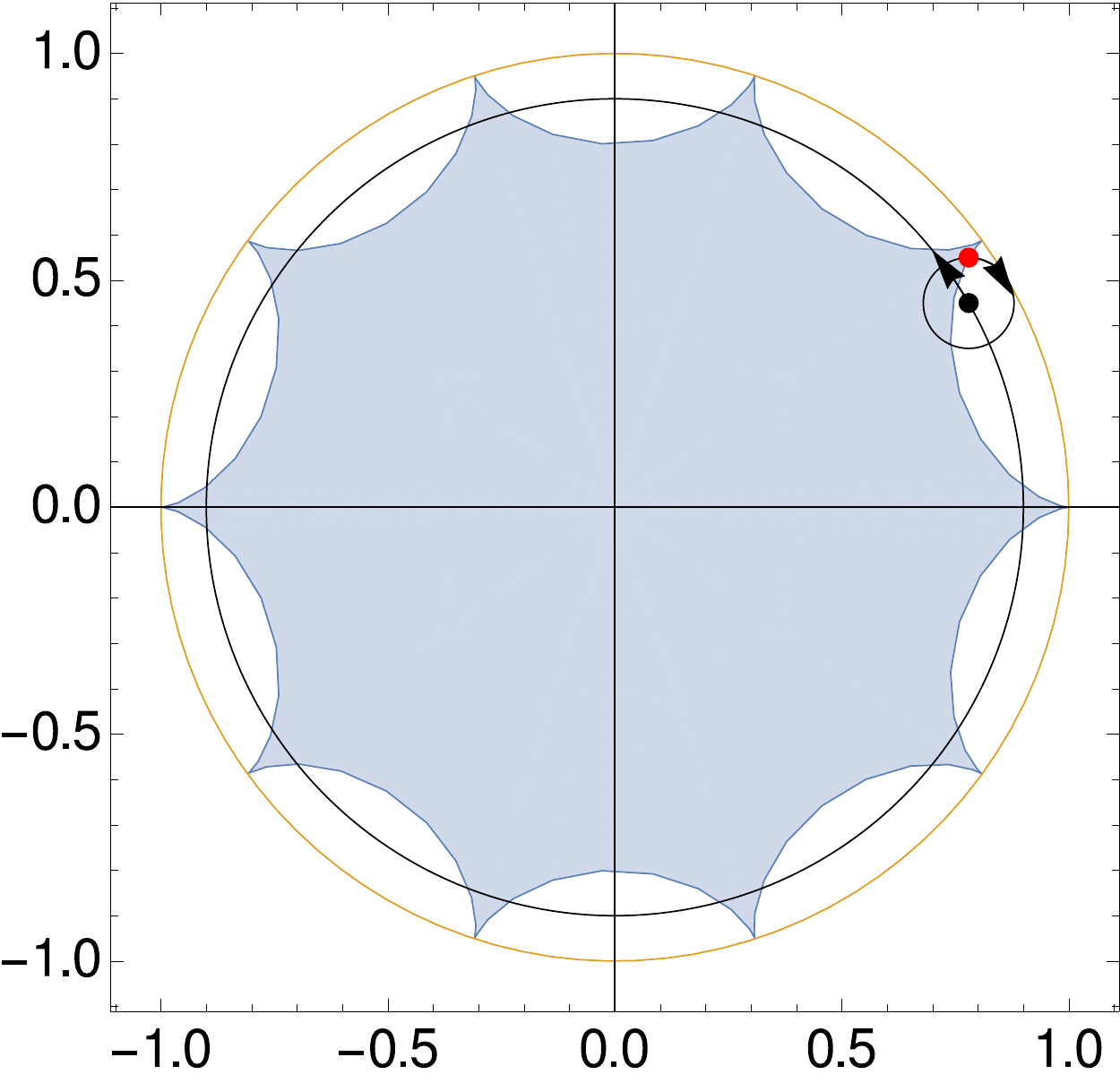} 
%    \ldots \\
\\
  };
\end{tikzpicture}  
\caption{Geometric origin and shape of the hypocycloids traced out by
  $h_{\nc}(\theta)$. The shaded region corresponds to allowed values
  of $\tr(U_{\bm x} U_{\bm y}^\dagger)/\nc$, i.e. the set of points in
  the complex plane reached by Eq.~\eqref{eq:sun-trace-formula}.}
  \label{fig:hypocyloid-boundary-const}
\end{figure}

The dipole correlator appears as an average over such configurations
and can be parametrized in terms of two real degrees of freedom
\begin{align}
  \label{eq:pom-odd}
  S_{\bm x \bm y}(Y) := & \langle\tr(U_{\bm x} U^\dagger_{\bm y})\rangle(Y)/\nc 
  \\ \notag = &
  1-P_{\bm x \bm y}(Y) + i O_{\bm x \bm y}(Y) 
  \\ \notag = &
  e^{-\cf ({\cal P}_{\bm x \bm y} + i {\cal O}_{\bm x \bm y})(Y)}
 \ ,
\end{align}
i.e{.}~either directly through real and imaginary parts ($1-P_{\bm x
  \bm y}$ and $O_{\bm x \bm y}$, respectively) or exponentially via a
logarithmic modulus and phase (${\cal P}_{\bm x \bm y}$ and ${\cal
  O}_{\bm x \bm y}$, respectively). Noting that complex conjugation
simply swaps the coordinates on $S$, $S_{\bm x \bm y}^* = S_{\bm y \bm
  x}$, implies that $P$ and ${\cal P}$ are symmetric, while $O$ and
${\cal O}$ are anti-symmetric under the exchange of $\bm x$ and $\bm
y$. This symmetry property links them to the pomeron and odderon
respectively.

\begin{figure}[tb]
  \centering
\begin{tikzpicture}
  \matrix(m)[matrix of math nodes,column sep=.6em]
   {
     \nc=3 & \nc = 4 \\
     \diagram[width=\mylength]{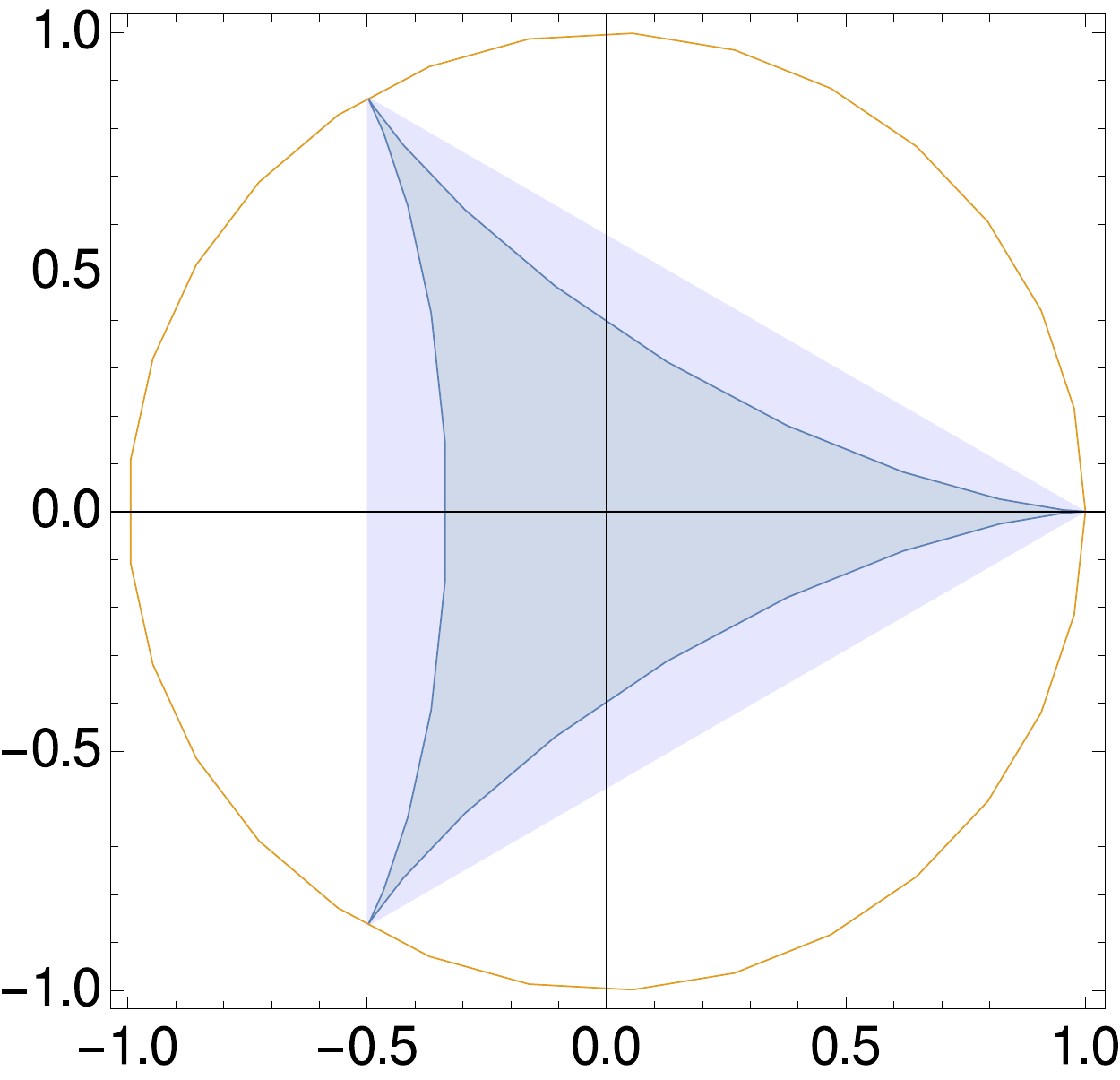} &
    \diagram[width=\mylength]{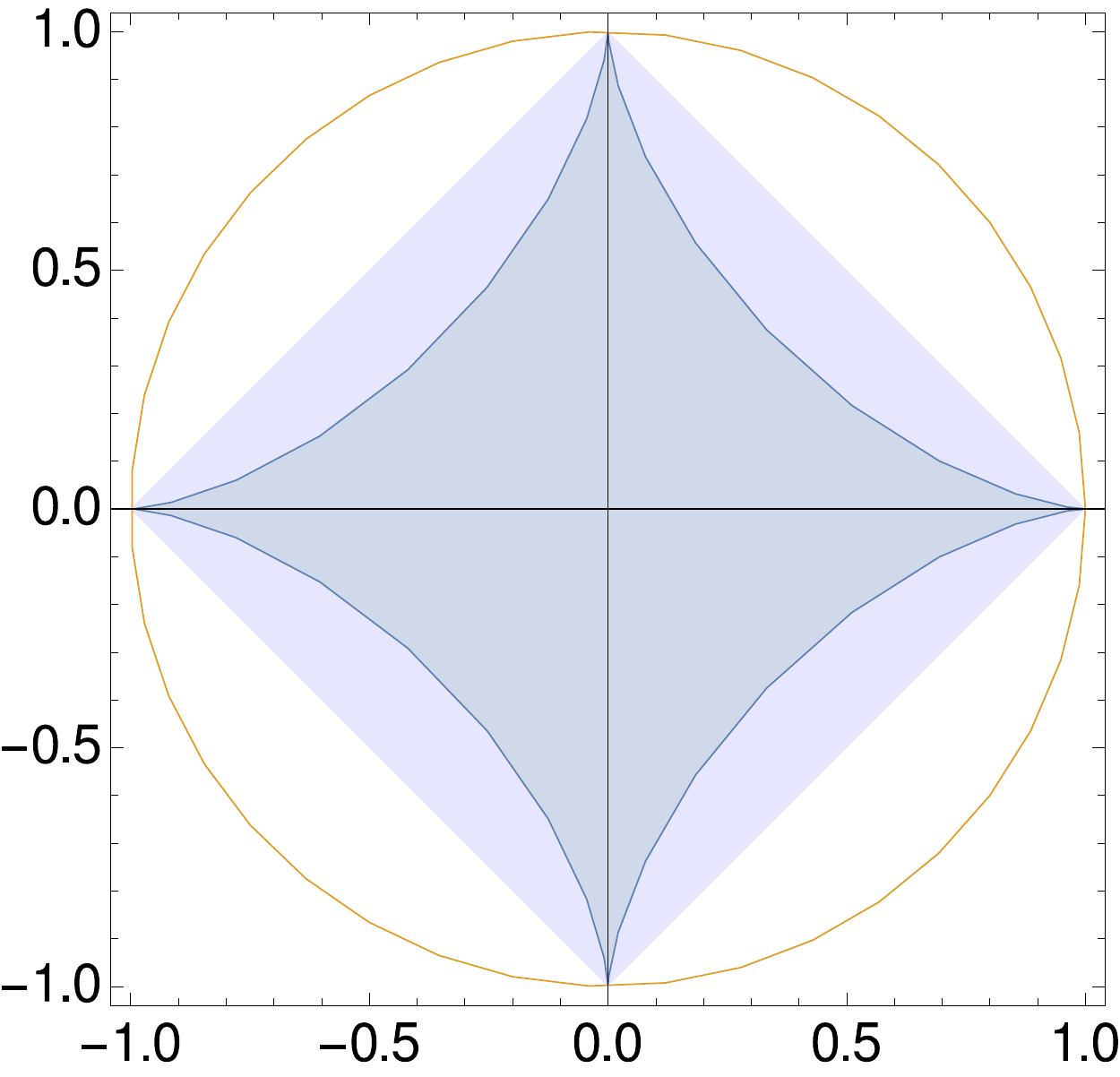} \\
 \nc = 5 & \nc = 10 & \\
    \diagram[width=\mylength]{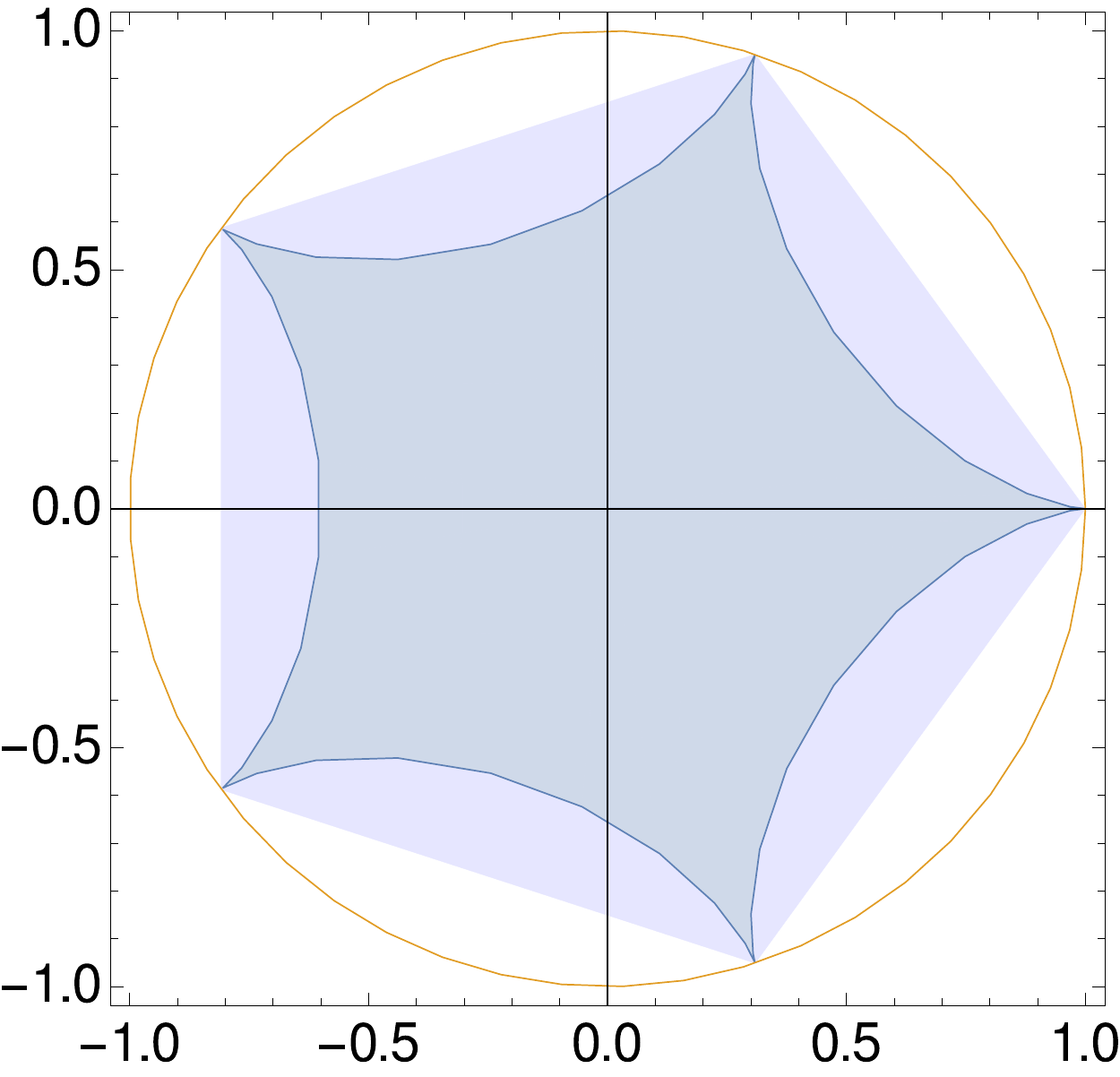} &
%    \ldots &
    \diagram[width=\mylength]{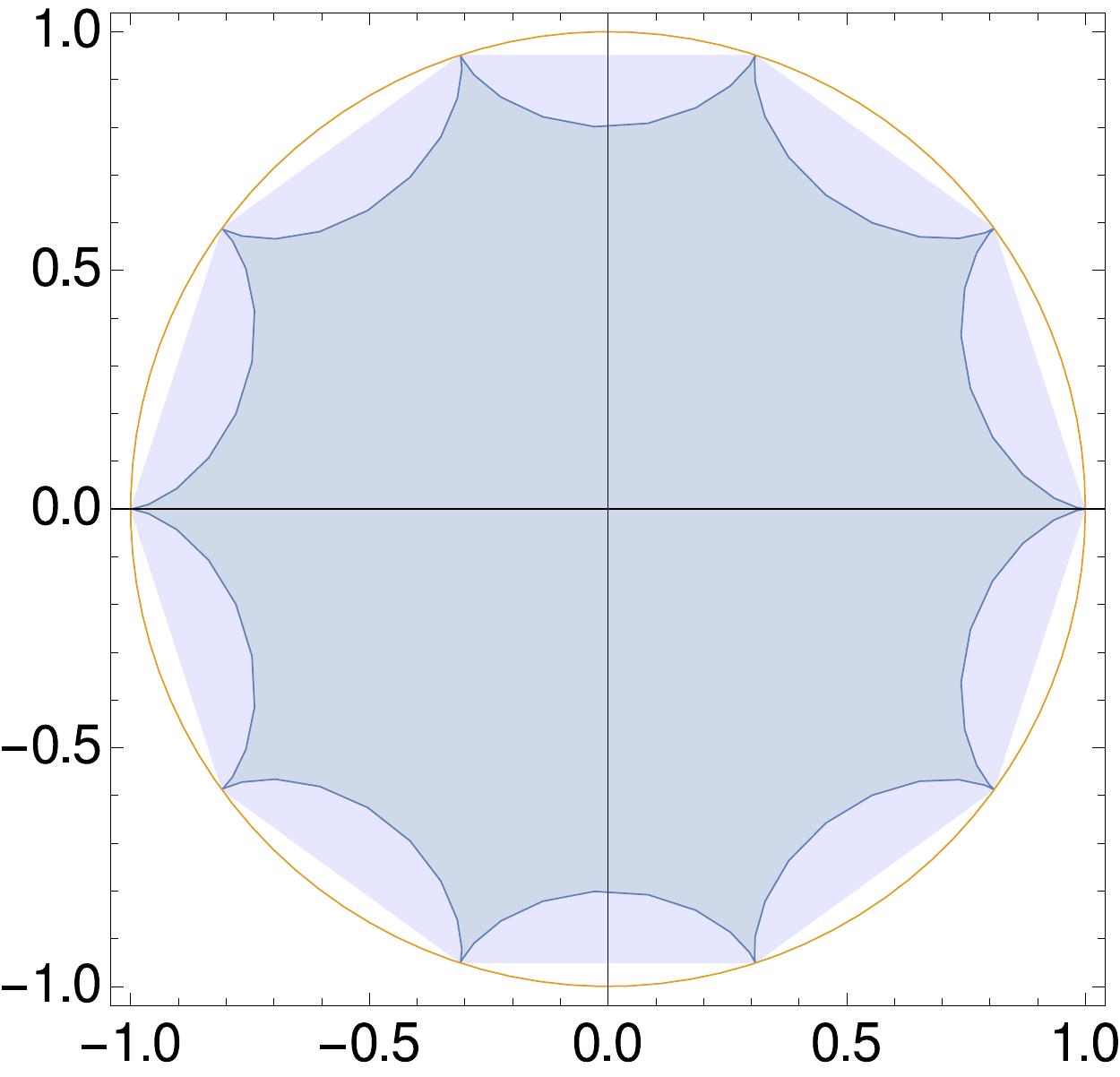} \\
%&
%    \ldots \\
  };
\end{tikzpicture}\caption{Averages group element traces must
  fall within a polygon connecting the $\nc$-th roots of unity. For
  $\nc\to\infty$ the polygon will approximate the unit circle.}
  \label{fig:hypocycloid-boundary-average}
\end{figure}

One striking observation is that nothing inherently prevents the real
part, the pomeron contribution, from taking negative values -- the
hypocycloids allow negative real parts. In fact the Wilson line dipole
correlators of Eq.~\eqref{eq:pom-odd} are averages of configurations
falling into the interior of the hypocycloid $h_{\nc}(\theta)$, and
thus are even slightly less constrained: such an average may fall
outside the hypocycloid but must remain inside a polygon connecting
the cusps, as illustrated for a few values of $\nc$ in
Fig.~\ref{fig:hypocycloid-boundary-average}. For $N_c\in\mathbbm{Z}
\ge 2$, where there is a well defined interior, the bounding polygon
can be parametrized by
\begin{align}
  \label{eq:poly-bound}
  p_{N_c}(\theta) 
  = 
  \frac{\cos(\frac{\pi}{N_c})}{%
    \cos(\text{mod}(\theta,\frac{2\pi}{N_c})-\frac{\pi}{N_c})} 
  e^{i\theta}
  \ ,
\end{align}
again with $\theta\in[0,2\pi[$.

Let us emphasize that this is not in contradiction with the bounds
observed for real correlators in earlier simulations. In fact,
consistent, real valued initial conditions for the $q\Bar q$ and
$q\Bar q g$ correlators of Eq.~\eqref{eq:stsaUcorr} during evolution
lie between the fixed points at $1$ and $0$ and the evolution equation
does not develop an imaginary part starting from a real initial
condition respecting these bounds.

This behavior is in fact a generic requirement on a consistent
evolution equation for Wilson line dipole correlators. To expose the
physics content of this statement, parametrize a point inside the
hypocycloid or inside its bounding polygon by a real factor $\rho \in
[0,1]$ and a point on the boundary $b_{N_c}(\theta)$ (with
$b_{N_c}(\theta)$ either given by $h_{N_c}(\theta)$ or
$p_{N_c}(\theta)$), so that the Wilson line correlator (the average)
takes the form
\begin{align}
  \label{eq:S-hypo}
  %S_{\bm x \bm y}(Y) := & 
  \langle\tr(U_{\bm x} U^\dagger_{\bm y})\rangle(Y)/\nc 
  =
  \rho_{\bm x\bm y}(Y) \ b_{\nc}(\theta_{\bm x\bm y}(Y)) 
 \ .
\end{align}
Like for the ingredients of Eq.~\eqref{eq:pom-odd} there are clear
symmetry requirements on $s$ and $\theta$ in Eq.~\eqref{eq:S-hypo}:
they must be symmetric and antisymmetric respectively under exchange
of $\bm x$ and $\bm y$. If the measurement is both translationally and
rotationally symmetric in the transverse plane (i.e. in the absence of
an additional transverse vector $\Hat{\bm s}$ to furnish the sign
change via a factor $\Hat{\bm s}\cdot(\bm x-\bm y)$), the contribution of any
antisymmetric function such as $\theta_{\bm x \bm y}$ \emph{must}
vanish. In that case Eq.~\eqref{eq:S-hypo} can be reduced to $S_{\bm x \bm
  y}(Y) = \rho_{\bm x\bm y}(Y) \in [0,1]$, i.e. the solutions are
restricted to the intersection of the hypocycloid with the
\emph{positive} real axis. (See also Sec~\ref{sec:jimwlk} where this
mechanism is demonstrated for JIMWLK ensembles.)

The need for additional directional information is a physics
requirement: to be able to see an odderon contribution in an
experiment one needs to break rotational symmetry in the transverse
plane such as is done by measuring a spin asymmetry in STSA. The total
cross section, by contrast, averages over all directions $\Hat{\bm s}$
(in the average that forms the correlator) and thus forces $b_{N_c}\to
1$. In this case there is no scope for an average odderon contribution
to couple to the real part visible in Eq.~\eqref{eq:DIScorr}, it is
zero throughout.

As a mathematical constraint, one needs an imaginary part in the
initial condition for the \emph{average} to allow it to move away from
the interval $[0,1]$ and, in particular, for the real part to turn
negative. (Note that this does not imply that individual
configurations may not fall onto the negative real axis, see again
Sec~\ref{sec:jimwlk} for explicit examples.)

To conclude this section, we note that the perturbative limit $r\to 0$
takes $\langle\tr(U_{\bm x} U^\dagger_{\bm y})\rangle(Y)/\nc \to 1$
and thus is associated strong correlations and the trivial center
element $U_{\bm x} U^\dagger_{\bm x} =\mathbbm{1}$. The origin on the
other side corresponds to $\langle\tr(U_{\bm x} U^\dagger_{\bm
  y})\rangle(Y)/\nc \to 0$ and thus the long distance limit where 
the Wilson lines are uncorrelated. The
remaining center elements (at least for $N_c = 3$) correspond to
maximally anticorrelated configurations. 

If such configurations are not present with considerable weight, the
averages will not have any chance of being pulled outside the
hypocycolid into the remainder of then enveloping polygon. The
perturbatively motivated initial conditions discussed in
Sec.~\ref{sec:constr-init-cond} below do not lead to such behaviour
despite the presence of noticeable anticorrelation in one of the
examples. 

% Experience from
% lattice QCD associates the remaining non-trivial center elements with
% non-perturbative configurations. In the absence of non-perturbative
% contributions --which we can only accommodate in the initial
% conditions as the evolution equation is purely perturbative-- we are
% led to expect that the vicinity of the non-trivial center elements is
% only weakly populated.

% This has consequences for the behavior of correlators $S_{\bm x \bm
%   y}(Y)$: We expect that it requires a notable anti-correlated
% contribution for the correlators to fall outside the hypcycoloid for
% which, to date, we have no evidence in the JIMWLK context.

\section{Constraints on the initial conditions}
\label{sec:constr-init-cond}

The physics expectations for the total cross section in the absence of
an imaginary part severely restrict the form of the initial
condition. The simplest assumption, based on exponentiating the $\bm
r^2$ behavior of leading-order light cone perturbation theory, leads
to the well known Golec-Biernat Wüsthoff~\cite{Golec-Biernat:1998js}
ansatz $S_{\bm x\bm y} = e^{-(\bm x-\bm y)^2 \qs^2/4}$ where $\qs$ is
the GB-W saturation scale. The MV model modifies this with a
logarithmic correction in the exponent, and evolution at leading order
will carry any of these into a scaling form entirely imposed by the
nonlinear form of the evolution equation. For our discussion here, all
of these forms are suitable since at leading order (see, however
\cite{Lappi:2015fma,Iancu:2015vea,Iancu:2015joa,Lappi:2016fmu})
evolution will quickly readjust these to take on the features of the
scaling form.

If we allow for a non-zero odderon admixture the choice of a
physically meaningful initial condition for the pair of $\mathcal P$
and $\mathcal O$ needs some thought. At short distance, light cone
perturbation theory leads to $\mathcal P_{\bm x\bm y} \propto |\bm
x-\bm y|^2$ and $\mathcal O_{\bm x\bm y} \propto |\bm x-\bm
y|^3$~\cite{Kovchegov:2003dm}, but the symmetry properties imposed by
complex conjugation %and evidenced in Eq.~(\ref{PandOGT})
require the presence of an additional transverse vector $\Hat{\bm s}$
ad discussed in Sec.~\ref{sec:group-theory-constr-corr}.  We thus
expect a short distance $r\ll 1/\qs$ behavior of the form
\begin{align}
  \label{eq:PO-pert}
 \mathcal P_{\bm x\bm y} 
  \propto \bm r^2  ,
\hspace{1cm}
  \mathcal O_{\bm x\bm y}\propto 
  \kappa \rt^2   \rt \cdot\Hat{\bm s} = 
 \kappa r^3   \Hat{\rt} \cdot\Hat{\bm s} 
,
%  \hspace{1cm}\text{where}\hspace{1cm} \bm r=\bm x-\bm y
%  \ .
\end{align}
where $\bm r=\bm x-\bm y$ and $\Hat{\rt} = \bm r/r$.  If we measure
both contributions in~\eqref{eq:PO-pert} in the same units $\kappa$
serves to parametrize the normalization of the odderon relative to the
pomeron amplitude. We will see in the following, that it is
constrained by the group theory limits on the scattering amplitude in
the dilute limit $r\to 0$.

As discussed in Sec.~\ref{sec:group-theory-constr-corr}, we can impose
the group theory constraints on the amplitude at two levels of
rigor. Physically we would expect that the average amplitude itself is
within the group manifold. In principle, it is also possible that the
average amplitude is within the polygon region defined by linear
combinations of amplitudes in the group.

In the first, more physical case, this leads in the $r\to0$ limit to
the constraint
\begin{equation}
 \mathcal O_\rt \leq \frac{\sqrt{2}}{3}\frac{\nc-2}{\sqrt{\nc-1}}
(\mathcal P_\rt)^{3/2}.
\end{equation}
Assuming that the pomeron has the perturbative behavior 
\begin{equation}\label{eq:pomparam}
\mathcal P_{\rt} 
  \approx \rt^2 \qs^2/4 
\end{equation}
and parametrizing the odderon with a general power law as
\begin{equation}\label{eq:oddparam}
\mathcal O_{\rt} 
  \approx \kappa (r \qs/2)^\alpha,
\end{equation}
this leads to the constraint $\alpha\geq 3$. Thus, the result
$\mathcal O_{\rt} \sim r^3$ of Ref.~\cite{Kovchegov:2003dm} indeed
corresponds to the mildest $r$-dependence allowed by the group theory
constraint. Assuming now the power $\alpha=3$ we get the limit
\begin{equation}
\kappa \leq  \frac{\sqrt{2}}{3}\frac{\nc-2}{\sqrt{\nc-1}} 
\quad \underset{\nc=3}{=} \quad 
\frac{1}{3}.
\end{equation}
We want to stress the remarkable nature of this result. With linear
BFKL evolution the magnitudes of the pomeron and odderon amplitudes
are only limited by phenomenology. The interpretation of the
scattering amplitude in terms of the Wilson line immediately places a
non-perturbative mathematical upper limit on the odderon amplitude.

Strictly speaking the averages of $\mathsf{SU}(\nc)$-matrices need not
be inside the group. Thus, in principle the upper limit for the
odderon follows from restricting the expectation value of the
amplitude to lie inside the $\nc$-sided polygon with corners at the
$\nc$-roots of unity.  This leads in the limit $r\to 0$ to
\begin{equation}
 \mathcal O_\rt \leq \frac{\sin\frac{2\pi}{\nc}}{1-\cos\frac{2\pi}{\nc}}
\mathcal P_\rt
\end{equation}
For amplitudes parametrized as in \nr{eq:pomparam}
and~\nr{eq:oddparam}, this leads to the less stringent limit $\alpha
\geq 2$. For the limiting power $\alpha=2$ the odderon amplitude is
constrained to $\kappa(\alpha=2) \leq
\frac{\sin\frac{2\pi}{\nc}}{1-\cos\frac{2\pi}{\nc}}$. For the value
$\alpha=3$, any value of $\kappa$ satisfies this more lax version of
the group theory constraint sufficiently close to $r=0$. From larger
dipole sizes one does obtain an upper limit on $\kappa$, but this
limit is universal in the same way, i.e. it depends on the functional
form at larger $r$.

\begin{figure}[tb]
  \centering
\resizebox{\linewidth}{!}{
\begin{tikzpicture}
  \matrix(m)[matrix of nodes,column sep=.5em,ampersand replacement=\u]
   {
     {\resizebox{!}{2.5mm}{$\mathsf{Re} \text{ \& } \mathsf{Im} \text{ of }  e^{-r^2Q_0^2+i \frac13 r^3Q_0^3}$} }
  \u 
 {\resizebox{!}{2.5mm}{$\mathsf{Re} \text{ \& } \mathsf{Im} \text{ of }  e^{-r^2Q_0^2+i .98 r^3Q_0^3}$}}
     \\
      \diagram[height=5cm]{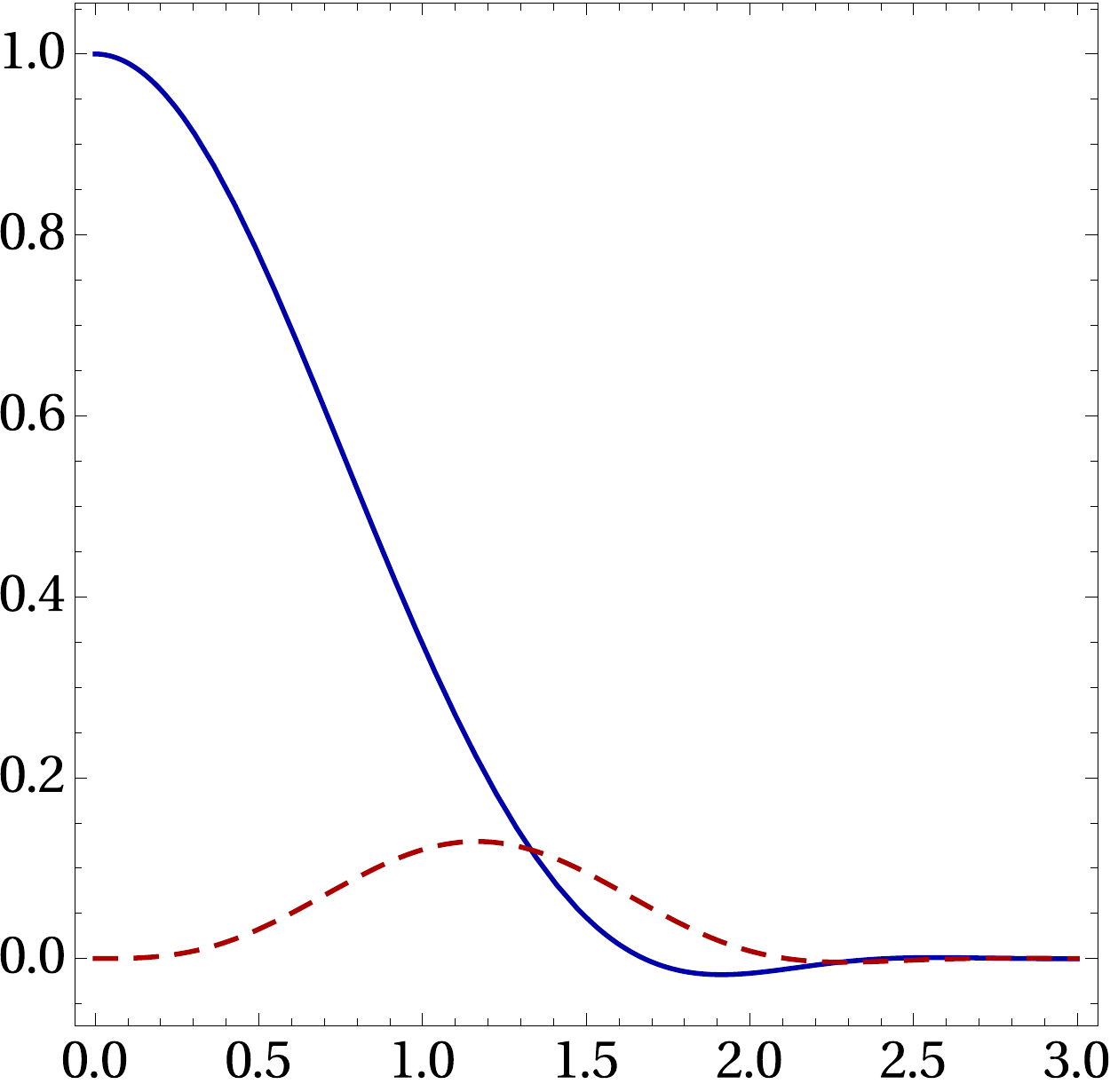}  \u
      \diagram[height=5cm]{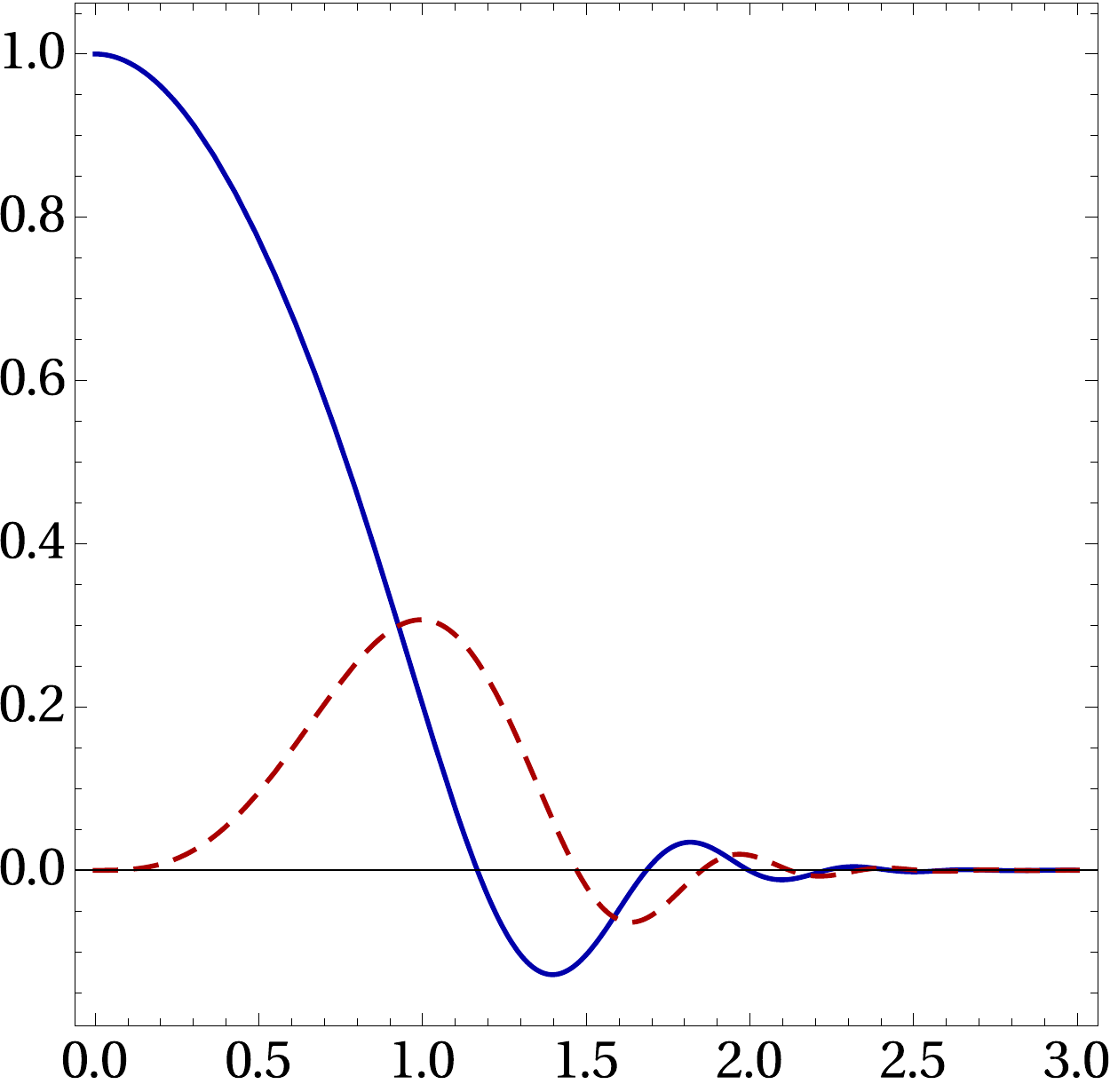} 
    \\
    {\resizebox{!}{2mm}{$r Q_0$}} \u 
    {\resizebox{!}{2mm}{$r Q_0$}}
    \\
  };
\end{tikzpicture}
}
 
\caption{Perturbatively motivated initial conditions for both real and
  imaginary parts (solid blue and red dashed, respectively) of
  $\langle\tr(U_{\bm x} U_{\bm y}^\dagger)\rangle(Y_0)/\nc = e^{-r^2
    Q_0^2+i\kappa r^3Q_0^3 \Hat{\bm r}\cdot\Hat{\bm s}}$ at $\Hat{\bm
    r}\cdot\Hat{\bm s}=1$ assuming no extreme anti-correlations to
  drive the correlator outside the hypocycloid (left) and relaxing
  this condition (right). In both cases, real and imaginary parts show
  modulation only near $Q_s(Y_0)$.}
  \label{fig:exp-initial}
\end{figure}

\begin{figure}[tb]
  \centering
\resizebox{\linewidth}{!}{
\begin{tikzpicture}
  \matrix(m)[matrix of nodes,column sep=.5em,ampersand replacement=\u]
   {
     {\resizebox{!}{2.5mm}{$\mathsf{Re} \text{ \& } \mathsf{Im} \text{ of }  e^{-r^2Q_0^2}(1+i \frac13 r^3Q_0^3)$} }
  \u 
 {\resizebox{!}{2.5mm}{$\mathsf{Re} \text{ \& } \mathsf{Im} \text{ of }  e^{-r^2Q_0^2}(1+i .98 r^3Q_0^3)$}}
     \\
      \diagram[height=5cm]{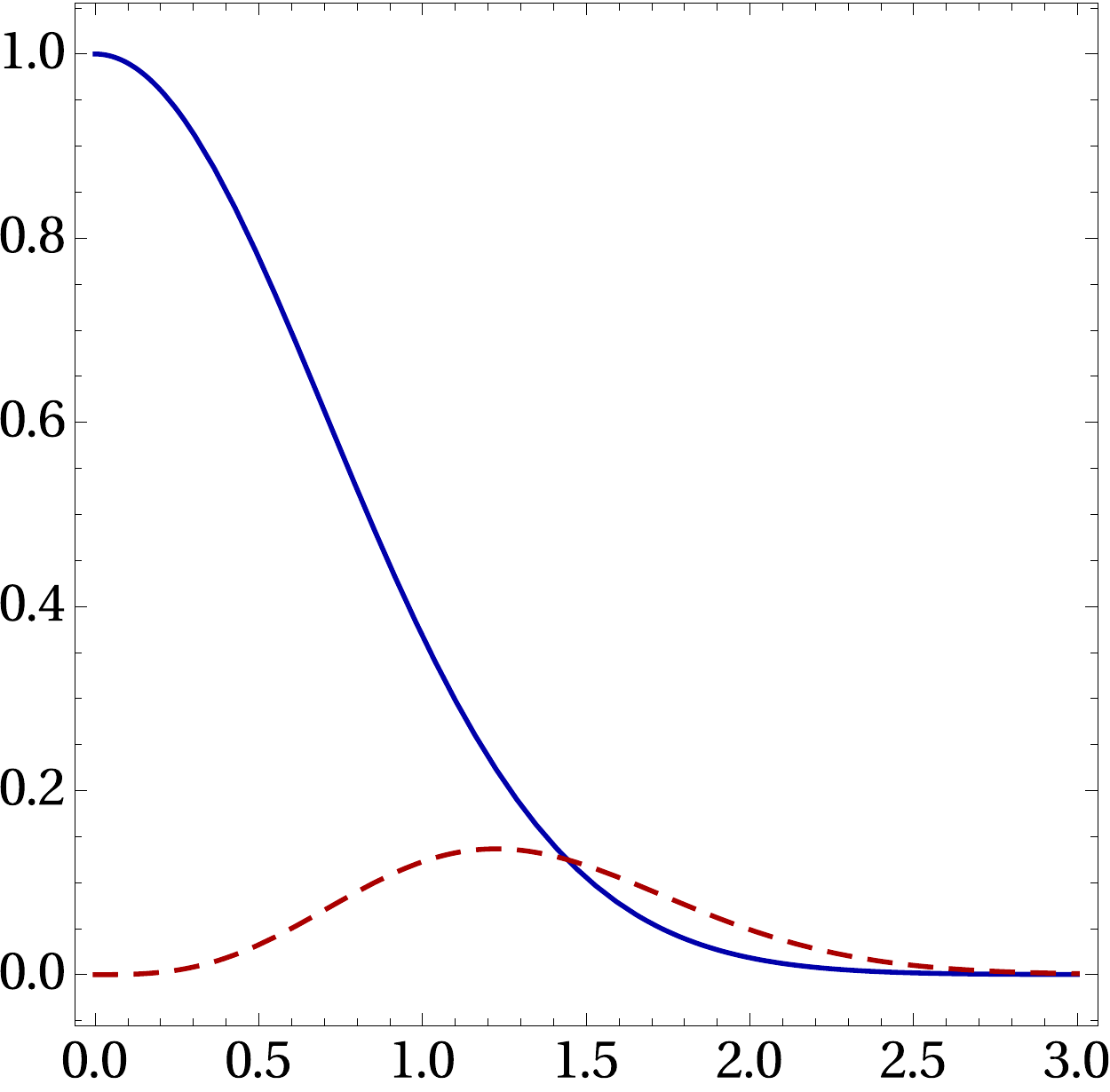}  \u
      \diagram[height=5cm]{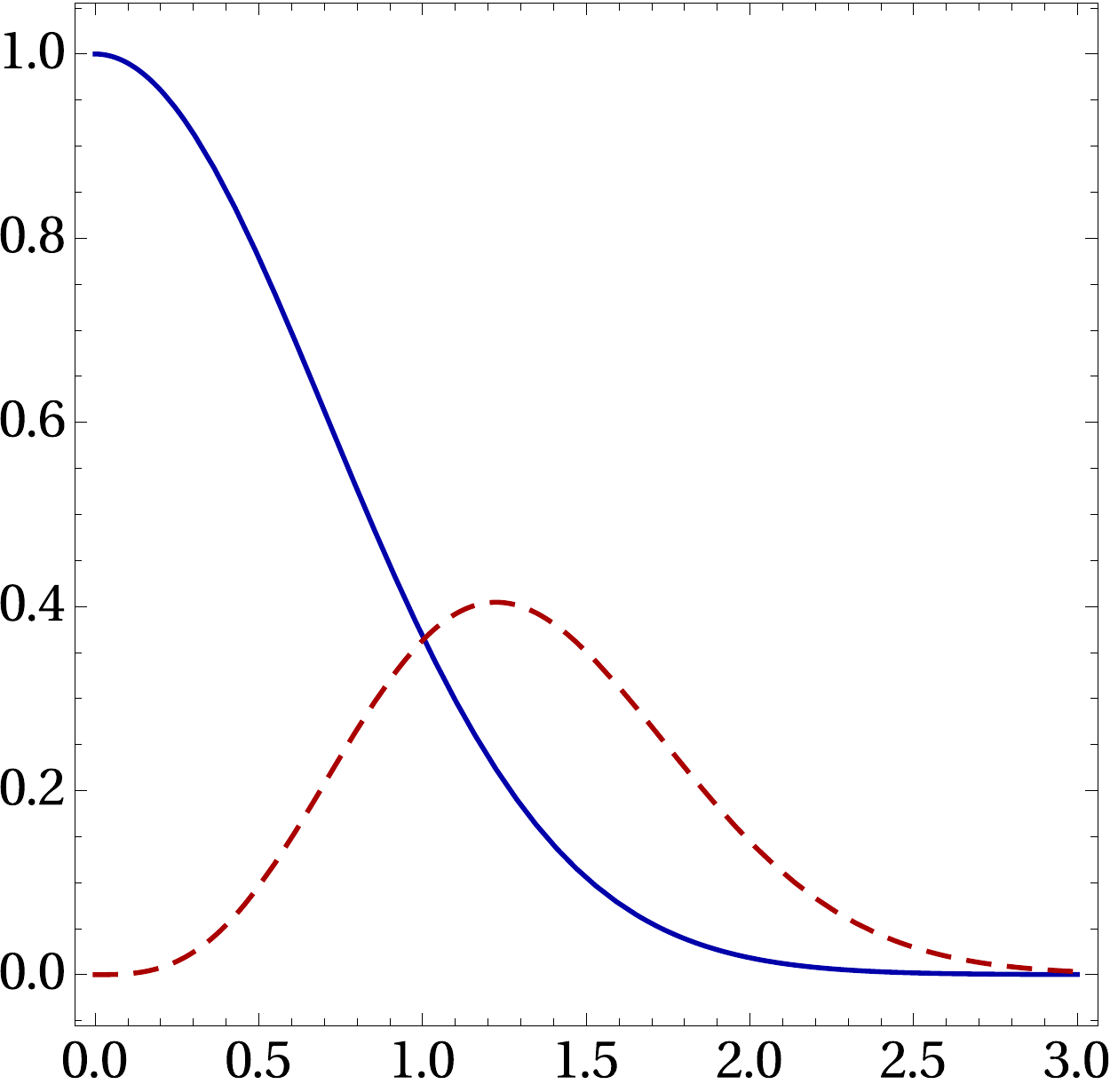} 
    \\
    {\resizebox{!}{2mm}{$r Q_0$}} \u 
    {\resizebox{!}{2mm}{$r Q_0$}}
    \\
  };
\end{tikzpicture}
}
 
\caption{Perturbatively motivated initial conditions for both real and
  imaginary parts (solid blue and red dashed, respectively) of
  $\langle\tr(U_{\bm x} U_{\bm y}^\dagger)\rangle(Y_0)/\nc = e^{-r^2
    Q_0^2+i\kappa r^3Q_0^3 \Hat{\bm r}\cdot\Hat{\bm s}}$ at $\Hat{\bm
    r}\cdot\Hat{\bm s}=1$ assuming no extreme anti-correlations to
  drive the correlator outside the hypocycloid (left) and relaxing
  this condition (right). In both cases real and imaginary parts show
  modulation only near $Q_s(Y_0)$.}
  \label{fig:ser-initial}
\end{figure}

As practical initial conditions for evolution including the odderon,
we suggest two possible extensions of the GB-W parametrization:
\begin{align}
\label{eq:exp-initial}
  \langle\tr(U_{\bm x} U_{\bm
    y}^\dagger)\rangle(Y_0)/\nc = e^{-r^2 Q_0^2/4+i \kappa r^3Q_0^3/8 \Hat{\bm r}\cdot\Hat{\bm s}}
\end{align}
or
\begin{align}
\label{eq:ser-initial}
  \langle\tr(U_{\bm x} U_{\bm
    y}^\dagger)\rangle(Y_0)/\nc = e^{-r^2 Q_0^2/4}(1+i \kappa r^3Q_0^3/8 \Hat{\bm r}\cdot\Hat{\bm s}) \ .
\end{align}
Exponentiating the real part ensures that both the short- and
long-distance behavior are qualitatively correct: the $\bm x-\bm
y\to\bm 0$ limit produces $1$ and the limit $\bm x-\bm y\to\bm \infty$
produces $0$. We have no similar bias for or against exponentiating
the imaginary contribution, but the two choices have very different
consequences: Ansatz~\eqref{eq:exp-initial} leads to anti-correlations
in the real part while~\eqref{eq:ser-initial} does not -- see
Figs.~\ref{fig:exp-initial} and~\ref{fig:ser-initial}. This is
qualitatively different, but neither option can be excluded on purely
theoretical grounds.

\begin{figure}[tb]
  \centering
\resizebox{\linewidth}{!}{
\begin{tikzpicture}
  \matrix(m)[matrix of math nodes,column sep=.5em,ampersand replacement=\u]
  { e^{-r^2+i \kappa r^3} %& e^{-r^2+i \kappa r^2} 
    \u e^{-r^2}(1+i \kappa r^3)
    \\
    \diagram[height=5cm]{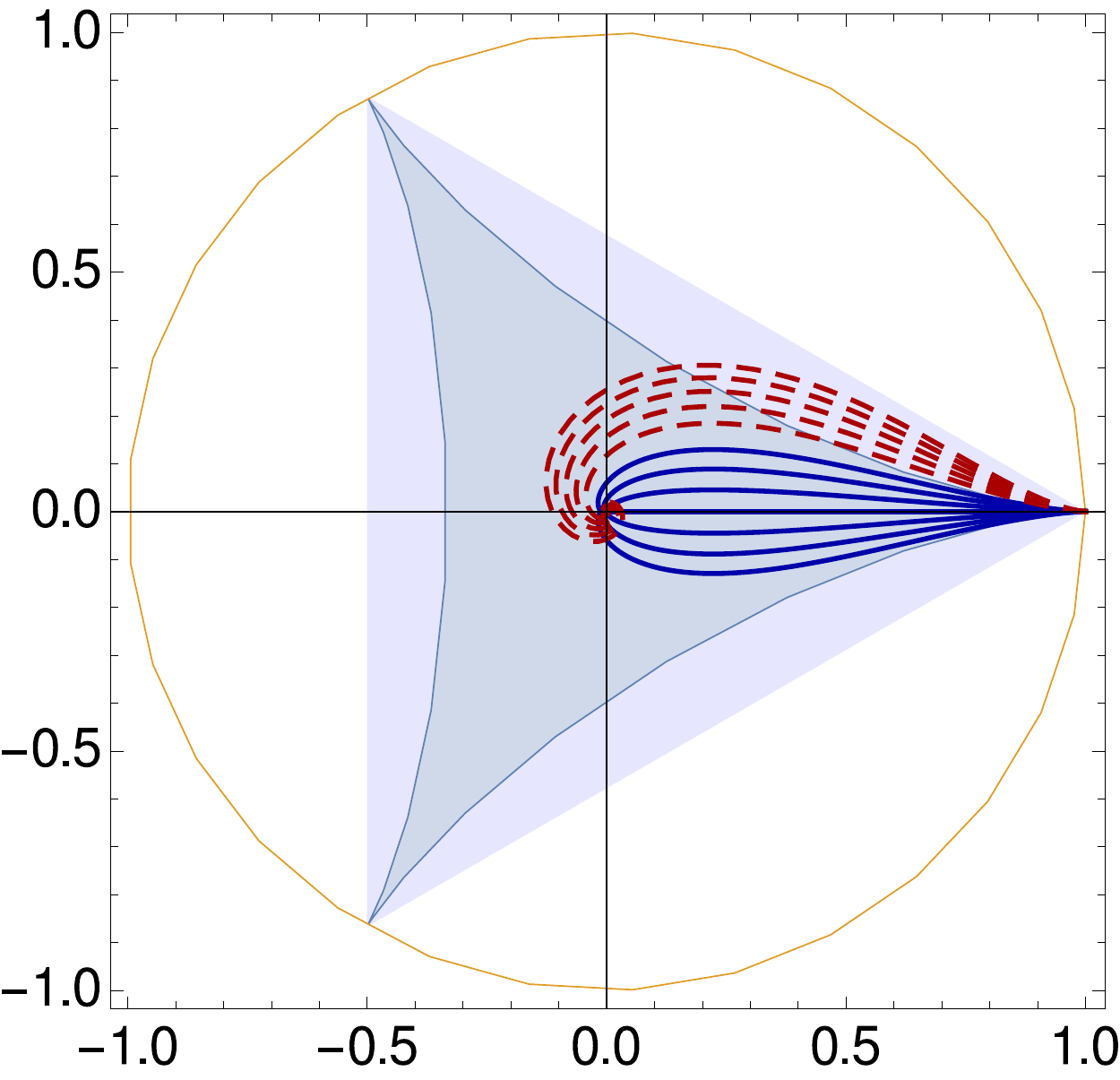}
    %&
    %\diagram[width=.3\textwidth]{figs_final/odderon-trace-power-constr}
    \u
    \diagram[height=5cm]{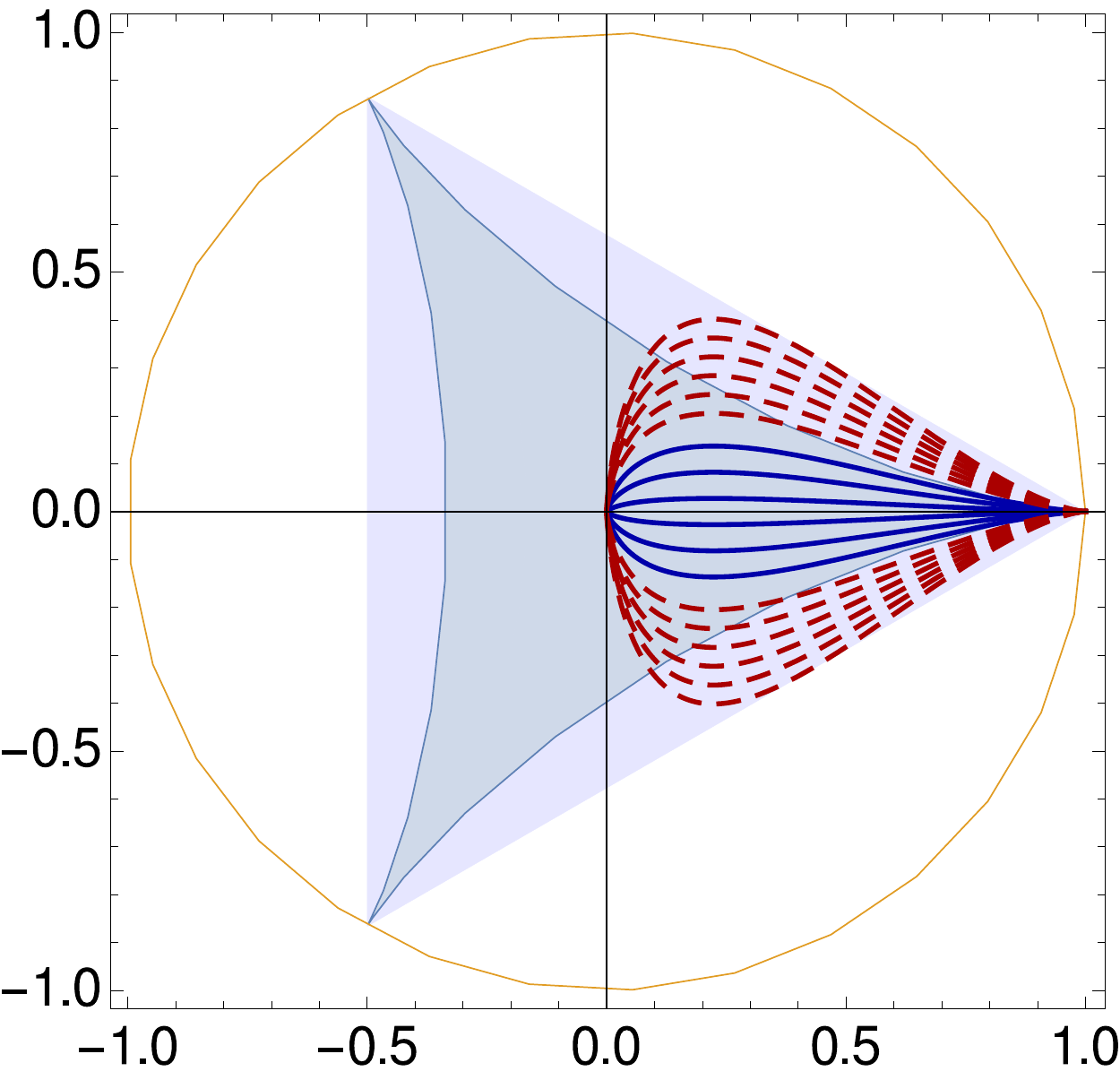}
    \\
  };
\end{tikzpicture}
}

\caption{Perturbatively motivated initial conditions for pomeron +
  odderon configuration averages. Lowest order perturbative
  calculations yield $r^2$ and $r^3$ behavior for small $r=|\bm x-\bm
  y|$ in the pomeron and odderon channel respectively. The plots show
  how different initial conditions seeded with this behavior at small
  $r$ traverse the complex plane. In the parametrizations shown
  $\kappa$ parametrizes the ratio of characteristic scales in the
  pomeron and odderon sector. To fall into the allowed region
  $|\kappa|$ must be small ($|\kappa| \le \frac13$, blue; $\kappa \le
  .98$, dashed red and blue), which limits the possible size of the pomeron
  contribution in the initial condition, see
  Figs.~\ref{fig:exp-initial} and~\ref{fig:ser-initial}.  }
  \label{fig:init-pom-odd-constr}
\end{figure}

The initial conditions \nr{eq:exp-initial} and~\nr{eq:ser-initial} are
visualized in Fig.~\ref{fig:init-pom-odd-constr}. As discussed above,
in order for the average amplitude to stay within the group manifold,
we must have $\kappa\leq 1/3$. For the functional form of
\eq\nr{eq:ser-initial} the parametrization stays within the triangle
allowed for linear combinations of $\mathsf{SU}(\nc=3)$ Wilson lines
for
\begin{equation}
\kappa \leq \frac{
2\sqrt{\frac{2}{3}}\left(
e^{W\left(-3e^{-3/2}/2\right) + 3/2}
\right)
}{
\left(
3 + 2 W\left(-3e^{-3/2}/2\right)
\right)^{3/2}
} 
\approx 
0.9867,
\end{equation}
where $W$ is the Lambert function, defined as the solution of
$z=W(z)e^{W(z)}$. We see that for the other parametrization
\nr{eq:exp-initial}, the amplitude stays within the triangular region
for the same values $\kappa$.

In both cases, $\kappa \lesssim 1$ so that real and imaginary
parts show modulation only near $Q_s(Y_0)$ and thus provide
perturbatively consistent starting points for evolution in a
calculation where $Q_s(Y_0)$ is assumed to be in the perturbative 
domain. Note that the size of the odderon peak (the maximum of
imaginary parts shown in red in Figs.~\ref{fig:exp-initial}
and~\ref{fig:ser-initial}) is severely limited by the bounds on
$\kappa$.
%  This is very different from the BFKL odderon in which there
% is only a phenomenological reason to assume a small odderon
% contribution in the initial condition. In the non-linear case
% \emph{both} the pomeron and odderon are subject to group constraints
% and severely limited in magnitude.

\begin{figure}[tb]
  \centering
\resizebox{\linewidth}{!}{
 \begin{tikzpicture}
    \node (tp) {
      $\displaystyle
    \tr(U_{\bm z} t^a U_{\bm z}^\dagger U_{\bm x} t^a U_{\bm y}^\dagger)
     $
  };
  \node[anchor=west] at (tp.east) {
    $\displaystyle =\Tilde U^{a b}_{\bm z}  \tr(t^a U_{\bm x} t^b U_{\bm y}^\dagger) $};
   \node (zx) at ($(tp.north east)+(2,.5)$) {$\cf \tr(U_{\bm x}  U_{\bm y}^\dagger)$};
   \node (xy) at ($(tp.south east)+(2,-.5)$) {$ \tfrac12\Tilde\tr(\Tilde U_{\bm z}\Tilde  U_{\bm x}^\dagger)$};
   \node (xzy) at ($(tp.east)+(5,0)$) {$\nc \cf$};
   \node (fierz) at ($(tp.south)+(0,-4)$) {$\displaystyle
     \frac12\tr(U_{\bm x} U_{\bm z}^\dagger) \tr(U_{\bm z} U_{\bm y}^\dagger)
     -\frac1{2 \nc} \tr(U_{\bm x} U_{\bm y}^\dagger)
     $};
   \node (fierzxy) at ($(zx.south)+(0,-3)$) {$\frac12( |\tr(U_{\bm x} U_{\bm z}^\dagger)|^2 -1)$};
   \draw[-stealth] (tp) |- node[above right]{$\bm z\mapsto \bm x$ or $\bm y$} (zx);
   \draw[-stealth] (tp) |- node[below right]{$\bm y\mapsto \bm x$} (xy);
   \draw[-stealth] (zx) -| node[above]{$\bm y\mapsto \bm x$} (xzy);
   \draw[-stealth] (xy) -| node[below]{$\bm z\mapsto \bm x$} (xzy);
   \draw[-stealth] (tp.south) -- node[left]{Fierz} (fierz);
   \draw[-stealth] (fierz.east) -| node[right] {$\bm y\mapsto \bm x$} (fierzxy);
   \draw (fierzxy) -- node[right]{$=$} (xy);
  \end{tikzpicture}
}
  
  \caption{Correlator relations between $q\bar q g$, $q\Bar q$ and $g g$. }
  \label{fig:correlator-relations}
\end{figure}

\section{Gaussian and higher-order exponential truncations}
 \label{sec:gt}

 The dependence of the Wilson lines on the factorization rapidity that
 separates the small- and large-$x$ degrees of freedom in the CGC
 formalism, is given by JIMWLK equation~\cite{Jalilian-Marian:1997xn,
   Jalilian-Marian:1997jx, Jalilian-Marian:1997gr,
   Jalilian-Marian:1997dw, JalilianMarian:1998cb, Weigert:2000gi,
   Iancu:2000hn, Iancu:2001md, Ferreiro:2001qy, Iancu:2001ad,
   Mueller:2001uk}, or equivalently by the Balitsky
 hierarchy~\cite{Balitsky:1995ub, Balitsky:1998kc, Balitsky:1998ya,
   Balitsky:2001re}. The Balitsky hierarchy is a coupled set of
 integro-differential equations for operators made from products of
 Wilson lines. The first equation of the hierarchy is
\begin{multline}
\label{eq:balitsky}
\frac{\ud}{\ud Y} 
\langle \tr(U_{\bm x}U^\dagger_{\bm y}) \rangle
= \frac{\as}{\pi^2} \int \ud^2 \bm z \, \Tilde{\mathcal{K}}_{\bm x\bm z\bm y} 
\\ \left\langle \Tilde U_{\bm z}^{ab} \, \tr(t^a U_{\bm x} t^b U_{\bm y}^\dagger) 
- \cf  \tr(U_{\bm x}U^\dagger_{\bm y}) \right\rangle ,
%\label{eq:balitsky}
\end{multline}
where $\Tilde{\mathcal{K}}_{\bm x\bm z\bm y} = \frac{(\bm x-\bm
  y)^2}{(\bm x-\bm z)^2 (\bm z-\bm y)^2}$.  It relates the rapidity
dependence of the dipole operator to a combination of dipole and
three-point operators. The equation is derived by considering the
emission of one soft gluon from the dipole; it is therefore not
surprising that the three-point function involved is the same as in
the STSA cross section~\nr{eq:stsaUcorr}. Using the Fierz identity,
Eq.~\eqref{eq:balitsky} becomes
\begin{align}
  \label{eq:bal-fierz}
  \frac{\ud}{\ud Y} &
\left\langle \frac{\tr(U_{\bm x}U^\dagger_{\bm y})}{\nc} \right\rangle
= \frac{\as}{\pi^2} \frac{\nc}2 
\int \ud^2 \bm z \, \Tilde{\mathcal{K}}_{\bm x\bm z\bm y} 
\\ \notag & \times \left\langle 
  \frac{\tr(U_{\bm x}U^\dagger_{\bm y})}{\nc} 
  \frac{\tr(U_{\bm x}U^\dagger_{\bm y})}{\nc} 
  -\frac{\tr(U_{\bm x} U_{\bm y}^\dagger)}{\nc} 
 \right\rangle .
\end{align}

This form is often used to truncate the hierarchy by factorizing 
\begin{align}
  \label{eq:BK-fact}
  \left\langle 
  \frac{\tr(U_{\bm x}U^\dagger_{\bm y})}{\nc} 
  \frac{\tr(U_{\bm x}U^\dagger_{\bm y})}{\nc} 
 \right\rangle
 \xrightarrow{\text{fact.}}
 \left\langle 
  \frac{\tr(U_{\bm x}U^\dagger_{\bm y})}{\nc}  
  \right\rangle\left\langle 
  \frac{\tr(U_{\bm x}U^\dagger_{\bm y})}{\nc} 
 \right\rangle
\end{align}
in the spirit of an independent scattering approximation for
dipoles. The resulting closed mean-field equation
\begin{align}
  \label{eq:bal-bk}
  \frac{\ud}{\ud Y} &
\left\langle \frac{\tr(U_{\bm x}U^\dagger_{\bm y})}{\nc} \right\rangle
= \frac{\as}{\pi^2} \frac{\nc}2 
\int \ud^2 \bm z \, \Tilde{\mathcal{K}}_{\bm x\bm z\bm y} 
\\ \notag & \times \left[
 \left\langle 
  \frac{\tr(U_{\bm x}U^\dagger_{\bm y})}{\nc}  
  \right\rangle\left\langle 
  \frac{\tr(U_{\bm x}U^\dagger_{\bm y})}{\nc} 
 \right\rangle
  -
 \left\langle 
\frac{\tr(U_{\bm x} U_{\bm y}^\dagger)}{\nc} 
 \right\rangle
\right]
\end{align}
is the Balitsky-Kovchegov
equation~\cite{Balitsky:1995ub,Kovchegov:1999yj}, and is a crucial
tool in practical phenomenological applications of the CGC formalism.

To develop a better understanding of the possible physics content of
the correlators in Eq.~\eqref{eq:stsaUcorr} and their JIMWLK evolution,
we now turn to a set of systematically extendable truncations of the
Balitsky hierarchy associated with the dipole operator, the simplest of
which is known as the Gaussian truncation.

The motivation and prototype of this truncation is a procedure applied
to the calculation of the gluon density in the MV model already
in~ Ref.\cite{Jalilian-Marian:1997xn} and heavily reused
since~\cite{Gelis:2001da, Blaizot:2004wu, Blaizot:2004wv,
  Marquet:2010cf, Dominguez:2011wm, Dominguez:2012ad}. The method
relies on the fact that Wilson lines are given as path ordered
exponentials in the gauge field as
\begin{equation}\label{eq:defU}
U_\xt = P \exp\left\{ i g \int \ud x^- A^{a,+}_{\ib{x},x^-} t^a\right\}
\end{equation}
and the assumption -- intrinsic to the MV model -- that the
correlators of the $A$ field in the exponent obey Gaussian statistics
with only a local correlation in the longitudinal coordinate
\begin{equation}\label{eq:nonlocalgauss}
g^2 \left< A^{a,+}_{\ib{x},x^-} A^{b,+}_{\ib{y},y^-}\right>
= \delta(x^--y^-) \delta^{ab} G_{x^-,\ib{xy}}
\end{equation}
The MV model further assumes a specific form of the correlation function
\begin{equation}
\partial_\ib{x}^2 \partial_\ib{y}^2 G_{x^-,\ib{xy}} = g^2\mu^2(x^-) \delta_\ib{xy} 
\end{equation}
but the latter is not a necessary ingredient for the truncations and
will not be assumed in the following.

One advantage of such a procedure is that all possible correlators
with any number of Wilson lines automatically obey all group
theoretical relations imposed in any possible coincidence limit, such
as the relationships listed in
Fig.~\ref{fig:correlator-relations}. More generally, we can achieve
this feature by parametrizing Wilson line correlators in a Gaussian
manner via
\begin{align}
  \label{eq:GT}
  %\hspace{-1.5cm}
  \langle \ldots \rangle(\eta) = \ \Bigl\langle 
    P_\eta e^{ \int\limits^\eta_{\eta_0} \ud \eta' \Bigl[
     \frac12 \int\limits_{\bm u\bm v}
    G_{\bm{u v}}(\eta') i\nabla_{\bm u}^a i\nabla_{\bm v}^a 
 \Bigr]} \ldots 
 \Bigr\rangle(\eta_0)
\ ,
\end{align}
where we have replaced $x^-$ by a general coordinate space
longitudinal (rapidity) variable $\eta \sim \ln x^-$.  Practically,
Eq.~\eqref{eq:GT} allows one to find a parametrization for any set of
correlators with consistent coincidence limits (such as listed in
Fig.~\ref{fig:correlator-relations}) by solving the functional
differential equation
\begin{align}  
 \label{eq:GT-diff}
 \frac{\ud}{\ud \eta} \langle F[ U ] &\rangle(\eta) 
 \\ \notag = &
   \frac{1}{2} \int\limits_{\ut,\vt}
 \left\langle G_\ib{uv}(X) \, \dd{a}{u} \dd{a}{v} 
   F \left[ U \right] \right\rangle(\eta)
\end{align}
to parametrize some \emph{singlet} observable $\langle F[U] \rangle$
in terms of the two-point correlator $G_\ib{uv}$. The simplest example
is the $q\Bar q$ correlator with $F[U] \to \tr(U_{\bm x}
U^\dagger_{\bm y})/\nc$ for which this functional equation indeed
turns into a closed differential equation. For more complicated
operators such as $q^2\Bar q^2$, the functional equation turns into a
set of coupled differential equations that need to be solved
simultaneously. The expectation values of several different Wilson
line operators with a Gaussian weight have been calculated in the
literature, e.g. in Refs.~\cite{Jalilian-Marian:1997xn,
  Blaizot:2004wu, Blaizot:2004wv, Dominguez:2008aa, Marquet:2010cf,
  Dominguez:2011wm, Iancu:2011nj, Dominguez:2012ad}.

The simplest generalization of the Gaussian truncation is what we will
call the 3-point exponential truncation. It includes the 2-point
contribution from the Gaussian truncation but adds all mathematically
independent three-point functions to the exponential. This can be
summarized by
\begin{multline}
  \label{eq:GT+}
  %\hspace{-1.5cm}
  \langle \ldots \rangle(\eta) =  \ \Biggl\langle 
    P_\eta \exp\Biggl\{ \int\limits^\eta_{\eta_0} \ud \eta' \Biggl[
    \frac12 \int\limits_{\bm u\bm v}
    G_{\bm{u v}}(\eta') i\nabla_{\bm u}^a i\nabla_{\bm v}^a 
%\right. \notag \\ & \left.
    \\ 
    \frac1{3!}
   \int\limits_{\bm u\bm v\bm w}
      G^d_{\bm{u v w}}(\eta')\ d^{a b c} 
    i\nabla_{\bm u}^a i\nabla_{\bm v}^b i\nabla_{\bm w}^c 
    \\ 
    \frac1{3!}
   \int\limits_{\bm u\bm v\bm w}
      G^f_{\bm{u v w}}(\eta')\ f^{a b c} 
    i\nabla_{\bm u}^a i\nabla_{\bm v}^b i\nabla_{\bm w}^c    
\\ 
+\text{4 pts }
 \Biggr]\Biggr\} \ldots (\eta_0)
 \Biggr\rangle
\ .
\end{multline}
As indicated there is, in principle, no obstruction to higher $n$-point
functions to this parametrization functional.

Note that in Eq.~\eqref{eq:GT+} both $G$ and $G^d$ are fully
symmetrical under exchange of any pair of transverse coordinates, while
$G^f$ has slightly more complicated features but will not enter any of
the correlators in Eqns.~(\ref{eq:DIScorr}) and~(\ref{eq:stsaUcorr})
so that we have no need to discuss it in any more detail.

We choose $G_{\bm u\bm u} =0$ and $G_{\bm u\bm u\bm
  u}=0$ as this simplifies some of the expressions below from the
outset. The correlators in \eq\nr{eq:balitsky}
% Eqns.~(\ref{eq:DIScorr}) and~(\ref{eq:stsaUcorr}) 
then take the form
\begin{subequations}
\label{eq:param-eqn}
 \begin{align}\label{eq:2pt}
  S_{\bm x\bm y} =   \frac{\langle \tr(U_{\bm x}U_{\bm y}^\dagger)\rangle}{\nc} = e^{-\cf \mathcal G_{\bm x\bm y}}\
  \end{align}
and
\begin{align}  \label{eq:3pt}
  \frac{\langle U_{\bm z}^{a b} \tr(t^a U_{\bm x} t^b U_{\bm y}^\dagger)\rangle}{2 \nc \cf} 
=
 e^{
-\left\{\left[\frac{\nc}2\left(
  \mathcal G_{\bm{x z}} + \mathcal G_{\bm{z y}} 
\right)
- \mathcal G_{\bm{x y}}\right)-  \cf\mathcal G_{\bm{x y}}
\right\}
}
  \end{align}
where
\begin{equation}
  \label{eq:calG}
  \mathcal G_{\bm x\bm y}(\eta) 
  = (\mathcal P + i \mathcal O)_{\bm x\bm y}
\end{equation}
\end{subequations}
whose real and imaginary parts are literally the $\mathcal P$ and
$\mathcal O$ introduced as parametrization functions of the complex
number $\langle\tr(U_{\bm x} U^\dagger_{\bm y})\rangle(Y)/\nc$. The
truncation procedure asserts that they also consistently parametrize
the $q\Bar q g$ correlator and provides an explicit expression of
$\mathcal G$ in terms of $G$ and $G^d$ and their respective initial
conditions at $\eta_0$:
\begin{subequations}
\label{PandOGT}
\begin{align}
  \mathcal P_{\bm x\bm y}(\eta) := & \int\limits_{\eta_0}^\eta \ud \eta' \  G_{\bm x\bm y} + \mathcal P_{\bm x\bm y}(\eta_0)= \mathcal P_{\bm y\bm x}(\eta)
\\ 
  i \mathcal O_{\bm x\bm y}(\eta) :=  &\frac{\cd}4 
  \int\limits_{\eta_0}^\eta \ud \eta' \left(
    G^d_{\bm{y x x}} - G^d_{\bm{y y x}} 
  \right) + i \mathcal O_{\bm x\bm y}(\eta_0)
  \notag \\ = &
   -i\mathcal O_{\bm y\bm x}(\eta).
\end{align}
\end{subequations}

Strikingly, the observables considered here do \emph{not} allow access
to $G_{\bm u\bm v\bm w}$ with all three coordinates independent
despite the fact that the $q\bar q g$ correlator features three
distinct coordinates. For $G_{\bm u\bm v\bm w}$ to occur with its full
coordinate dependence one needs for example a proton-like state which
remains outside the scope of this paper.

In the literature one often uses the same notation for the
factorization rapidity $Y$ and the longitudinal coordinate in the path
ordered exponential $\eta$ (or $\ln x^-$). With the physical
interpretation given above these are, however, not the same
quantity. The coordinate $x^-$ that gives the interpretation of the
Wilson line as a path ordered exponential in the color field is a
spatial coordinate along the trajectory of the path ordered
exponential. The local structure of the correlation function in the
longitudinal coordinate essentially imposes the Gaussian property on
the distribution of Wilson lines, since these are made up of
independent infinitesimal increments. The coordinate $x^-$ is,
however, not directly related to an experimental observable, but cross
sections always depend on Wilson lines integrated over $x^-$.

The rapidity $Y$, on the other hand, is a longitudinal \emph{momentum}
scale separating large- and small-$x$ degrees of freedom in the CGC
formalism. The scale $Y$ is a factorization scale in the
renormalization group evolution, and should be chosen according to the
typical longitudinal momentum scale in the studied scattering process.
A loose uncertainty principle argument states that for a momentum
space scale $Y$, the fields in the target are localized in a
coordinate space interval $\Delta x^- \sim \exp\{Y\}$. Thus,
identifying $\eta$ and $Y$ is indeed justified at the leading
logarithmic level (but not at higher orders in perturbation
theory). This identification is not needed or used anywhere in the
present calculation and we will keep the separate notation for these
two variables.

It is important to realize that Eq.~\eqref{eq:GT-diff} is a
parametrization equation that expresses singlet Wilson line
correlators in terms of a two-point function whose $Y$-dependence
needs to be derived \emph{separately} or is \emph{known} a priori.
Thus, the solutions of the parametrization equation,
e.g. \eqs\nr{eq:2pt} and \nr{eq:3pt}, have an unknown dependence on
the factorization rapidity $Y$ that will need to be derived from the
actual QCD dynamics.

One way to derive an equation for the dependence of the two-point
function $\mathcal{G}$ on the factorization rapidity $Y$ is to take
the \emph{solutions} of Eq.~\eqref{eq:GT-diff} and insert them into
the appropriate equations from the Balitsky hierarchy.  Every equation
of the hierarchy leads to a different equation for $\mathcal{G}$.
Choosing this equation to be the equation \nr{eq:balitsky} for the
expectation value of $\tr(U_{\bm x} U^\dagger_{\bm y})/\nc$ dipole
leads to \emph{a} Gaussian truncation of the JIMWLK hierarchy.

It is also possible to rewrite the parametrization
equation~\nr{eq:GT+} in terms of a more abstract longitudinal
coordinate, which can then be chosen to be equal to the evolution
rapidity. This procedure can be used to simplify the coupled evolution
equations for more complicated higher-point operators of Wilson
lines~\cite{Marquet:2010cf}, at the expense of losing the physical
interpretation of the longitudinal coordinate. Here we will
concentrate on the evolution equation for just the dipole operator,
and can easily remain with the physical interpretation of $\eta$ as
being related to $x^-$ and distinct from the momentum rapidity $Y$.

As indicated in Eq.~\eqref{PandOGT}, the effective two-point functions
\emph{do} show the required symmetry properties of $\langle\tr(U_{\bm
  x}U_{\bm y}^\dagger)\rangle/\nc$ under complex conjugation; one
obtains a consistent \emph{gauge invariant} truncation of the
associated Balitsky hierarchy with the evolution equation
\begin{align}
  \label{eq:evo-with-odd}
  \frac{\ud}{\ud Y}  {\cal G}_{\bm{x y}}%(Y)  
 = & \frac{\as}{\pi^2}\int  \ud^2 z\ \mathcal K_{\bm{x z y}} 
  \left(1-
      e^{-\frac{\nc}2 \left[
       {\cal G}_{\bm{x z}}+{\cal G}_{\bm{z y}}-{\cal G}_{\bm{x y}}
   \right]%(Y)
 }
 \right)
%\hspace{1cm}
%{\cal G}_{\bm{x y}}(Y) =(\mathcal P+i\mathcal O)_{\bm{x y}}(Y)
\end{align}
which obviously couples real and imaginary
parts. Equation~\nr{eq:evo-with-odd} generalizes the large-$\nc$
results of~\cite{Kovchegov:2003dm,Hatta:2005as} to finite $\nc$. 

The solutions to the parametrization equations~\eqref{eq:param-eqn} and
thus the evolution equation Eq.~(\ref{eq:evo-with-odd}) differ from their BK
counterpart for the total cross section in the Gaussian truncation only
through a non-vanishing imaginary part $i\mathcal O\neq 0$ appearing in $\mathcal G$.
% (this is
% one of the differences between our treatment and the discussion
% in~\cite{Kovchegov:2003dm,Hatta:2005as}). 
% TL removed: same is true for \cite{Kovchegov:2003dm,Hatta:2005as}
Indeed $i\mathcal O = 0$ and $\mathcal P\in [0,\infty]$ is a
consistent solution to this equation which leads to a successful
phenomenology for HERA data at small $x$~\cite{Kuokkanen:2011je}: If
the initial condition for ${\cal G}$ is real, the equation never
generates an imaginary part. This can be seen from the coupled
equations for the real and imaginary parts explicitly:
\begin{widetext}
  \begin{subequations}
\label{eq:evo-with-odd-reim}
\begin{align}
  \label{eq:evo-with-odd-re}
  \frac{\ud}{\ud Y}  {\cal P}_{\bm{x y}}(Y)  
 = & \frac{\as}{\pi^2}\int  \ud^2 \zt K_{\bm{x z y}} 
  \left[1-
      e^{-\frac{\nc}2 \left[
       {\cal P}_{\bm{x z}}+{\cal P}_{\bm{z y}}-{\cal P}_{\bm{x y}}
   \right](Y)}\cos\left(\frac{\nc}2 \left[
       {\cal O}_{\bm{x z}}+{\cal O}_{\bm{z y}}-{\cal O}_{\bm{x y}}
   \right](Y)\right)
 \right]
\\ \label{eq:evo-with-odd-im}
\frac{\ud}{\ud Y}  {\cal O}_{\bm{x y}}(Y)  
 = & \frac{\as}{\pi^2}\int \ud^2 \zt K_{\bm{x z y}} 
  \left[
      e^{-\frac{\nc}2 \left[
       {\cal P}_{\bm{x z}}+{\cal P}_{\bm{z y}}-{\cal P}_{\bm{x y}}
   \right](Y)}\sin\left(\frac{\nc}2 \left[
       {\cal O}_{\bm{x z}}+{\cal O}_{\bm{z y}}-{\cal O}_{\bm{x y}}
   \right](Y)\right)
 \right] .
\end{align}
\end{subequations}
\end{widetext}
% The odderon contribution  will naturally shrink no matter which
% initial condition is used for it.

\section{Evolution in the 3-point exponential truncation}
\label{sec:3pt}

There are many different ways to rewrite the equation before
implementing it numerically. The most prominent among these is the
possibility to map~\eqref{eq:evo-with-odd} into the BK equation, as
noted in~\cite{Weigert:2005us}. Let us briefly recapitulate how this
is done. Inserting the Gaussian correlator parametrizations of the
3-point exponential truncation~\eqref{eq:param-eqn} into the first
equation of the Balitsky hierarchy \eqref{eq:balitsky} and canceling
an overall factor of $\nc$, we get
\begin{align}
\label{eq:evo-with-odd-precursor}
\frac{\ud}{\ud Y} &
e^{-\cf \mathcal{G}_{\bm x\bm y}} 
= 
\frac{\as \cf }{\pi^2}
 \int \ud^2 \zt \, \Tilde{\mathcal K}_{\bm x\bm z\bm y}
\\ \notag  & \times
 \Bigl(
    e^{-\frac{\nc}2 \left[
       {\cal G}_{\bm{x z}}+{\cal G}_{\bm{z y}}-{\cal G}_{\bm{x y}}
   \right]%(Y)
 -\cf \mathcal{G}_{\bm x\bm y}} 
-e^{-\cf \mathcal{G}_{\bm x\bm y}} 
\Bigr)
\end{align}
From here it is straightforward to arrive at
Eq.~\eqref{eq:evo-with-odd} after canceling an overall factor
$e^{-\cf \mathcal{G}_{\bm x\bm y}}$ common to both sides.

To derive the relation between the BK equation and the Gaussian
truncation one starts by multiplying both sides of
Eq.~\eqref{eq:evo-with-odd} with $-\tfrac{\nc}2 e^{-\frac{\nc}2
  {\mathcal G}_{\bm x\bm y}}$, leading to the alternative form
 \begin{align}
\label{eq:BK-G}
\frac{\ud}{\ud Y} &
e^{-\frac{\nc}2 {\mathcal G}_{\bm x\bm y}} 
= 
\frac{\as}{\pi^2}\frac{\nc}2
 \int \ud^2 \zt \, {\mathcal K}_{\bm x\bm z\bm y}
\\ \notag  & \times
 \Bigl(
 e^{-\frac{\nc}2 \left[
     {\mathcal G}_{\bm{x z}}+{\mathcal G}_{\bm{z y}}
   \right]
 }
 -e^{-\frac{\nc}2 {\mathcal G}_{\bm x\bm y}} 
\Bigr)
\end{align}
for this evolution equation.
Now, after identifying
\begin{equation}
  \label{eq:tildeS}
  S_{\bm x\bm y}^{\text{BK}} :=e^{-\frac{\nc}2 {\mathcal G}_{\bm x\bm y}},
\end{equation}
it is manifest that Eq.~\eqref{eq:BK-G} is equivalent to the BK equation
\begin{align}
\label{eq:BK}
\frac{\ud}{\ud Y} &
S_{\bm x\bm y}^{\text{BK}}
= 
\frac{\as }{\pi^2}\frac{\nc}2
 \int \ud^2 \zt \, \Tilde{\mathcal K}_{\bm x\bm z\bm y}
%\\ \notag  & \times
 \Bigl(
 S_{\bm x\bm z}^{\text{BK}} S_{\bm z\bm y}^{\text{BK}}
 - S_{\bm x\bm y}^{\text{BK}}.
\Bigr)
\end{align}
Equations~\eqref{eq:evo-with-odd},~\eqref{eq:evo-with-odd-precursor},~\eqref{eq:BK-G}
and~\eqref{eq:BK} are all equivalent: they all determine the evolution
for the same function $\mathcal G$ and will lead to the same
$Y$-dependence provided we set the same initial condition on $\mathcal
G$. The difference between the Gaussian truncation and the large-$\nc$
BK equation (in the sense the term is commonly used) is in the
relation between the physical scattering amplitude and the solution of
the evolution equation. In the large-$\nc$ limit the solution of the
BK equation $S^{\text{BK}}$ is assumed to be the physical scattering
amplitude. In the finite-$\nc$ Gaussian truncation the physical
scattering amplitude is related to the fundamental two-point
correlator $\mathcal{G}$ by \eqref{eq:evo-with-odd} and to the
solution of the BK equation by \nr{eq:BK-G}. Thus the physical
scattering amplitude is obtained from the solution of the BK equation
as
\begin{align}
S_{\bm x\bm y} 
= \left\langle 
  \frac{\tr(U_{\bm x} U^\dagger_{\bm y})}{\nc} 
\right\rangle
= \left(S_\ib{xy}^{\text{BK}}\right)^{\frac{2\cf}{\nc}} .
\label{eq:gtvsbkodd}
\end{align}
Thus the real and imaginary parts of the physical dipole operator in
the Gaussian truncation $S_\ib{xy}$, expressed in terms of the real
and imaginary parts of the BK-evolved dipole, read
\begin{align}
\label{eq:bktogtre}
\mathsf{Re}\ { S_\ib{xy}} =& \left|S_\ib{xy}^{\text{BK}}\right|^{\frac{2\cf}{\nc}}
 \cos \left\{
\frac{2\cf}{\nc} \arctan \frac{\mathsf{Im}\ S_\ib{xy}^{\text{BK}}}{\mathsf{Re}\ S_\ib{xy}^{\text{BK}}}
\right\}
\\
\label{eq:bktogtim}
\mathsf{Im}\ { S_\ib{xy} } =& \left|S_\ib{xy}^{\text{BK}}\right|^{\frac{2\cf}{\nc}} 
\sin 
\left\{
\frac{2\cf}{\nc} \arctan \frac{\mathsf{Im}\  S_\ib{xy}^{\text{BK}}}{\mathsf{Re}\ S_\ib{xy}^{\text{BK}}}
\right\}.
\end{align}
Note that a purely real solution of the BK equation still gives a
purely real dipole expectation value, but the presence of an imaginary
part in the BK equation affects both the real and imaginary parts of
the physical dipole.

Separating the identity from the BK-equation dipole operator
$S^{\text{BK}}$, one gets the scattering amplitude
$N^{\text{BK}}_\ib{r}= 1-S^{\text{BK}}_\ib{r}$, where $\rt =
\xt-\yt$. In line with our earlier conventions in
Eq.~\eqref{eq:pom-odd} we denote its real and imaginary parts as
$P^{\text{BK}}_\ib{r} := \mathsf{Re}\ (N^{\text{BK}}_\ib{r})$, the
BK-\emph{pomeron} and $O^{\text{BK}}_\ib{r} := \mathsf{Im}\
(N^{\text{BK}}_\ib{r})$, the BK-\emph{odderon}.  In terms of these,
the evolution equation~\eqref{eq:BK} now becomes a set of two real
integro-differential equations
%--------------------
\begin{eqnarray}
\frac{\ud P^{\text{BK}}_\ib{r}}{\ud Y} &=& \frac{\as \nc}{2 \pi^2} 
\, \int \ud^2 \rtp \, \frac{\rt^2}{\rtp^2 \, \rtpp^2} 
\left( P^{\text{BK}}_{\ib{r}'} + P^{\text{BK}}_{\ib{r}''}
\right.
\nonumber \\ && \quad 
\left.
 - P^{\text{BK}}_\rt - P^{\text{BK}}_{\rtp} P^{\text{BK}}_{\rtpp} 
+ O^{\text{BK}}_{\rtp} O^{\text{BK}}_{\rtpp} \right) 
\label{eq:final_re} 
\\
\frac{\ud O^{\text{BK}}_\rt}{\ud Y} &=& \frac{\as \nc}{2 \pi^2} \, \int \ud^2 \rtp \,
 \frac{\rt^2}{\rtp^2 \, \rtpp^2} 
\left( O^{\text{BK}}_{\rtp} + O^{\text{BK}}_{\rtpp} - O^{\text{BK}}_\ib{\rt} 
\right.
\nonumber \\ && \quad 
\left.- P^{\text{BK}}_{\rtp} O^{\text{BK}}_{\rtpp} - O^{\text{BK}}_{\rtp} P^{\text{BK}}_{\rtpp} \right),
\label{eq:final_im}
\end{eqnarray}
where $\rtpp \equiv \rt-\rtp$.  These equations are identical to those
derived in Refs.\cite{Kovchegov:2003dm,Hatta:2005as} in the
large-$\nc$ limit. As discussed above, what changes at finite $\nc$ in
the Gaussian truncation is the relation between the solution of these
equations and the physical scattering amplitude, which is now given by
Eqs.~\nr{eq:bktogtre} and~\nr{eq:bktogtim}.

Recall that from the definition of $S_\rt$ it follows that $S_\rt^* =
S_{-\rt}$. This in turn imposes separate symmetry properties on the
real and imaginary parts of $\mathcal G = \mathcal P + i\mathcal O$
which are correctly reproduced by our truncation in
Eq.~\eqref{PandOGT}. Via~\eqref{eq:tildeS} they directly imply that
${S^{\text{BK}}_\rt}^* = S^{\text{BK}}_{-\rt}$. Thus the real and
imaginary parts of the amplitude are odd or even under reflections:
\begin{eqnarray}
\label{eq:Psymm}
 P^{\text{BK}}_{-\rt} &=&  P^{\text{BK}}_{\rt}
\\
\label{eq:Osymm}
 O^{\text{BK}}_{-\rt} &=&  -O^{\text{BK}}_{\rt} .
\end{eqnarray}
For the linear BFKL part of the equation it is particularly convenient
to decompose the solution to the evolution equation in terms of
eigenfunctions of the kernel, both in $|\rt|$ and in azimuthal angle.
For the nonlinear case it is more convenient to continue working in
$r$-space, but we can still perform a Fourier series expansion in the
azimuthal angle.  The symmetry \eqref{eq:Psymm},~\eqref{eq:Osymm}
dictates that the pomeron and odderon can only have even or odd
harmonics, respectively:
\begin{eqnarray} \label{eq:pomharmonics}
 P^{\text{BK}}_{\rt} &=&  \sum_{n=0}^\infty P^{\text{BK}}_{2n}(r)\cos(2n\varphi_\rt)
\\ \label{eq:oddharmonics}
 O^{\text{BK}}_{\rt} &=&  \sum_{n=0}^\infty O^{\text{BK}}_{2n+1}(r)\cos((2n+1)\varphi_\rt),
\end{eqnarray}
where $\varphi_\rt$ is the angle of the vector $\rt$ with respect to
an (arbitrary) reaction plane.  Based on the known BFKL dynamics in
the linear regime, we expect the small azimuthal harmonics to dominate
in the high energy regime.  Our working hypothesis here is that the
same is true also in the nonlinear case.  This allows us to
efficiently study the equations by truncating the series and only
keeping the lowest harmonics. We can control the error made in this
approximation ex post by calculating the rapidity dependence of the
first neglected term in the series from the ones that are kept.

A quick examination of the equations \eqref{eq:final_re}
and~\eqref{eq:final_im} reveals that the usual azimuthal
$P^{\text{BK}}_{0}(r)$ is a fixed point of the equation. Including the
lowest odderon harmonic $O^{\text{BK}}_{1}(r)$ will generate a second
pomeron $P^{\text{BK}}_{2}(r)$ through the $\left({O^{\text{BK}}}(r)\right)^2$ term
in \eqref{eq:final_re}, which again will generate a higher
$O^{\text{BK}}_{3}(r)$ harmonic through the nonlinear
$P^{\text{BK}}O^{\text{BK}}$ coupling in \eqref{eq:final_im}. In this
way, including an odderon will automatically generate an infinite
tower of higher harmonics in both the odderon and the pomeron
amplitudes. In principle, it could thus be possible that, through this
coupling, an odderon component would have an observable signal in a
P-even observable such as dijet correlations at an
electron-ion-collider~\cite{Dumitru:2015gaa}.

\begin{figure*}[tb]
\begin{center}
\includegraphics[width=0.49\textwidth]{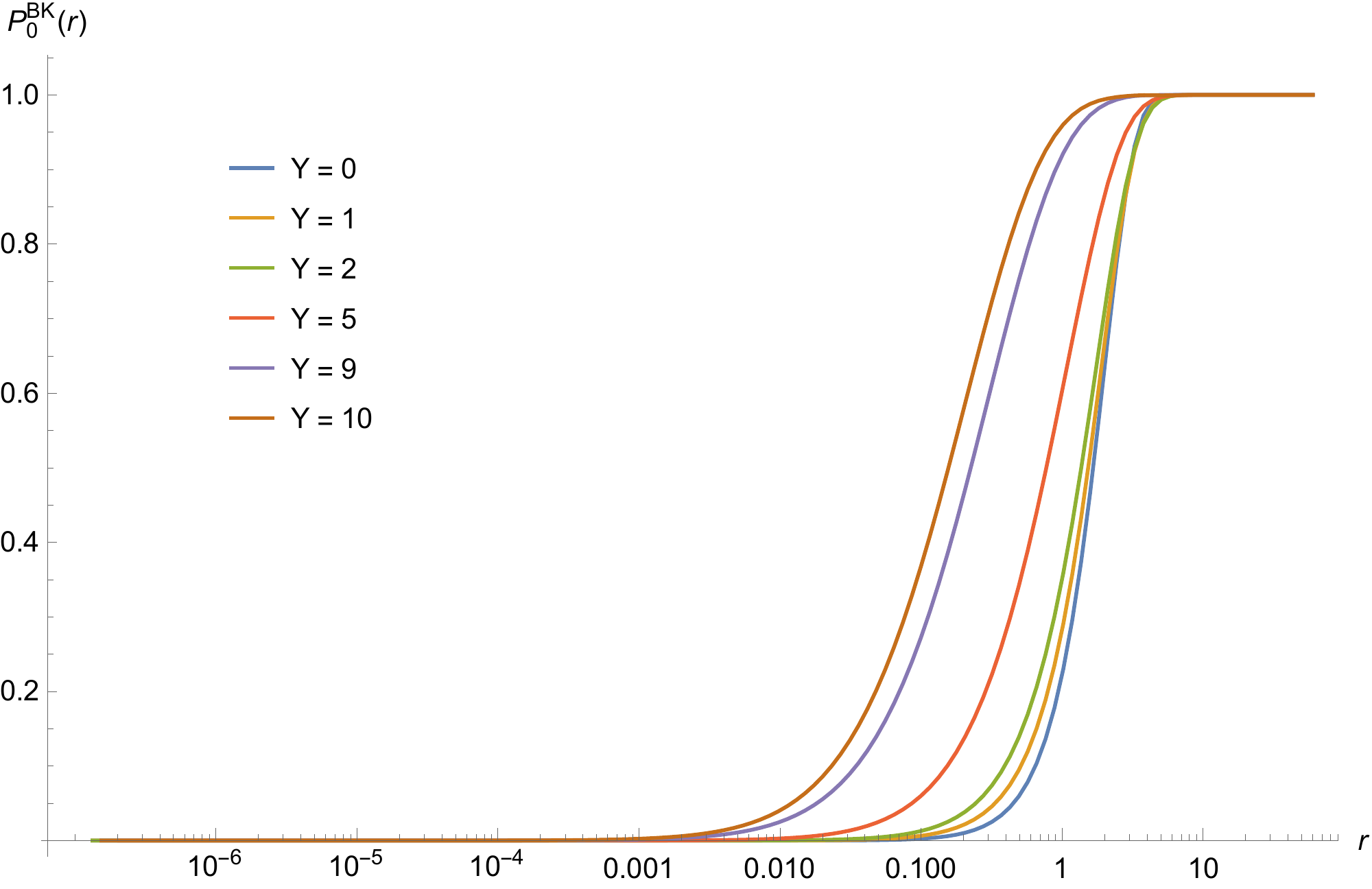}
\includegraphics[width=0.49\textwidth]{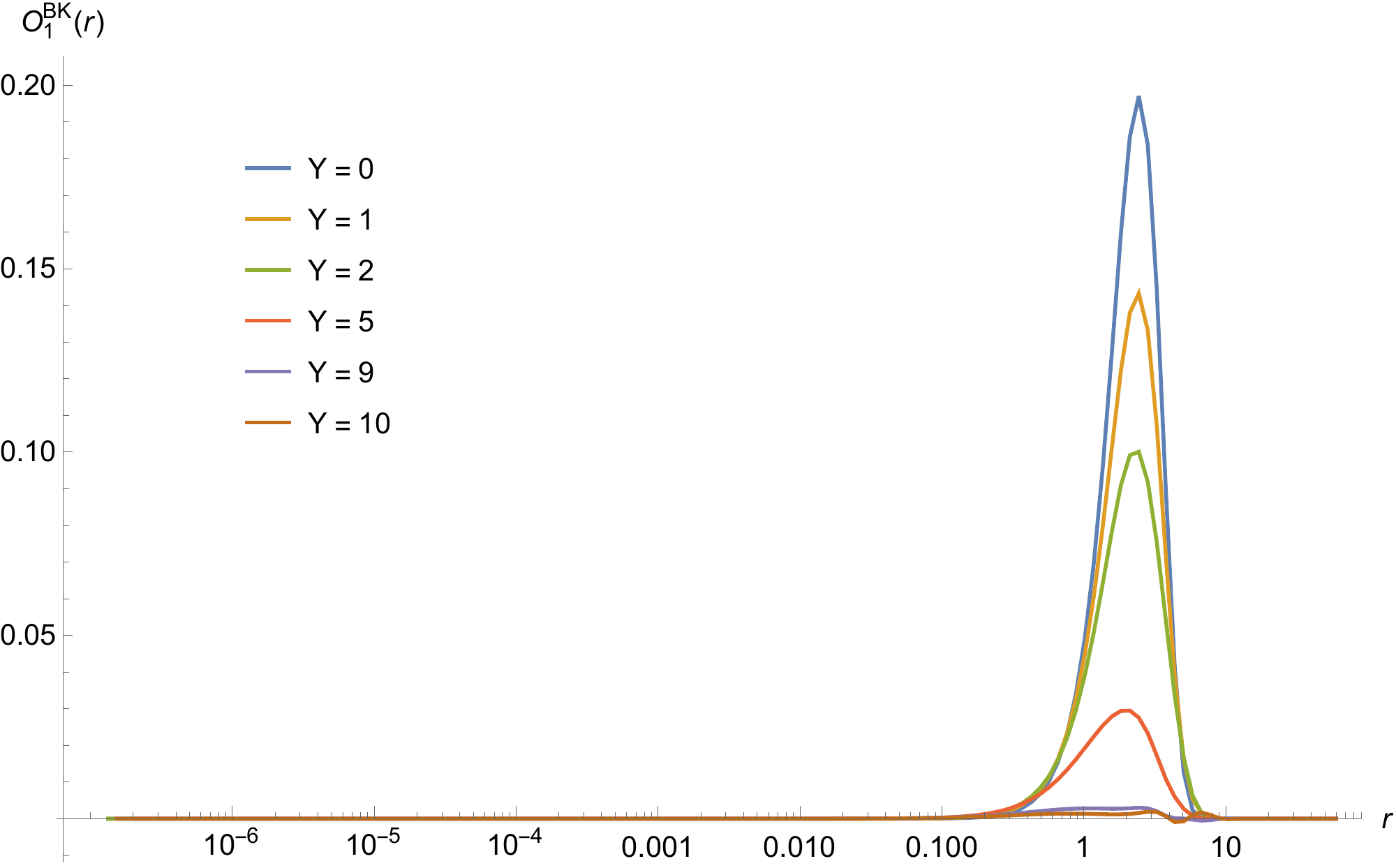}
\caption{Evolution of the pomeron (left) and odderon (right) amplitudes according to \eqs\nr{eq:trunc_re} and~\nr{eq:trunc_im} with the initial condition \nr{eq:ser-initial} with $\kappa=1/3$.}
\label{fig:evol_BK}
\end{center}
\end{figure*}

\begin{figure}[tb]
\begin{center}
\includegraphics[width=0.49\textwidth]{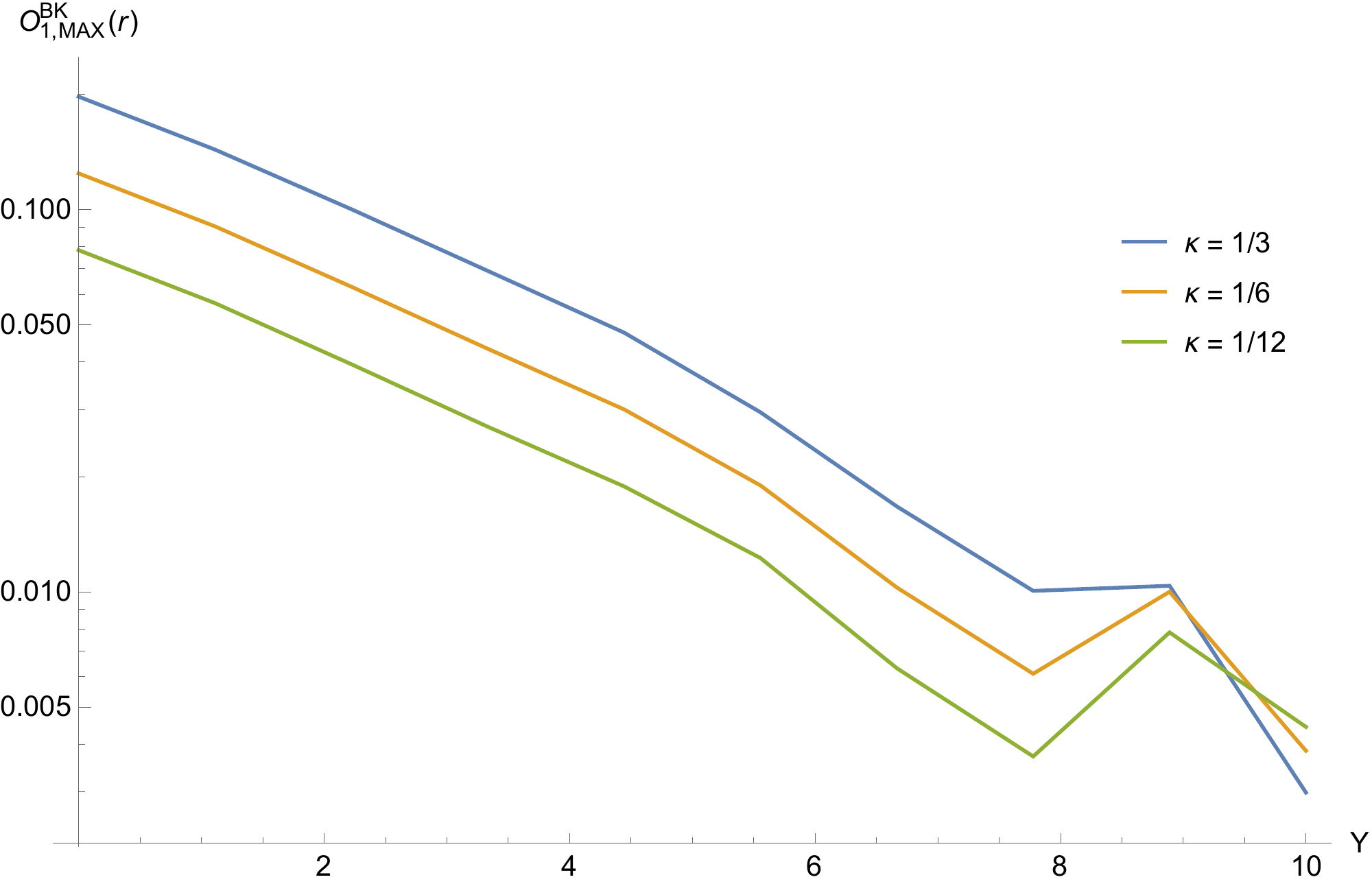}
\caption{Plot of the height of the odderon peak (as shown in Fig. \ref{fig:evol_BK}) as a function of rapidity. Three different values
$\kappa=1/3, 1/6$ and $1/12,$ are shown.
%  $Q_{s\mathcal{O}0}^2$ are considered, relative to $Q_{s0}^2 = 0.2$.
% {\bf Change to $\kappa=1/3, 1/6$ and $1/12,$ i.e. 
% $Q_{s0}^2 = 1,$ $Q_{s\mathcal{O}0}^2 = (1/3)^{2/3},(1/6)^{2/3},(1/12)^{2/3} $
% }
}
\label{fig:evol_im}
\end{center}
\end{figure}
%--------------------
\begin{figure}[tb]
\begin{center}
\includegraphics[width=0.49\textwidth]{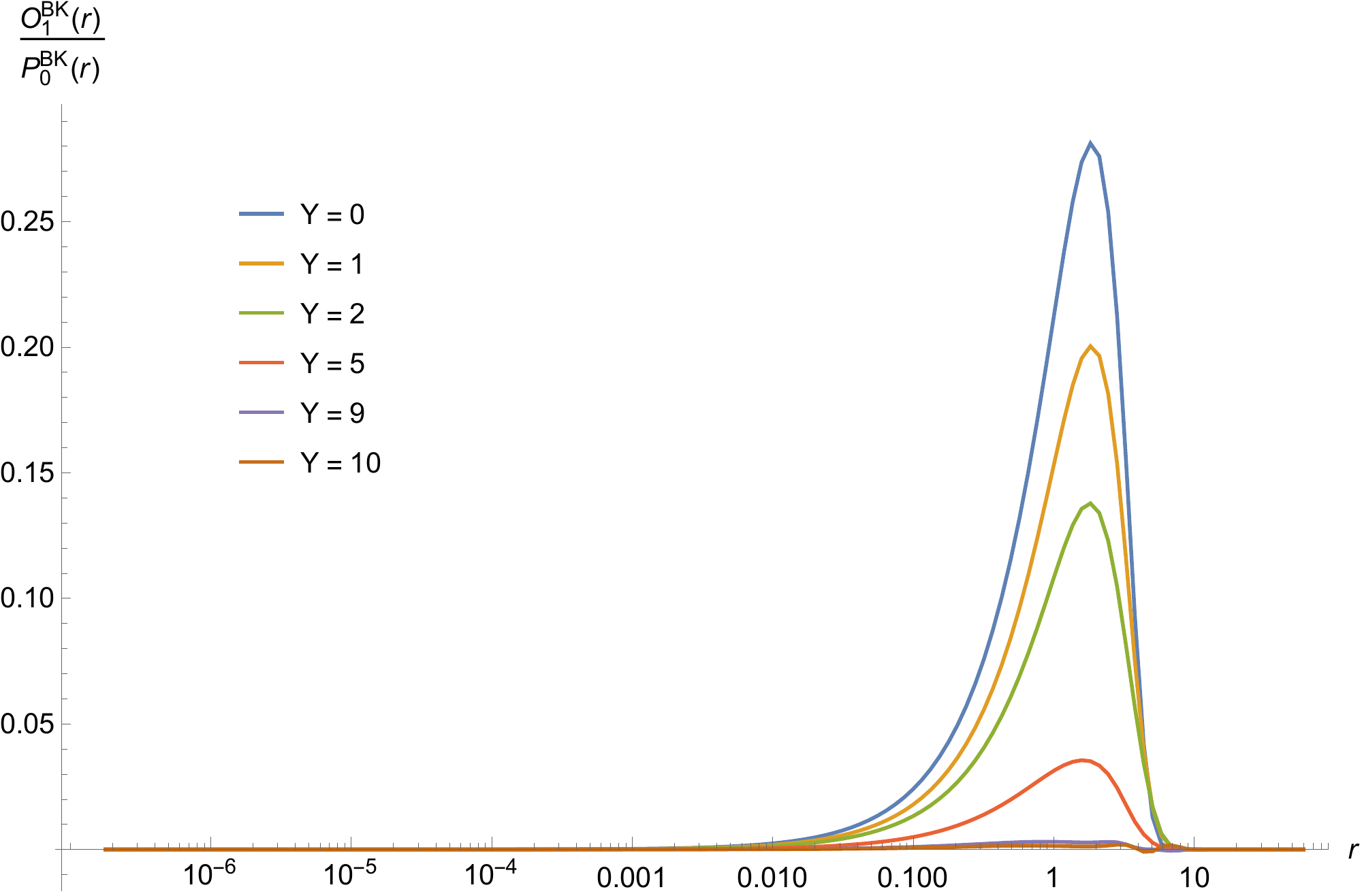}
\caption{Ratio of $\mathcal{O}(r)/\mathcal{P}(r)$  for $\kappa=1/3$ as a function of $r$ at three different rapidities.
}
% {\bf this now has the inconsistent in initial condition $\mathcal{O}\sim r^2$, because the ratio at $Y=0$ is a constant. Either redo with the $r^3$ initial condition  and
%  $Q_{s0}^2=1$, $Q_{s\mathcal{O}0} = Q_{s0}/\sqrt[3]{3}$ i.e. $\kappa=1/3$ or remove figure}
\label{fig:ratio}
\end{center}
\end{figure}
%--------------------

For this first numerical study we have truncated the series to the
smallest harmonics of both the pomeron and odderon,
i.e. $P^{\text{BK}}_{0}(r)$ and $O^{\text{BK}}_{1}(r)$, by dropping
the $ O^{\text{BK}}_{\rtp} O^{\text{BK}}_{\rtpp}$ term from the
pomeron evolution equation \nr{eq:final_re}. With this truncation, the
pomeron amplitude stays rotationally invariant and the odderon
equation \nr{eq:final_im} can be solved as such with only the lowest
harmonic $O^{\text{BK}}_{1}(r)$. To see this explicitly, note that
with only the azimuthally symmetric component in the pomeron
amplitude, we can write the odderon equation as

\begin{multline}
 \frac{\ud O^{\text{BK}}_1(r) \cos\theta}{\ud Y} = \frac{\as \nc}{2 \pi^2} \, \int \ud^2 \rtp \,
 \frac{\rt^2}{\rtp^2 \, \rtpp^2} 
\\ \times
\left( O^{\text{BK}}_1(r')\cos\theta' + O^{\text{BK}}_1(r'')\cos\theta'' - O^{\text{BK}}_1(r) \cos\theta
\right.
 \\ 
\left.- P^{\text{BK}}_0(r') O^{\text{BK}}_1(r'')\cos\theta''
 - O^{\text{BK}}_1(r') P^{\text{BK}}_0(r'')\cos\theta' \right),
\end{multline}
where $\theta$,$\theta'$ and $\theta''$ are the angles of $\rt$,$\rtp$ and $\rtpp$ with respect to the $x$-axis. 
In practice we can solve \eq\nr{eq:final_im} by choosing the vector $\rt$ to  lie on the $x$-axis, with thus $\cos \varphi_{\rt}=1$. 
To show explicitly that this is the case, we use
$(r'')^2 = r^2+(r')^2 -2rr'\cos(\theta-\theta')$ and $\cos\theta'' = (r\cos \theta-r'\cos\theta')/r''$ to write this as
\begin{multline}\label{eq:odd1intermediate}
 \frac{\ud O^{\text{BK}}_1(r) \cos\theta}{\ud Y} = \frac{\as \nc}{2 \pi^2}  \int \ud r' \ud \theta' r'
 \frac{r^2}{(r')^2 (r'')^2)} 
\\ \times
\bigg[ \left( 1-P^{\text{BK}}_0(r'') \right) O^{\text{BK}_1(r')} \cos \theta'
\\
+ \left(1-P^{\text{BK}}_0(r') \right) O^{\text{BK}}_1(r'')
\frac{r\cos \theta-r'\cos\theta'}{r''}
\\
- O^{\text{BK}}_1(r) \cos\theta
\bigg].
\end{multline}
We then take $\phi=\theta'-\theta$ as a new integration variable and
use the identity $\cos(\theta') = \cos \phi \cos\theta-\sin\phi\sin
\theta$. Now the terms that are proportional to $\sin \theta$ are also
proportional to $\sin \phi$ times an even function of $\phi$ and
vanish upon integration over $\phi$, leaving every term on the
r.h.s. of \eq\nr{eq:odd1intermediate} proportional to $\cos
\theta$. Thus, as discussed earlier, with this approximation of a
$\theta$-independent pomeron amplitude the equation for the
$\cos\theta$-harmonic of the odderon closes. We can cancel the
$\cos\theta$ from \eq\nr{eq:odd1intermediate}, and we are left with
the truncated set of equations
\begin{eqnarray}
\frac{\ud P^{\text{BK}}_0(r)}{\ud Y} &=& \frac{\as \nc}{2 \pi^2} 
\, \int \ud^2 \rtp \, \frac{\rt^2}{\rtp^2 \, \rtpp^2} 
\left( P^{\text{BK}}_0(r') + P^{\text{BK}}_0(r'')
\right.
\nonumber \\ && \quad 
\left.
 - P^{\text{BK}}_0(r) - P^{\text{BK}}_0(r') P^{\text{BK}}_0(r'') 
 \right) 
\label{eq:trunc_re} 
\\
\label{eq:trunc_im}
 \frac{\ud O^{\text{BK}}_1(r)}{\ud Y} &=& \frac{\as \nc}{2 \pi^2}  \int \ud r' \ud \phi r'
 \frac{r^2}{(r')^2 (r'')^2)} 
\\ \nonumber && \times
\bigg[ \left(1-P^{\text{BK}}_0(r'') \right) O^{\text{BK}_1(r')} \cos \phi
\\ \nonumber &&
+ \left(1-P^{\text{BK}}_0(r') \right)O^{\text{BK}}_1(r'')
\frac{r-r'\cos\phi}{r''}
\\ \nonumber &&
- O^{\text{BK}}_1(r) 
\bigg],
\end{eqnarray}
with $(r'')^2 = r^2+(r')^2 -2rr'\cos\phi$.

We will now proceed to numerically solve \eqs\nr{eq:trunc_re}
and~\nr{eq:trunc_im}. We will parametrize the initial condition as in
\eq\nr{eq:ser-initial}:
 \begin{align}
 \left. P^{\text{BK}}_0(r'') \right|_{Y=0} &= 
 1 - \exp \left\{ \frac{- Q_{0}^2 r^2}{4} \right\}
 \label{eq:icp} \\
   \left. O^{\text{BK}}_1(r)  \right|_{Y=0} 
   &= 
 - \kappa \exp \left\{ \frac{- Q_{0}^2 r^2}{4} \right\} 
 \left(\frac{Q_{0}^3 r^3}{8} \right),
  \label{eq:ico}
 \end{align}
 with the maximal value $\kappa = 1/3$.  Figure~\ref{fig:evol_BK}
 shows the resulting amplitudes. For the pomeron part one sees the
 familiar ``traveling wave'' solution moving towards smaller dipoles
 with rapidity. The odderon amplitude, on the other hand, merely
 decreases in magnitude but its characteristic dipole size scale does
 not decrease. This behavior is quantified further in
 Figs.~\ref{fig:evol_im} and~\ref{fig:ratio}, showing the height of
 the odderon amplitude peak as a function of rapidity and the ratio of
 the (BK) odderon to the (BK) pomeron amplitude.

\begin{figure}[tb]
\begin{center}
\includegraphics[width=0.49\textwidth]{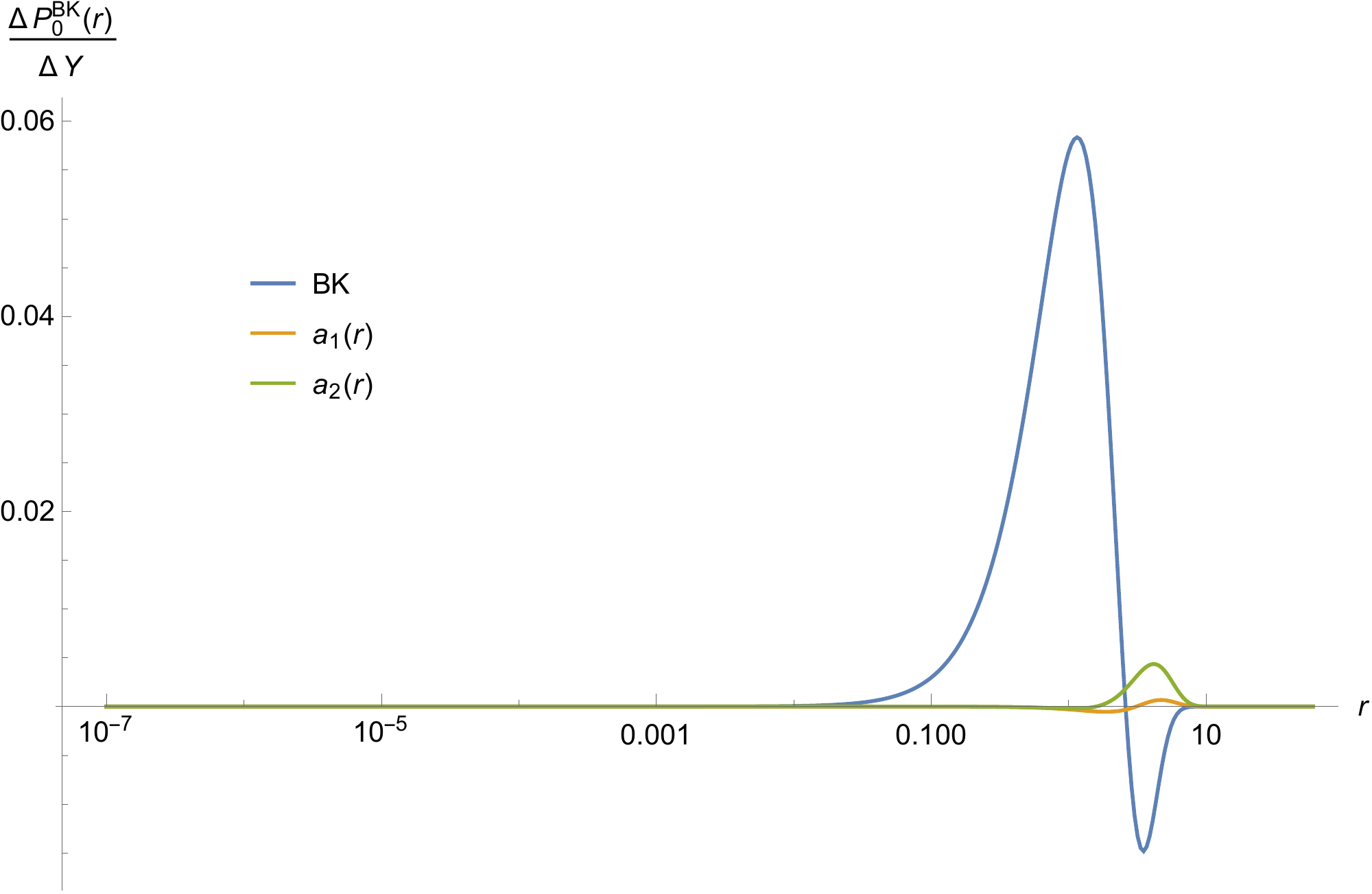}
\caption{Contribution to $\Delta \mathcal{P}_\ib{r}/\Delta Y$ in one
  step in rapidity from: the BK equation, $a_1(r)$ and $a_2(r)$, with
  the maximal odderon amplitude $\kappa=1/3$.  }
\label{fig:coeffs}
\end{center}
\end{figure}

In the calculations presented above, we have neglected the odderon
squared term in the evolution equation. This can be justified by the
fact that since it has not yet been unambiguously observed
experimentally, the odderon amplitude can be expected to be small.
Also expectations based on the linear evolution equation would lead to
an odderon amplitude that decreases as a function of
rapidity~\cite{Kovchegov:2003dm,Hatta:2005as}. Within the truncation
of the harmonic series \nr{eq:pomharmonics}, \nr{eq:oddharmonics} one
can estimate the size of this approximation by evaluating the
contribution of the first neglected term on one time step, i.e. the
contribution of the odderon squared term to the evolution equation of
the pomeron. As discussed previously, the square of the odderon term
$O^{\text{BK}}_1(r) \cos \theta$ gives both a $\theta$-independent and
a $\cos 2 \theta$ contribution to the evolution equation of the
pomeron. We denote the coefficients of these by $a_1$ and $a_2$,
i.e. we write
\begin{equation}
 \frac{\ud P^{\text{BK}}_\rt}{\ud Y} = [\textnormal{BK}] + a_1(r) + a_2(r) \cos(2 \theta_{\rt}) .
\end{equation}
We can now compare the odderon terms to the rotationally invariant
solution. Figure~\ref{fig:coeffs} shows the initial condition for the
fairly large value of $\kappa=1/3$. It can be seen that the odderon
squared terms are negligible in the small-$r$ region that drives the
evolution. The $\theta$-independent $a_1$-term is particularly small,
while $a_2$ is slightly larger. We conclude that the main effect of
the nonlinear odderon term is not to modify the evolution of the
rotationally invariant $ P^{\text{BK}}_0(r)$ amplitude, but to
introduce a small $\cos 2\theta$ term into the pomeron.

\section{Full leading order JIMWLK evolution with odderon admixture}
\label{sec:jimwlk}

Since the evolution equation~\eqref{eq:evo-with-odd}
or~\eqref{eq:evo-with-odd-reim} is the result of a truncation of the
full functional JIMWLK evolution, one may legitimately ask if the
truncation deviates quantitatively or even qualitatively from a full
JIMWLK simulation. This can be performed at leading order or at
partial next-to-leading order with running coupling corrections
included. Since in this paper we are only interested in qualitative
behavior we have, for simplicity, chosen the former option.

What is quite remarkable is that the only thing that needs to change
to perform a simulation run is the initial condition. The Langevin
simulation governing JIMWLK evolution is carried out on a square
transverse grid. A simulation for the total cross section starts from
an initial condition that treats both of the principal directions of
the transverse plane in the same way. To obtain an odderon admixture
one must break this lattice remnant of rotational invariance and
introduce a bias towards one of the directions into the initial
condition -- the code used for evolution needs no modification at all.

In this work we investigate the properties of the parity-odd initial
states with the following simple setup: we use a
lattice of size $L^2$ with periodic boundary conditions. 
First we generate an ensemble of standard parity-even initial states with
a probability distribution 
\begin{equation}
  P(\mathsf{Re} \ \tr(U_{\bm x} U^\dagger_{\bm y}) ) \propto \exp(-(x-y)^2/4R^2),
\end{equation}
using the methods described in \cite{Rummukainen:2003ns}.  This
generates configurations with saturation scale $\qs \sim 1/R$.  The
expectation value of the imaginary part of $\langle \tr(U_{\bm x}
U^\dagger_{\bm y}) \rangle/\nc$ vanishes and its real part falls into
the interval between $0$ and $1$. Such an ensemble is a
suitable starting point for a simulation without an odderon admixture.

    To introduce an odderon contribution one needs to generate an
    imaginary part. A convenient way of doing so is to consider a
    ``potential''
\begin{equation}
  V[U](\bm x) = \alpha \sum_{y} \mathsf{Im}\ \tr({\uu{x} \uud{y}}) 
  \,f(\bm x-\bm y) ,
\end{equation}
where $\alpha$ is a small real parameter and $f$ is an odd function of
$\bm r$ that breaks the symmetry between the two coordinate
directions:
\begin{equation}
  f(\bm x) = x_1 e^{-x^2/4R^2} \frac{(x_1^2 - (L/2)^2)(x_2^2 - (L/2)^2)}{L^4}
\ .
\end{equation}
The last part of the expression merely ensures that $f(x)$ vanishes at
the boundaries $x_i = \pm L/2$ so that no discontinuities arise there.

We use a left derivative to define the force induced by the
``potential'' $V$ as
\begin{equation}
  F_a(x) = 
  -\alpha \sum_y\mathsf{Re}\ \tr(\lambda_a \uu{x}\uud{y}) f(\bm x-\bm y),
\end{equation}
and make an update $\uu{x} \rightarrow e^{i\alpha F_a(x)\lambda_a} \uu{x}$.
This update is repeated a few times for all sites $\bm x$.  The magnitude
of the parity-odd contribution can be modified by adjusting the constant
$\alpha$ and the number of update steps.

In Figs.~\ref{fig:JIMWLK-sim-no-odd},~\ref{fig:JIMWLK-sim-weak-odd}
and~\ref{fig:JIMWLK-sim-strong-odd} we show the behavior of the real
and imaginary parts of $\tr(\uu{x}\uud{y})$ as measured on $L^2 =
128^2$ lattices, using a) none, b) mild and c) very large odderon
contributions, respectively.  Let us discuss the three cases in turn:

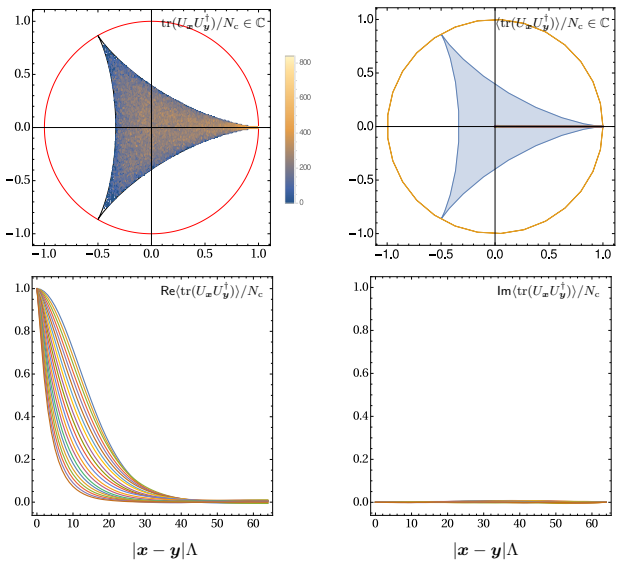
\begin{figure}[tb]
  \centering
\resizebox{\linewidth}{!}{
\begin{tikzpicture}
 \matrix(m)[matrix of math nodes,align=right,column sep=.6em,ampersand replacement=\u]
   { \diagram[height=4cm]{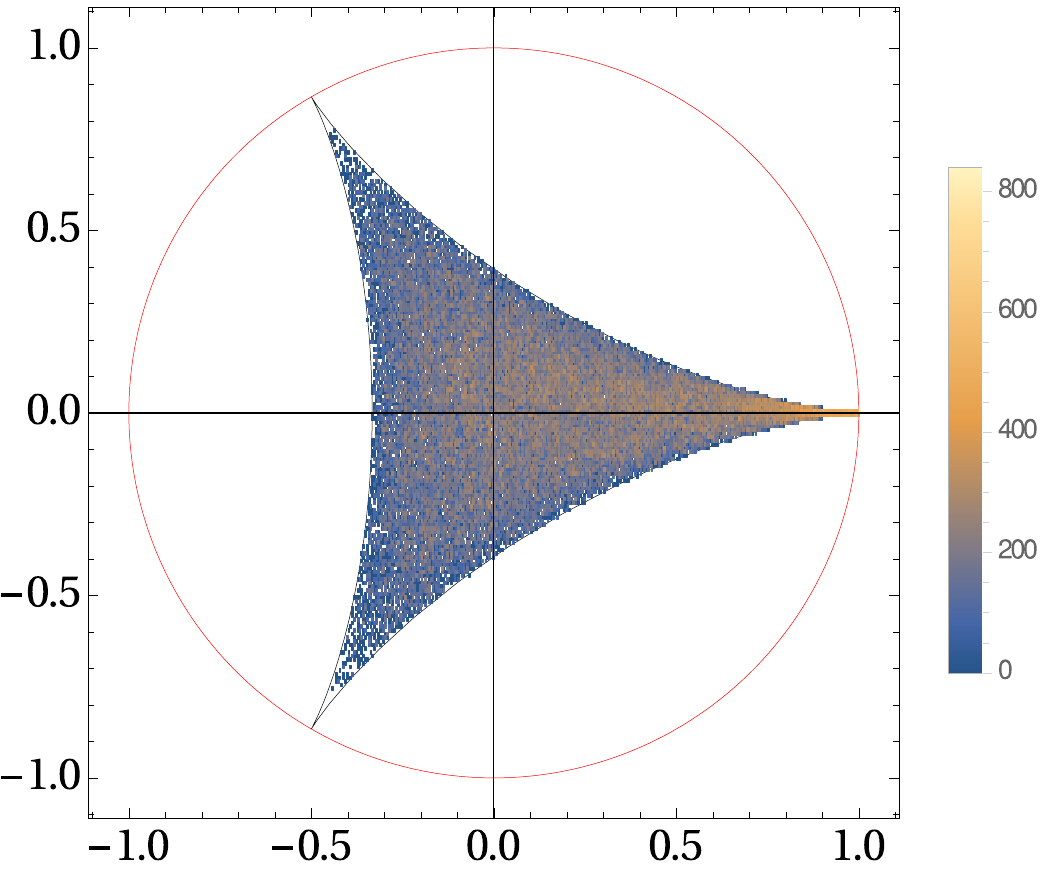} \u
     \diagram[height=4cm]{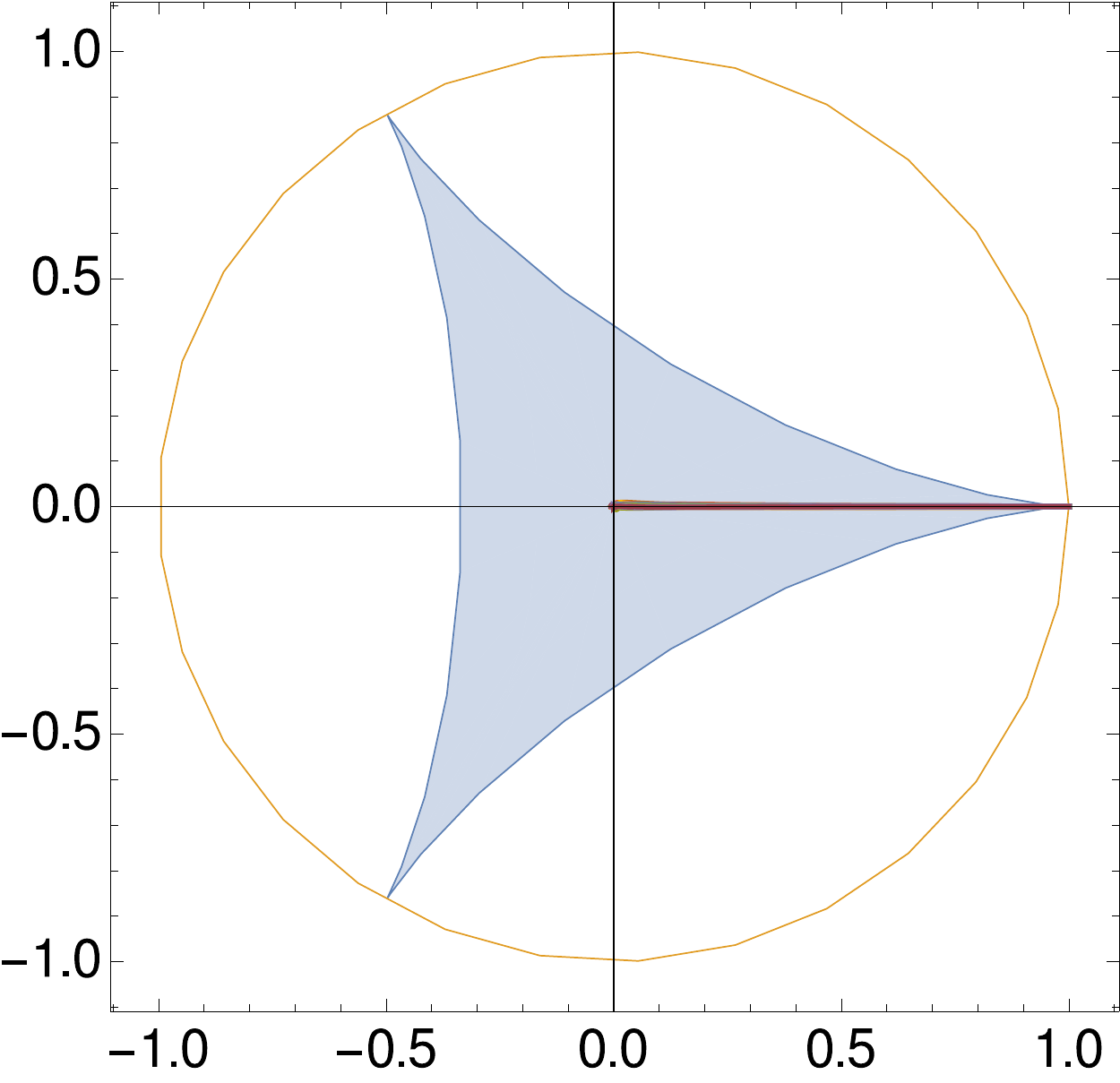} \\
     {\diagram[height=4cm]{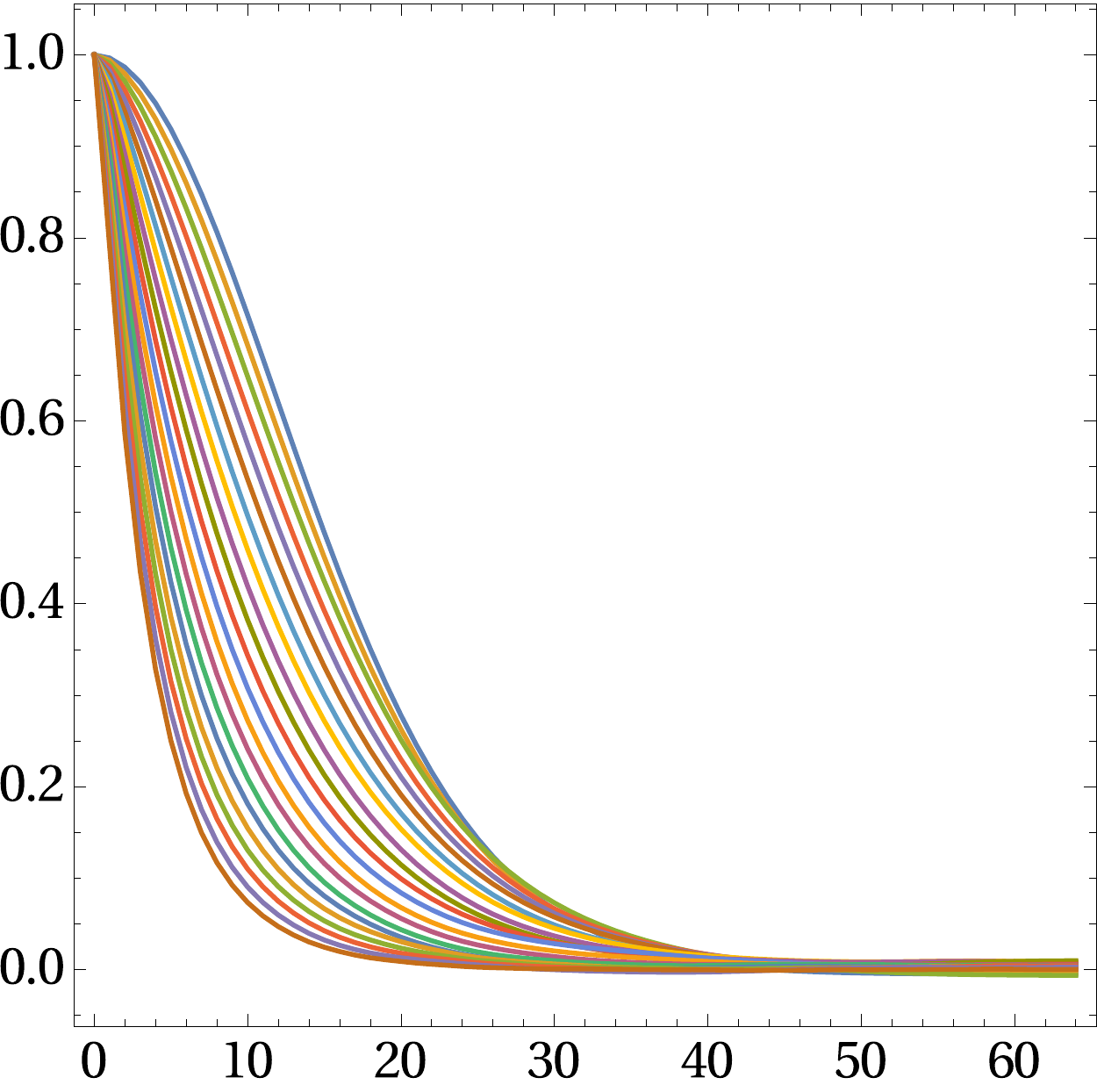}\hspace{5.5mm}} \u
     \diagram[height=4cm]{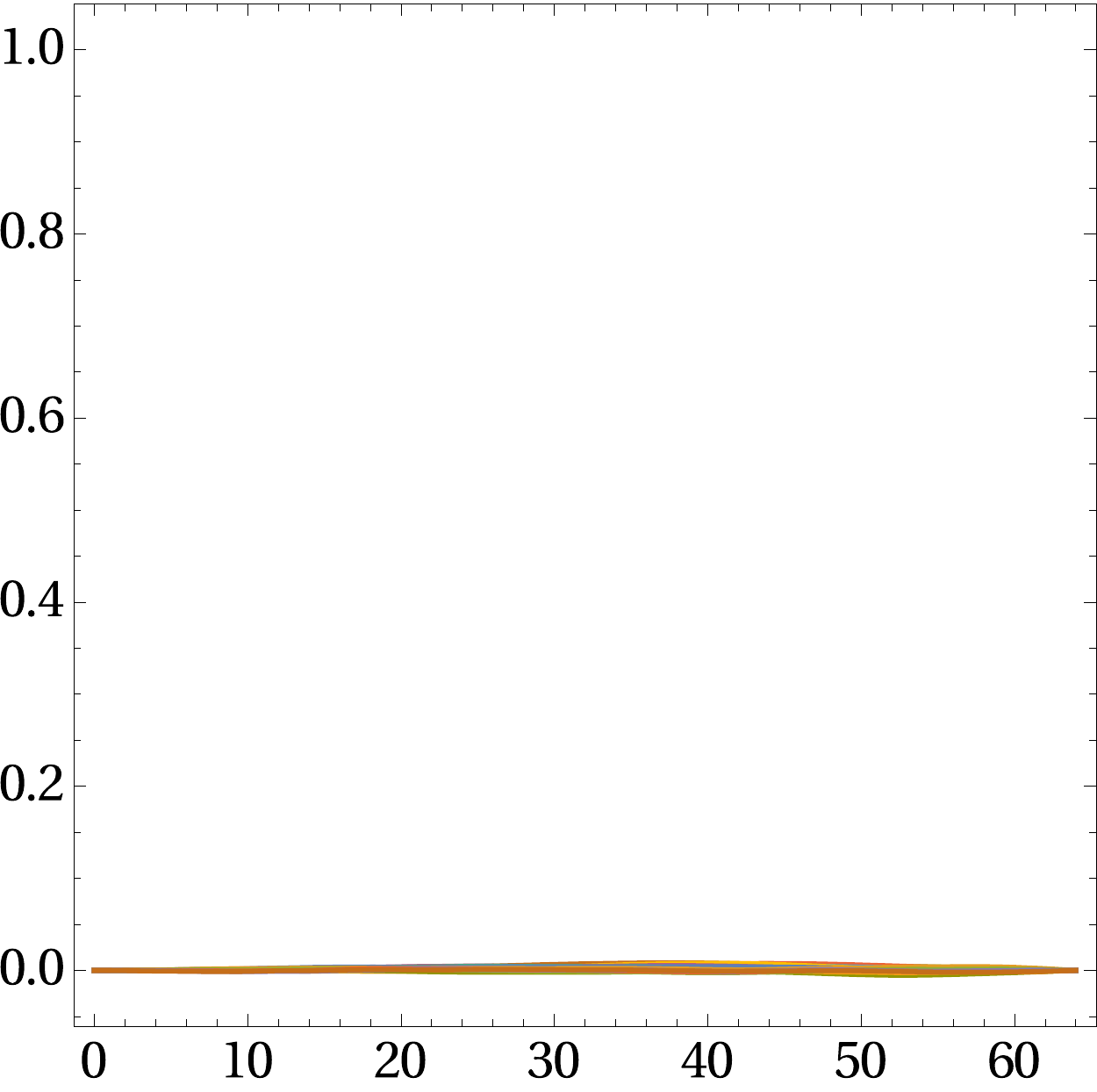} \\
     };
\node[anchor=east] at ($(m-1-1.north east)+(-.9,-.35)$) {\resizebox{!}{.2cm}{$\tr( U_{\bm x} U^\dagger_{\bm y})/\nc\in \mathbb C$}};
 \node at ($(m-1-2.north east)+(-1.2,-.35)$) {\resizebox{!}{.2cm}{$\langle \tr( U_{\bm x} U^\dagger_{\bm y})\rangle/\nc\in \mathbb C$}};
  \node[anchor=east] at ($(m-2-1.north east)+(-.8,-.35)$) {\resizebox{!}{.2cm}{$\mathsf{Re} \langle \tr( U_{\bm
    x} U^\dagger_{\bm y})\rangle/\nc$}};
\node at ($(m-2-2.north east)+(-1.2,-.35)$) {\resizebox{!}{.2cm}{$\mathsf{Im} \langle \tr( U_{\bm
    x} U^\dagger_{\bm y})\rangle/\nc$}};
\node at ($(m-2-1.south)+(0,-.2)$) {\resizebox{!}{.2cm}{$|\bm x-\bm y|\Lambda$}};
\node at ($(m-2-2.south)+(0,-.2)$) {\resizebox{!}{.2cm}{$|\bm x-\bm y|\Lambda$}};
\end{tikzpicture}
}
\caption{Leading-order Langevin simulations with curves for different
  $Y$ values at $\nc=3$ without an odderon admixture. The top row
  shows a density histogram of individual trace values (left) and
  averages (right) for configurations pulled from the ensembles
  through some $Y$ range. The dipole $\tr(U_{\bm x} U^\dagger_{\bm
    y})/\nc$ becomes $1$ in the short-distance limit. This is why the
  distribution exhibits a strong maximum there. Note that the
  configurations cover much of the allowed range with only a small
  fraction falling near the maximally anti-correlated corners at
  $e^{\pm i 2\pi/3}$. The density distribution is symmetric under
  reflection about the real axis. The averages fall into the real
  interval $[0,1]$, despite the fact that many individual
  configurations show negative real parts. The bottom row shows real
  and imaginary parts of the correlator averages for a number of $Y$
  values. The real part exhibits the familiar approach to scaling
  (curves move ``left'' with increasing $Y$), and an imaginary part
  that is zero within good accuracy with only small fluctuations
  visible.}
  \label{fig:JIMWLK-sim-no-odd}
\end{figure}

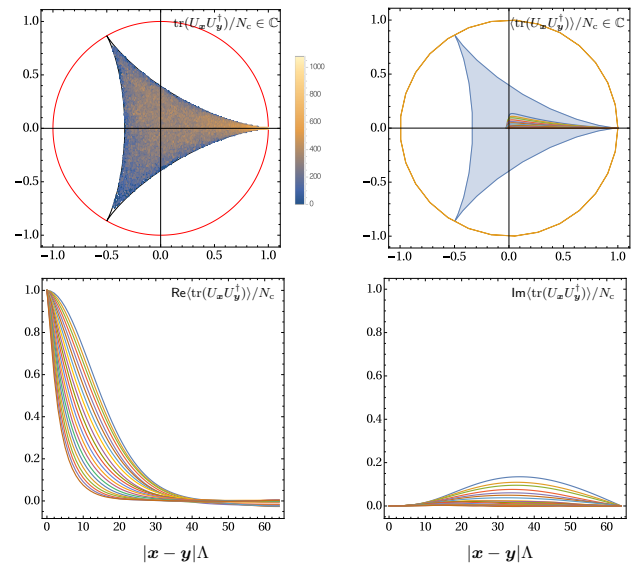
\begin{figure}[tb]
  \centering
\resizebox{\linewidth}{!}{\begin{tikzpicture}
 \matrix(m)[matrix of math nodes,align=right,column sep=.6em,ampersand replacement=\u]
   { \diagram[height=4cm]{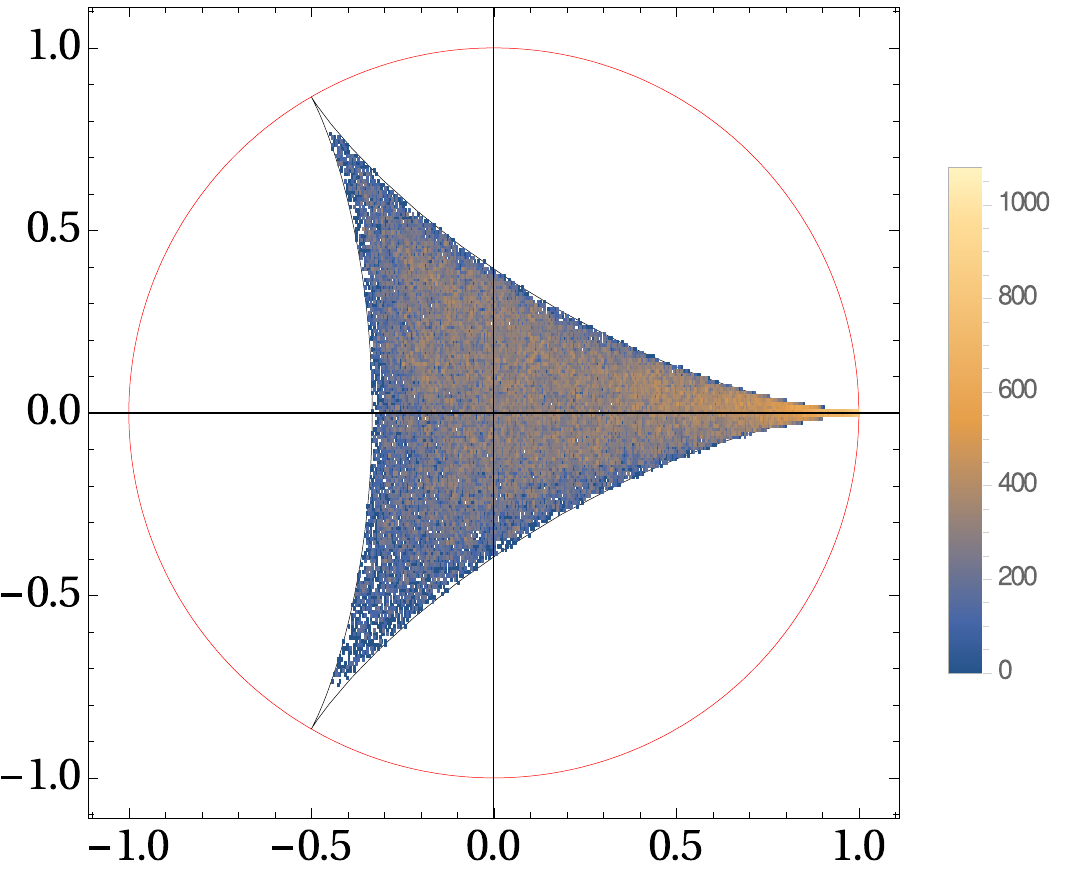} \u
     \diagram[height=4cm]{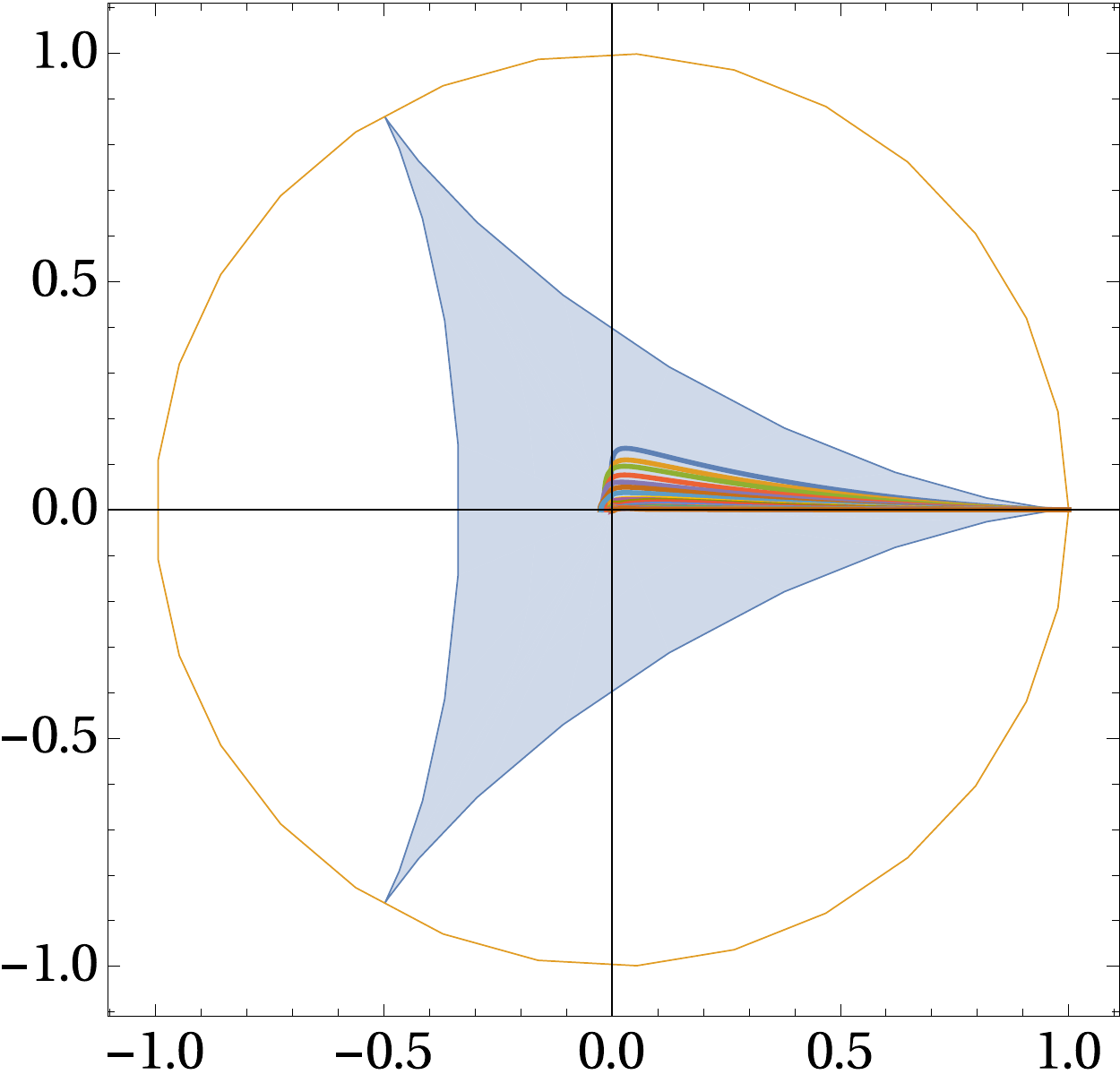} \\
     {\diagram[height=4cm]{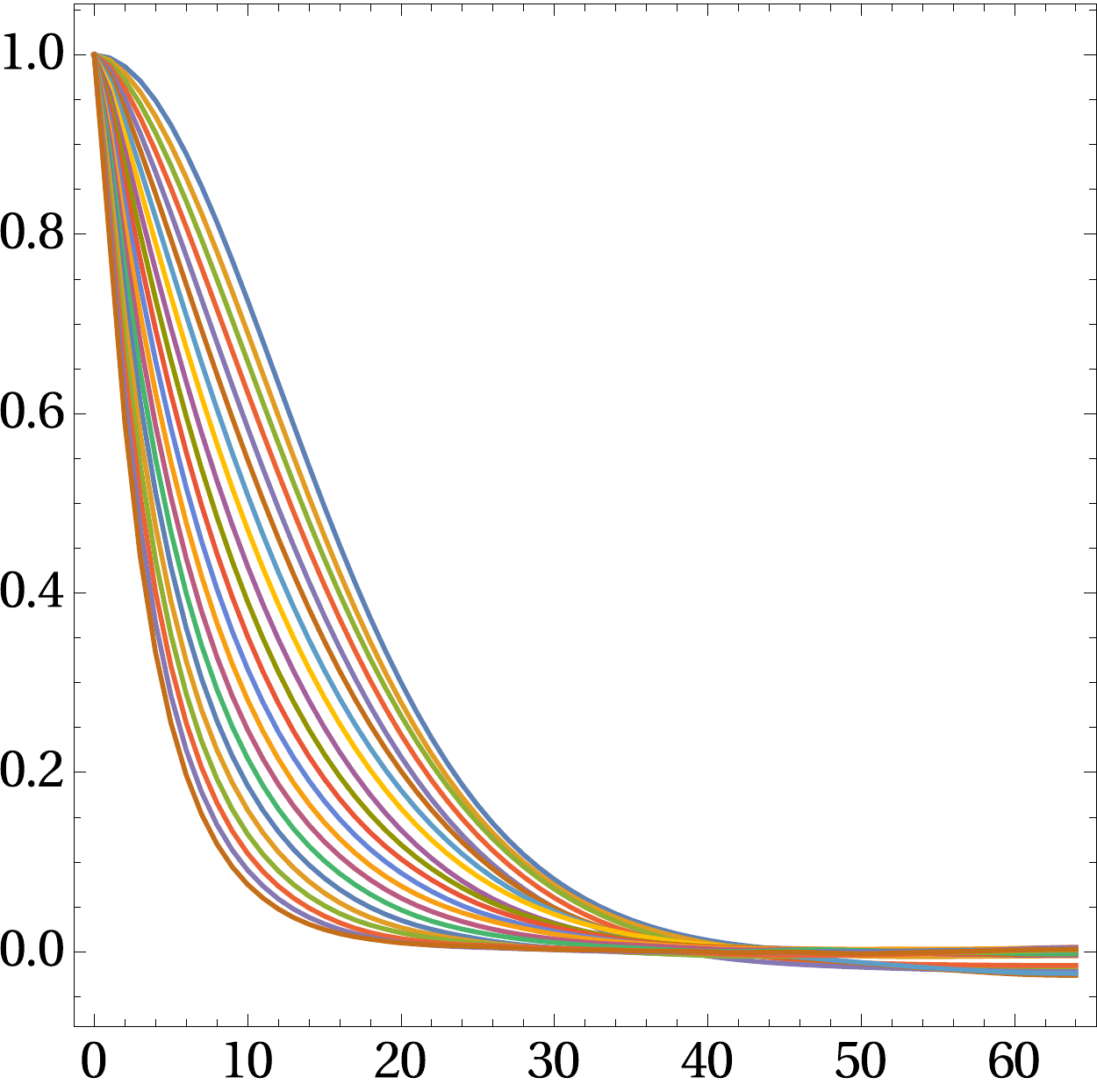}\hspace{5.5mm}} \u
     \diagram[height=4cm]{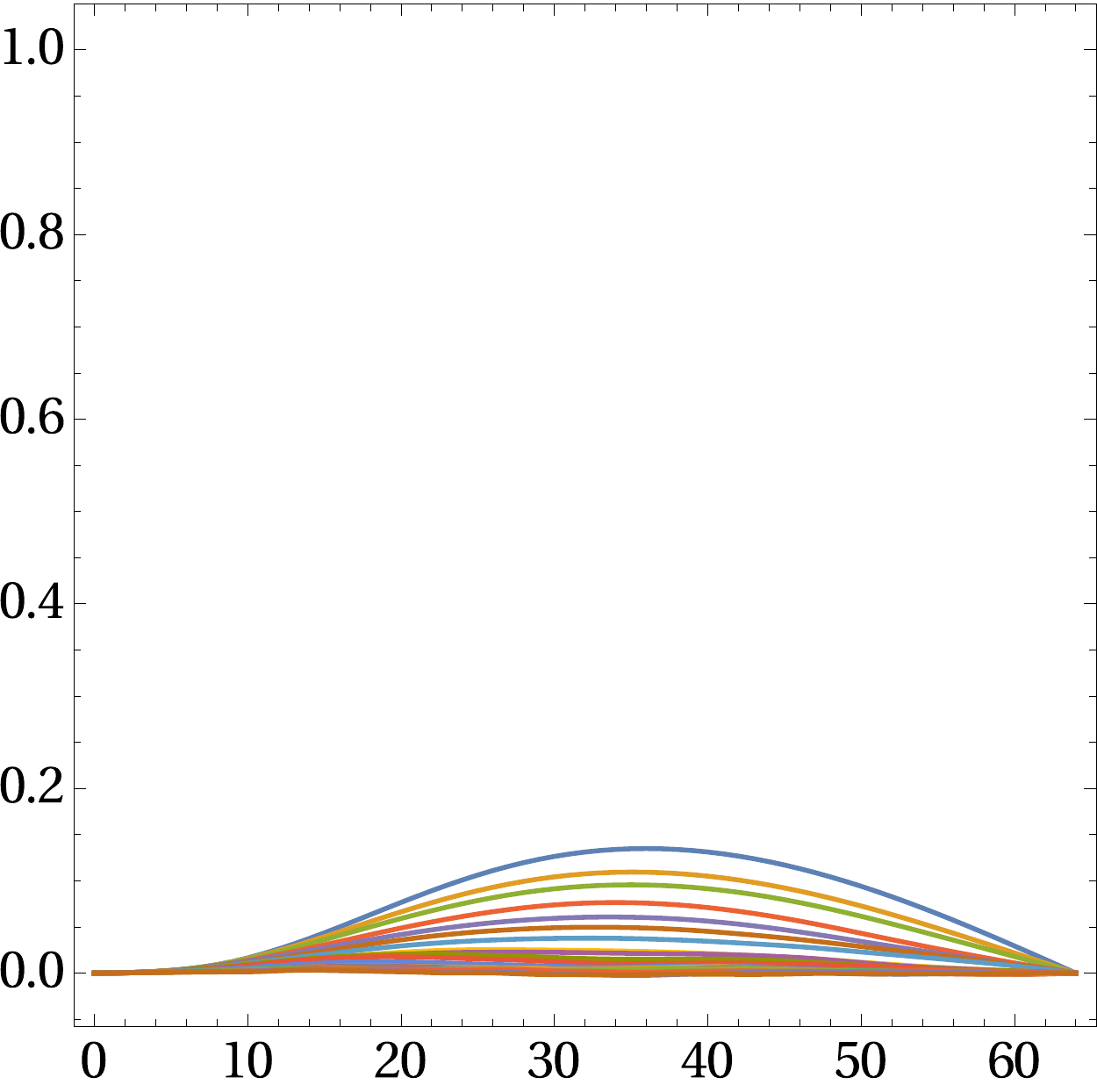} \\
     };
\node[anchor=east] at ($(m-1-1.north east)+(-.9,-.35)$) {\resizebox{!}{.2cm}{$\tr( U_{\bm x} U^\dagger_{\bm y})/\nc\in \mathbb C$}};
 \node at ($(m-1-2.north east)+(-1.2,-.35)$) {\resizebox{!}{.2cm}{$\langle \tr( U_{\bm x} U^\dagger_{\bm y})\rangle/\nc\in \mathbb C$}};
  \node[anchor=east] at ($(m-2-1.north east)+(-.8,-.35)$) {\resizebox{!}{.2cm}{$\mathsf{Re} \langle \tr( U_{\bm
    x} U^\dagger_{\bm y})\rangle/\nc$}};
\node at ($(m-2-2.north east)+(-1.2,-.35)$) {\resizebox{!}{.2cm}{$\mathsf{Im} \langle \tr( U_{\bm
    x} U^\dagger_{\bm y})\rangle/\nc$}};
\node at ($(m-2-1.south)+(0,-.2)$) {\resizebox{!}{.2cm}{$|\bm x-\bm y|\Lambda$}};
\node at ($(m-2-2.south)+(0,-.2)$) {\resizebox{!}{.2cm}{$|\bm x-\bm y|\Lambda$}};
\end{tikzpicture}
}
\caption{Leading-order Langevin simulations with curves for different
  $Y$ values at $\nc=3$ with a moderate odderon admixture. The top row
  shows a density histogram of individual trace values (left) and
  averages (right) for configurations pulled from the ensembles
  through some $Y$ range. Note that as in
  Fig.~\ref{fig:JIMWLK-sim-no-odd}, the configurations cover much of
  the allowed range with only a small fraction falling near the
  maximally anti-correlated corners at $e^{\pm i 2\pi/3}$. The density
  distribution now shows a small bias towards positive real parts that
  leads to a nontrivial imaginary part in the averages. These move
  towards the real axis as $Y$ increases. The bottom row shows real
  and imaginary parts of the correlator averages for a number of $Y$
  values. The real part now shows a small amount of anti-correlation
  which is erased quickly. The overall trend is an approach to scaling
  behavior very similar to that of the odderon free simulation of
  Fig.~\ref{fig:JIMWLK-sim-no-odd}. The imaginary part is small and
  erased in place as $Y$ increases. The two contributions behave in a
  qualitatively different manner: approach to scaling for the real
  part (the pomeron) decay for the imaginary part (the odderon).}
  \label{fig:JIMWLK-sim-weak-odd}
\end{figure}

\begin{figure}[tb]
  \centering
\resizebox{\linewidth}{!}{
\begin{tikzpicture}
 \matrix(m)[matrix of math nodes,align=right,column sep=.6em,ampersand replacement=\u]
   { \diagram[height=4cm]{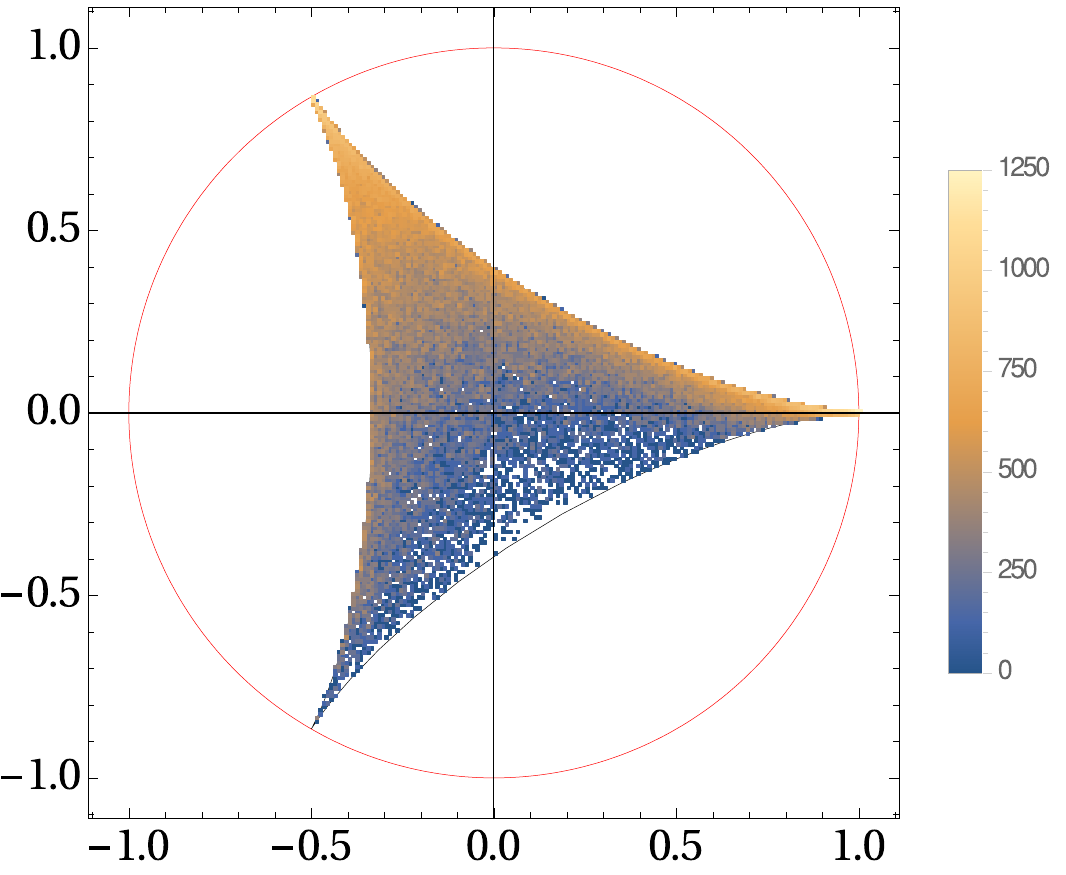} \u
     \diagram[height=4cm]{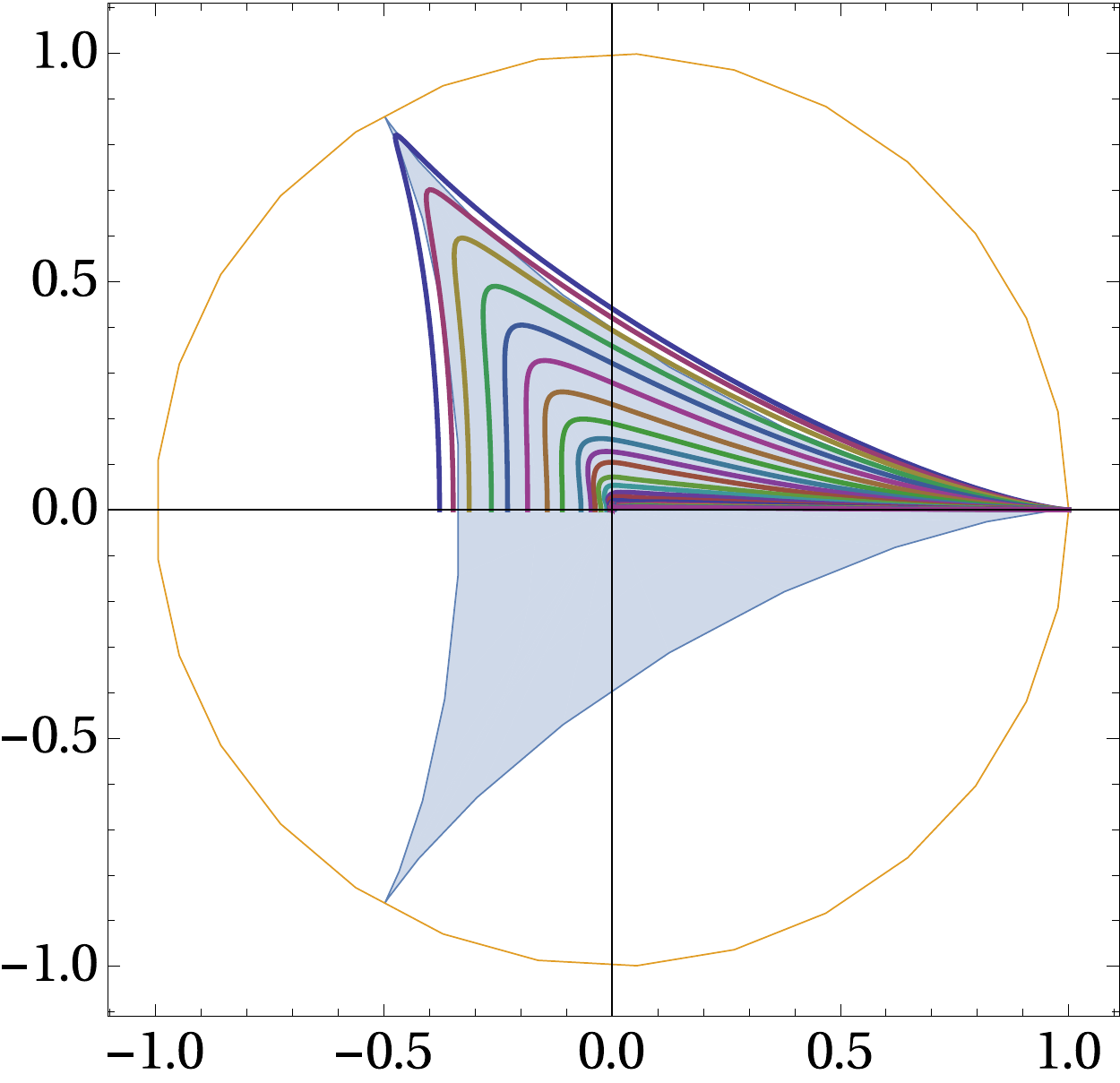} \\
     {\diagram[height=4cm]{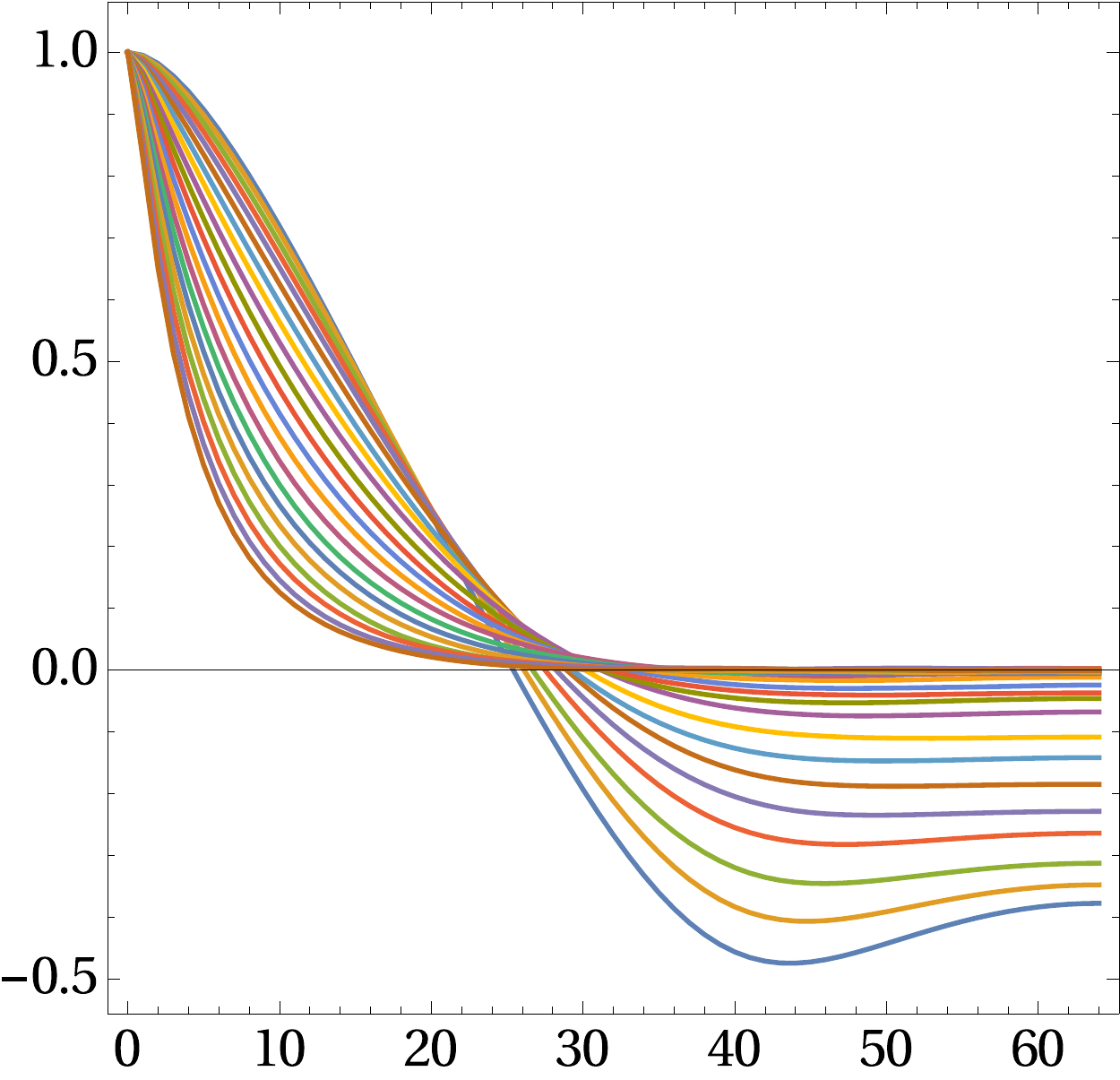}\hspace{5.5mm}} \u
     \diagram[height=4cm]{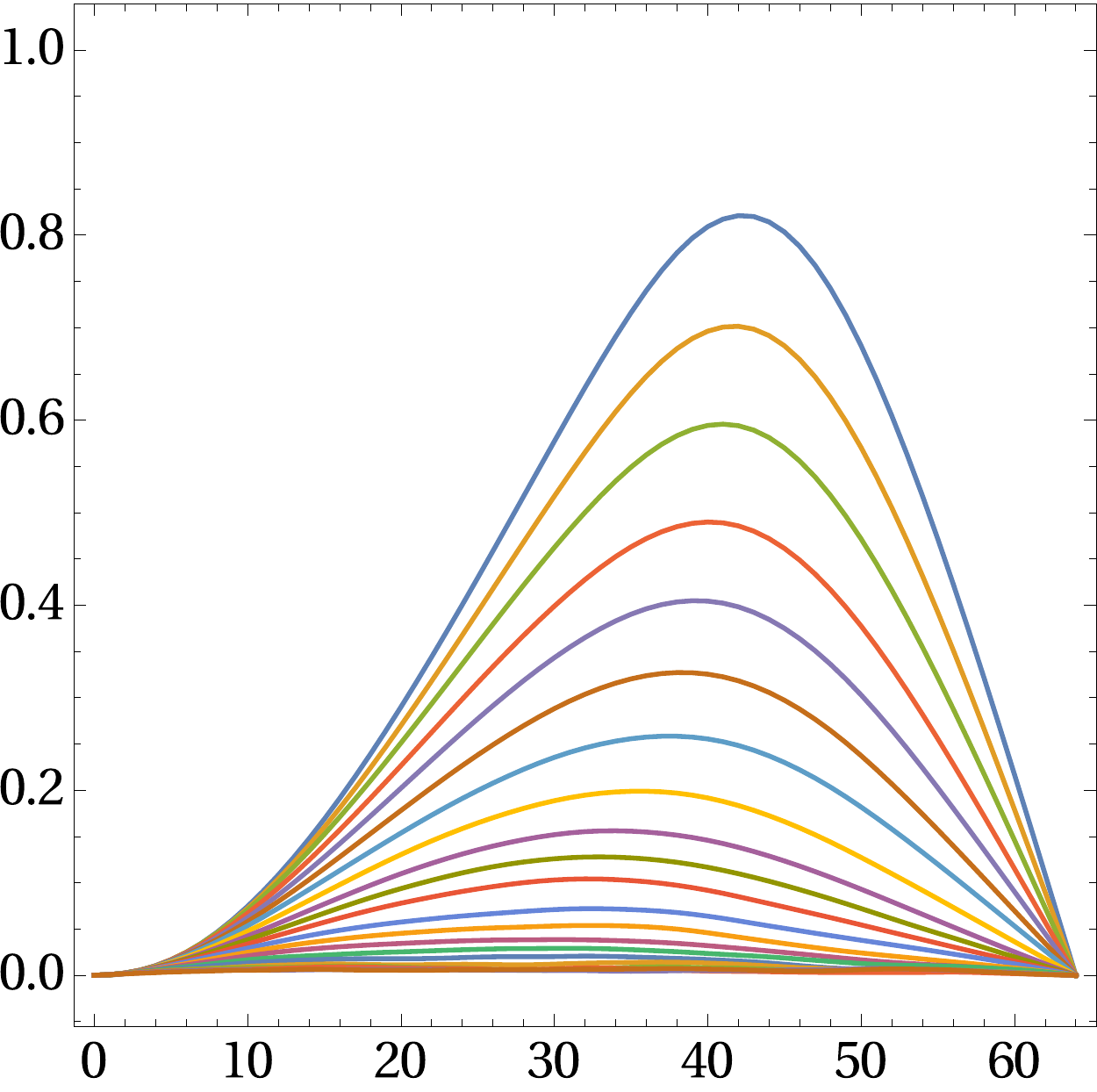} \\
     };
\node[anchor=east] at ($(m-1-1.north east)+(-.9,-.35)$) {\resizebox{!}{.2cm}{$\tr( U_{\bm x} U^\dagger_{\bm y})/\nc\in \mathbb C$}};
 \node at ($(m-1-2.north east)+(-1.2,-.35)$) {\resizebox{!}{.2cm}{$\langle \tr( U_{\bm x} U^\dagger_{\bm y})\rangle/\nc\in \mathbb C$}};
  \node[anchor=east] at ($(m-2-1.north east)+(-.8,-.35)$) {\resizebox{!}{.2cm}{$\mathsf{Re} \langle \tr( U_{\bm
    x} U^\dagger_{\bm y})\rangle/\nc$}};
\node at ($(m-2-2.north east)+(-1.2,-.35)$) {\resizebox{!}{.2cm}{$\mathsf{Im} \langle \tr( U_{\bm
    x} U^\dagger_{\bm y})\rangle/\nc$}};
\node at ($(m-2-1.south)+(0,-.2)$) {\resizebox{!}{.2cm}{$|\bm x-\bm y|\Lambda$}};
\node at ($(m-2-2.south)+(0,-.2)$) {\resizebox{!}{.2cm}{$|\bm x-\bm y|\Lambda$}};
\end{tikzpicture}
}
\caption{Leading-order Langevin simulations with curves for different
  $Y$ values at $\nc=3$ with a maximized odderon admixture. The layout
  repeats that of
  Figs.~\ref{fig:JIMWLK-sim-no-odd} and~\ref{fig:JIMWLK-sim-weak-odd}.
The density
  histogram now shows a second maximum at $e^{i 2\pi/3}$, a strong
  anti-correlating distortion of the initial condition. This manifests
  itself in averages that (initially) push outside the hypocycloid into
  the triangle connecting the 3rd roots of unity, as discussed in the
  text. With this come strong anti-correlations in the real parts in
  the initial condition. These features that are extreme in the
  initial conditions nevertheless are erased during evolution which
  again approaches scaling form. The imaginary parts are maximized in
  the initial condition beyond what we expect to be physical, but
  the overall behavior is the same as for a moderate odderon
  admixture: the odderon contribution is erased in place.}
  \label{fig:JIMWLK-sim-strong-odd}
\end{figure}
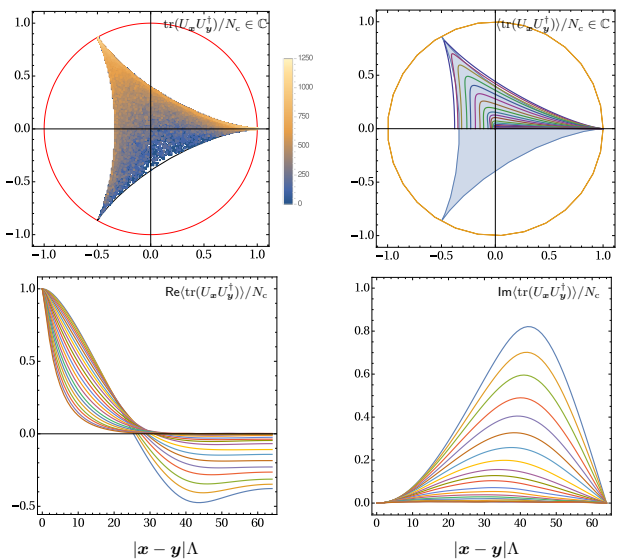

\paragraph{No odderon in the initial state:}

The ensemble is generated to follow Eq.~\eqref{eq:final_re} with no
distortion applied so that the average
\begin{align}
  \label{eq:pomeron-simul}
  \langle \tr(U_{\bm x}
  U^\dagger_{\bm y}) \rangle(Y)/\nc \in[0,1] \text{ for all $Y$.}
\end{align}
The imaginary part vanishes in the initial condition and none is
generated during evolution.  This can likely be traced back to the
adjoint nature of the Wilson lines ``dressing'' the Gaussian noise in
the Langevin version of the leading-order JIMWLK equation.  Note that
this holds despite the fact that individual configurations $\tr(U_{\bm
  x} U^\dagger_{\bm y})/\nc$ occur anywhere inside the hypocycloid
allowed by the constraints discussed in
Sec.~\ref{sec:group-theory-constr-corr}. Plots illustrating this
simulation are shown in Fig.~\ref{fig:JIMWLK-sim-no-odd}. The top left
displays a density histogram of configurations $\tr(U_{\bm x}
U^\dagger_{\bm y})/\nc$ which almost fill the whole allowed
region. Note that the density is lowest at the anti-correlated cusps
corresponding to $e^{\pm i 2\pi/3}\mathbbm{1}$, and that the
distribution is symmetric under a reflection along the real axis. As a
consequence, Eq.~\eqref{eq:pomeron-simul} is satisfied for all $Y$, as
shown top right. The plots in the second row show real and imaginary
parts as a function of dipole size.\footnote{Units can only be
  assigned after a data fit, which is not the goal of this
  publication.} The real part shows the familiar decay of the
correlation length $\rs(Y)$ and will develop scaling behavior if
allowed to evolve far enough.

\paragraph{A moderate odderon admixture to the initial state:} Plots
illustrating this simulation are shown in
Fig.~\ref{fig:JIMWLK-sim-weak-odd}. The density histogram for
individual configurations $\tr(U_{\bm x} U^\dagger_{\bm y})/\nc$ shows
a bias towards a positive imaginary part, which is confirmed by the
averages $\langle \tr(U_{\bm x} U^\dagger_{\bm y}) \rangle(Y)/\nc$
shown top right. Along with the appearance of an imaginary
part, one observes that the real part shows negative values --
anti-correlations appear. This simulation is qualitatively close to
what we have discussed in the context of the truncations with a
perturbative boundary condition, although no effort was made to create
a literal $r^3$ behavior for the initial condition.

The striking feature of this is the qualitatively different behavior
seen in the $Y$ dependence of real and imaginary parts. While the real
part does exhibit a small anti-correlation for $r \gtrsim \rs \equiv 1/\qs$, the evolution of
$\rs$ and the approach to scaling are affected very little by the
presence of an imaginary part. The imaginary part is \emph{not}
characterized by a moving $Y$-dependent scale; it shows a single,
clearly developed maximum that remains largely at the same distance
scale as $Y$ changes. The dominating feature of evolution is that the
height of the maximum shrinks -- the odderon is erased in place.

\paragraph{A maximized odderon admixture in the initial state:}
Plots illustrating this simulation are shown in
Fig.~\ref{fig:JIMWLK-sim-strong-odd}. The initial condition is
maximally distorted and the distribution of trace values has developed
a second maximum in the maximally anti-correlated region at $e^{i
  2\pi/3}\mathbbm{1}$.

This is an extreme case that we have included to illustrate  the
features of JIMWLK evolution. 
%Since one of the prerequisites of the
%derivation of JIMWLK evolution is that all contributions come from
%perturbatively short distances, we expect that there is no experiment
%where JIMWLK is applicable in which a non-perturbative mechanism is
%available to initialize such an initial condition. 
Due to the extreme initial condition, the dipole averages start to
fall outside the hypocycloids and the lattice is too small for the
real part to reach zero at large distances. Nevertheless, the behavior
of the imaginary part still mirrors that of the realistic odderon
admixture: the odderon does not move, it decays in place. We conclude
that both the scale shift for the real part and the fixed scale decay
for the imaginary part of the dipole correlator are genuine features
of JIMWLK evolution, irrespective of the details of the initial
condition.

\section{Discussion}
\label{sec:disc}
In this paper we have shown how to derive the high energy evolution
equations for the odderon amplitude using a consistent 3-point
truncation of the Balitsky hierarchy. In the large-$\nc$ limit our
solution recovers that of~\cite{Kovchegov:2003dm,
  Hatta:2005as}. Decomposing the amplitudes in terms of Fourier
harmonics yields an infinite series of coupled equations. Due to the
nonlinear relation between the real and imaginary parts of the
physical scattering amplitude, and the solutions of the corresponding
BK equation (see \eqs\nr{eq:bktogtre} and~\nr{eq:bktogtim}), the
presence of any odderon component introduces an angular dependence at
all harmonics $n$ into the scattering amplitude. This correlation
vanishes in the large-$\nc$ limit and would therefore not have been
accessible previously. This coupling between the odderon amplitude and
higher harmonics could allow for a quantitative experimental access to
the odderon component in multi-particle correlations in future precise
DIS experiments~\cite{Accardi:2012qut}.

By truncating the harmonic series to the first nontrivial terms for
both the pomeron and odderon parts, one gets a closed nonlinear
equation for the energy dependence of the odderon amplitude. We have
presented the first numerical solution to this equation available in
the literature. We have then completely independently confirmed these
results with a numerical lattice solution of the full JIMWLK equation
with an initial condition containing an odderon component. Both of
these numerical calculations have confirmed the earlier analytical
conjectures based on the linear BFKL limit~\cite{Kovchegov:2003dm,
  Hatta:2005as}, showing that the odderon amplitude decreases with
increasing collision energy. This observation justifies the truncation
of the higher harmonic terms used in the BK-like numerical evaluation
(the JIMWLK simulation needs no such truncation).

The odderon appears in full JIMWLK evolution \emph{only} by preparing
its initial conditions in a way that breaks rotational invariance in
the transverse plane. This mirrors directly what is happening in a
measurement process: The ensemble we average over needs to break
rotation invariance in both cases. In an experiment this can be
achieved by measuring polarizations, as is done in an STSA
measurement.  The total cross section as an average over a fully
symmetric unbiased set of events will not be able to couple to the
odderon at all. (Not even through mixing of real and imaginary parts
during evolution -- there are no average imaginary parts to begin with
and none are generated during evolution.) A targeted observable on the
other hand, like STSA, can give access to the imaginary parts of
Wilson line correlators directly. Once a preselected ensemble of
events generates such imaginary parts in the average, they also impact
the real parts and may even trigger anti-correlations there. The
mechanism for this mixing in full JIMWLK lies in the non-linear nature
of the evolution equation. Judging from the agreement between full
JIMWLK evolution and the numerical results in the 3-point exponential
trunction it would appear that this nonlinear mechanism is well
captured in the truncated theory.

The coupling of the odderon component may not have a strong effect on
high energy asymptotical behavior even of targeted observables -- as
we have demonstrated, the odderon still decays with energy even beyond
the linear BFKL approximation. At realistic collider energies the
question is still open, but only if we consider tailored experiments
of sufficient accuracy. If the presence of an odderon contribution
comes with anti-correlations in the real parts in its initial
condition this might open new avenues to access them
experimentally.
% \\ \smash{\includegraphics[height=3cm]{figs_final/out}}
%  \smash{\includegraphics[height=3cm]{figs_final/odd-1-reim}}

\begin{acknowledgments}
  We are grateful to H. M\"antysaari for help with BK numerics.
  T.~L.\ is supported by the Academy of Finland, projects 267321 and
  273464. A.R. has been supported by a scholarship from the Center for
  International Mobility, Finland.  K.R. is supported by the Academy
  of Finland project 267286.  H.W. is supported by NRF
  under CPRR grant nr 90509.
% This work used computing resources from
% CSC -- IT Center for Science in Espoo, Finland.
\end{acknowledgments}

\bibliography{spires}
% \bibliography{master}
\bibliographystyle{JHEP-2modlong}

\end{document}